\documentclass[aps,prd,nofootinbib,showpacs,amsmath,amssymb,notitlepage,10pt,preprintnumbers]{revtex4-1} 

\usepackage{graphicx}
\usepackage{amsfonts}
\usepackage{amssymb}
\usepackage{mathtools}
\usepackage{dsfont}
\usepackage{longtable}
\usepackage{dcolumn}
\usepackage{multirow}
\usepackage{bm}		
\usepackage{float}
\usepackage{ragged2e}

\newcommand{\tabref}[1]{Table \ref{tab:#1}}		
\newcommand{\figref}[1]{Fig.\ \ref{fig:#1}}			
\newcommand{\sectref}[1]{Sec.\ \ref{sect:#1}}		
\newcommand{\appref}[1]{Appendix \ref{app:#1}}	
\newcommand{\lb}{\left}						
\newcommand{\rb}{\right}						
\newcommand{\eten}[1]{\times 10^{#1}}			
\newcommand{\up}[1]{\textsuperscript{#1}}		

\begin{document}

\title{Gamma-ray constraints on dark-matter annihilation \\ to electroweak gauge and Higgs bosons}
\author{Michael A.\ Fedderke} \email{mfedderke@uchicago.edu}
\author{Edward W.\ Kolb} \email{Rocky.Kolb@uchicago.edu} 
\author{Tongyan Lin} \email{tongyan@kicp.uchicago.edu}
\author{Lian-Tao Wang} \email{liantaow@uchicago.edu}

\affiliation{Enrico Fermi Institute and Kavli Institute for Cosmological Physics, the University of Chicago, Chicago, Illinois \ \ 60637-1433 }

\begin{abstract}
Dark-matter annihilation into electroweak gauge and Higgs bosons results in $\gamma$-ray emission.   We use observational upper limits on the fluxes of both line and continuum $\gamma$-rays from the Milky Way Galactic Center and from Milky Way dwarf companion galaxies to set exclusion limits on allowed dark-matter masses. (Generally, Galactic Center $\gamma$-ray line search limits from the Fermi-LAT and the H.E.S.S.\ experiments are most restrictive.)  Our limits apply under the following assumptions: a) the dark matter species is a cold thermal relic with present mass density equal to the measured dark-matter density of the universe; b) dark-matter annihilation to standard-model particles is described in the non-relativistic limit by a single effective operator ${\cal O} \propto J_{DM}\cdot J_{SM}$, where $J_{DM}$ is a standard-model singlet current consisting of dark-matter fields (Dirac fermions or complex scalars), and $J_{SM}$ is a standard-model singlet current consisting of electroweak gauge and Higgs bosons; and c) the dark-matter mass is in the range 5 GeV to 20 TeV.   We consider, in turn, the 34 possible operators with mass dimension 8 or lower with non-zero $s$-wave annihilation channels satisfying the above assumptions. Our limits are presented in a large number of figures, one for each of the 34 possible operators; these limits can be grouped into 13 classes determined by the field content and structure of the operators. We also identify three classes of operators (coupling to the Higgs and $SU(2)_L$ gauge bosons) that can supply a 130 GeV line with the desired strength to fit the putative line signal in Fermi data, while saturating the relic density and satisfying all other indirect constraints we consider.
\end{abstract}

\pacs{98.70.Cq, 95.35.+d, 95.30.Cq, 95.55.Ka, 95.85.Ry, 98.35.Jk, 95.85.Pw}

\date{\today}

\maketitle

\section{Introduction}
\label{sect:intro}

Until the nature of dark matter and dark energy is understood, the remarkable success of the standard model of cosmology in accounting for observations will be less than completely satisfying.  The most popular explanation for dark matter (DM) is that it is an early-universe relic in the form of an undiscovered, electrically neutral, non-hadronic, stable species of elementary particle.  Of the many possibilities for the origin and nature of this purported new elementary particle, one that is particularly amenable to experimental and observational tests is that the new species is a \textit{cold thermal relic.}  

The cold thermal relic scenario assumes that the new particle species was established in local thermodynamic equilibrium when the temperature of the universe was larger than the mass of the particle.  The processes establishing the initial equilibrium abundance are assumed to be the annihilation of the new species into standard-model (SM) particles and the production of the new species from initial SM-particle states.  Then, as the universe cools to temperatures below the dark-matter mass, the relative abundance of the new species (relative to, say, the entropy density) ``freezes out'' as the rate of processes keeping the species in equilibrium falls below the expansion rate of the universe.  The relative freeze-out abundance, and therefore the present mass density, is thus related to the cross section of dark-matter annihilation into SM particles.

Observationally the ratio of the present average dark-matter mass density to the critical density is well determined to be $\Omega_{DM}h^2\simeq0.12$ \cite{Planck1303}. The requirement that this mass density is saturated by the mass density of the cold thermal relic determines the annihilation cross section (and, in some cases, also the mass) of the new species.  Since the mass and annihilation cross section required are both of order the weak scale, the cold thermal relic is known as a weakly interacting massive particle, or WIMP. 

A key feature of the WIMP hypothesis is the requisite DM--SM coupling, and its close relationship to the present dark-matter mass density. Indeed, it is knowledge of the coupling of WIMPs to SM particles that provides the experimental and observational avenues that could lead to the discovery of the nature of the dark matter:  detection of WIMPs through their present-day scattering with SM particles ($\mathrm{WIMP}+\mathrm{SM}\rightarrow \mathrm{WIMP}+\mathrm{SM}$), known as ``direct detection''; detection of WIMPs through the observation of the present-day annihilation products of WIMPs into SM particles ($\mathrm{WIMP}+\mathrm{WIMP}\rightarrow \mathrm{SM}+\mathrm{SM}$), known as ``indirect detection''; and production and detection of WIMPs at particle colliders ($\mathrm{SM}+\mathrm{SM}\rightarrow \mathrm{WIMP}+\mathrm{WIMP}$). 

While the WIMP--SM coupling is a common refrain in all WIMP scenarios, there are a tremendous number of variations on this theme.  One possibility is that the WIMP is not the only new particle at the weak scale, and these additional new particles play a crucial role in the annihilation, production, or scattering of WIMPs.  In the other limit that the masses of any additional new particles are much greater than momenta in the processes of interest, we can integrate out the additional states and describe the WIMP--SM interaction in terms of an effective field theory (EFT).   We will work in this ``Maverick'' limit. 

Even within this context there are a large number of possibilities.  The first study assumed that the WIMP couples to SM fermions \cite{Beltran:2008xg}.  In this case direct detection \cite{Beltran:2008xg} and collider searches \cite{Beltran:2010ww,Goodman:2010yf,Goodman:2010ku,Fox:2011pm,Fox:2012ee,Rajaraman:2011wf,
Fox:2011fx,Fox:2011pm, ATLAS:2012ky,ATLAS-CONF-2012-147,Chatrchyan:2012me,CMS-PAS-EXO-12-048,Lin:2013sca} offer the best possibility for testing the models. 

Another class of possibilities which admits an EFT description is that dark matter couples to (SM) diboson final states of electroweak gauge and Higgs bosons \cite{CKW}. Here we consider DM--SM couplings of the form $\Lambda^{-n} J_{DM}\cdot J_{SM}$, where $J_{DM}$ is a standard-model-singlet current consisting of dark-matter fields (assuming Dirac fermions or complex scalars), $J_{SM}$ is a standard-model-singlet current consisting of electroweak gauge and Higgs bosons, and $\Lambda$ is the mass scale (``suppression scale'') of the effective field theory.  Moreover, we only consider the case that the dark-matter field is a singlet under $SU(2)_L \times U(1)_Y$. 

\citet{CKW} (hereafter ``CKW'') listed all such operators with diboson final states up to mass dimension 8, and presented detailed calculations of dark-matter annihilation cross sections to all possible final states, including $\gamma\gamma$, $\gamma Z$, $\gamma h$, $ZZ$, $Zh$, $W^+W^-$, $hh$ and $\bar{f}f$.\footnote{These cross sections are listed in Tables VI-XX of CKW. We will refer to an operator studied in CKW by a table number, and an operator number: for instance, operator VI-3 is the third operator of Table VI in CKW, in this case, $\phi^\dagger\phi W^a_{\ \mu\nu}W^{a\, \mu\nu}$.  Our notation for field operators will follow the notation in CKW.} In order to consistently describe this list of possible channels, they required $J_{SM}$ to be invariant under $SU(2)_L \times U(1)_Y$.  Of the 50 such operators they enumerated, they demonstrated that 34 of them have at least one $s$-wave annihilation channel for at least some values of the dark-matter mass $M$, and hence could produce a detectable indirect detection signal. ($p$-wave annihilation is suppressed by $v^2 \sim 10^{-6}$ relative to $s$-wave annihilation at the present epoch.)

DM annihilation to diboson final states leads in all cases to prompt production of $\gamma$-rays, including possibly monochromatic photons. This leads naturally to the prospect that there could be non-negligible present-day DM-annihilation photon fluxes emanating from regions of high dark-matter density: in particular, from the Galactic Center of the Milky Way, or from Milky Way companion dwarf galaxies. The aim of the present work is to utilize the experimentally measured upper bounds on these photon fluxes to place constraints on those EFT operators which have non-negligible $s$-wave annihilation cross sections.

More precisely, considering in turn each of the 34 EFT operators
mentioned above, we will derive constraints on the allowed dark-matter
mass $M$ assuming that the WIMP annihilates only via the operator in
question, and that it is a cold thermal relic which accounts for 100\%
of the measured DM density in the universe. We use the following
analyses to set limits: the Galactic Center (GC) Fermi-LAT line-search
limits for photon energies 5 to 300 GeV presented in
Ref. \cite{FERMI1305}; the H.E.S.S.\ GC line-search limits for photon
energies 500 GeV to 20 TeV presented in Ref. \cite{HESS1301}; the
analysis of the GC Fermi-LAT inclusive flux limits for photon energies
0.1 to 100 GeV presented in \citet{Hooper1209}; and the Fermi-LAT
Milky Way dwarf companion limits for photon energies 0.2 to 100 GeV
presented in Ref. \cite{FERMI1108}.  In addition, many of the
operators have one or more annihilation modes with monochromatic
photons; we investigate whether these operators can fit the possible
$\gamma$-ray line at $E_\gamma \approx$ 130 GeV observed in the
Fermi-LAT data, as first reported in \citet{Weniger1204}. In cases
where a reasonable fit to the photon line is possible, we apply the
limits presented in \citet{Wacker1207} for the ratio of
continuum-to-line emission for DM masses of 125 to 150 GeV.

Our limits are presented in a large number of figures, one for each of the 34 possible operators, in \sectref{results}.  The limits on the operators depend on available annihilation channels, which are determined by both the field content and structure of the operator, and the dark-matter mass. Based on qualitative differences in their predicted indirect detection signals, we group the operators into 13 classes, which are summarized in \tabref{classification} in \sectref{conclusion}. Overall, we observe that line searches give the most constraining limits for operators whose indirect detection signal has a significant contribution from one or more photon lines. Other limits, such as that from the continuum photons, are weaker but still useful for lighter dark matter masses. Although some operators are quite severely constrained  by the photon fluxes over a large region of parameter space, for the majority of the operators we find that most of the parameter space is still open or is only weakly constrained. Among all the operators considered here, we have also identified three classes which can simultaneously give the correct thermal relic abundance, account for the potential 130 GeV line signal, and be consistent with all the constraints from indirect searches which we have considered. More importantly, our results demonstrate that it is possible to set interesting limits on dark matter annihilations by combining various channels. They can also be used to infer additional predictions and learn about the nature of dark matter if a signal is observed in a particular channel. 

We also identify an interesting effect for fermion DM coupling via pseudo-vector $J_{DM}$ where the $p$-wave annihilation cross section is enhanced (relative to the $s$-wave) at large $M$ by coupling to the longitudinal mode of a massive vector boson in the final state. This actually has the effect, since saturation of $\Omega h^2$ fixes the total cross section at freeze-out, of suppressing the $s$-wave (i.e. present-day) annihilation cross section thereby alleviating constraints on these operators for large $M$.

Since there have been many works that discuss indirect-detection limits, it is appropriate for us to discuss why our analysis is new.  Our starting point is the assumption that low-energy WIMP annihilation processes can be described by an effective field theory with the assumptions discussed above.  Working within that framework, for each possible operator, the full $SU(2)_L \times U(1)_Y$ gauge invariance of $J_{SM}$ forces us to consider all possible final states.\footnote{For some operators, gauge invariance leads to the inclusion of SM fermions in the final state when $s$-channel SM vector boson exchange is involved.} Moreover, the requirement of gauge invariance determines the relative strength of possible annihilation channels. We expect the existence of multiple annihilation channels to offer much more discriminating power in the analysis. A related earlier work is \citet{Cotta:2012nj}, where a more limited set of operators (without including the Higgs and without requiring full $SU(2)_L$ gauge invariance) was considered. In addition, some of these operators have been discussed in Refs.\ \cite{Weiner:2012cb,Weiner:2012gm,Rajaraman:2012fu}, with an emphasis on explaining the 130 GeV line as well as a possible additional line at around 114 GeV \cite{Rajaraman:2012db}.

We emphasize that, in general, the EFT approach must be taken with some caution; the case at hand is no exception. Every relevant process involving WIMPs has a characteristic scale for the momentum transfer, and the accuracy of the EFT approach can be measured by the comparison between this scale and the suppression scale of the effective operators \cite{Shoemaker:2011vi,Busoni:2013lha,Buchmueller:2013dya}. In this sense, the approach would become less accurate in describing the freeze-out and present-day DM annihilation if the suppression scale is close to the dark-matter mass. As it happens, in order to give the correct thermal relic abundances, some of the operators we study are forced to have $\Lambda \sim M$ and, in principle, one therefore needs to delve into possible UV completions in these cases.  This caveat notwithstanding, given the large number of possible operators and final states we consider, we find it useful to use `na\"ively' the effective operator approach to give rough estimates.  Considering the large systematic astrophysical uncertainties in the indirect detection measurements we utilize, a more accurate description within a UV complete model would not change our conclusions qualitatively. 

The remainder of this paper is arranged as follows. In \sectref{prelim} we discuss some theoretical background on the WIMP relic density, DM halo profiles, annihilation photon fluxes, and continuum photon spectra from various DM annihilation channels. In \sectref{limits} we discuss the individual experimental limits we have worked with, summarizing how they were obtained and how we have interpreted them. The large number of plots giving the results of our work are given in Figs.\ \ref{fig:TabVI_OP1} to \ref{fig:TabXX_OP1} in \sectref{results}. We discuss the results and conclude in \sectref{conclusion}. There are a number of appendices: \appref{ICS} discusses in some detail the contribution to the photon spectrum due to inverse Compton scattering.  \appref{systematics} summarizes the magnitude of other systematic errors we did not take into account. Finally, \appref{exposure} discusses some further analysis we have performed on one of the literature sources we have consulted for the case where there are two partially resolved lines in the photon spectrum.

\section{Preliminaries}
\label{sect:prelim}

\subsection{Relic density}
\label{sect:relic}
It is a well known phenomenon \cite{EarlyUniverse} that a massive particle species capable of annihilation into less massive species, and which is initially in thermodynamic equilibrium in the early universe, cannot maintain an equilibrium abundance to arbitrarily low temperature as the universe expands. Eventually, annihilations become too infrequent\footnote{If the annihilation rate is $\Gamma \sim n \sigma v$ where $n$ is the DM number density and $\sigma$ is the annihilation cross section, the condition is roughly that $\Gamma < H$, where $H$ is the Hubble constant \cite{EarlyUniverse}.} owing to the dilution of the particles in the expanding volume of the universe, and creation of a pair of the new particle species becomes rare because it becomes exponentially rare to have collisions with sufficient center-of-mass energy.  This leads to a ``freeze out'' of the abundance of the new particle species to some relic abundance.  We shall assume that the measured average dark matter density of the universe, $\Omega_{DM} h^2 = 0.12$ \cite{Planck1303}, is entirely ascribable to the WIMP, which we will assume to be either a complex scalar or a Dirac fermion.

The phenomenon of thermal freeze-out is well described by the Boltzmann equation \cite{EarlyUniverse}, and given the necessary initial data and annihilation cross sections as a function of temperature, one can simply solve this equation numerically to find the relic abundance. There does, however, exist an approximate method which has the benefit of expressing the relic abundance in closed form. We shall assume that a non-relativistic approximation to the annihilation cross section can be written in the form $\langle \sigma v \rangle_{NR} = a + b v^2$, where $a$ and $b$ are constants, $v^2=6T/M$, and $\langle \cdots\rangle$ indicates the thermal average.\footnote{Although $\sigma$ is by itself the annihilation cross section, we shall for reasons of brevity frequently also refer to the quantity $\sigma v$ as the annihilation cross section.  Furthermore, we will only be interested in the cross section in the non-relativistic limit so we will drop the subscript ``$NR$.''  Annihilations relevant for indirect detection occur at very low velocity, so when we write $\sigma v$ for indirect detection we mean the velocity-independent part of the non-relativistic cross section ($a$ in the non-relativistic expansion $\sigma v = a + b v^2$).}  The DM relic density can then be given to \textit{ca}.\ 5\% accuracy by \cite{EarlyUniverse,CKW}
\begin{subequations}
\label{eq:relic_density}
\begin{align}
\Omega_{DM} h^2 & =  \frac{1.04 \times 10^9 \text{GeV}^{-1}}{M_{\text{Pl}} \sqrt{ g_{*}(x_F)}} \frac{x_F}{( a + 3b/x_F )} \times \lb\{ \begin{array}{ll} 1 & \text{for self-conjugate DM} \\ 2 & \text{for non-self-conjugate DM} \end{array} \rb. , \\
  & x_F \equiv \frac{M}{T_F} = \ln \lb[ c(c+2) g \sqrt{\frac{45}{8}} \frac{M M_{\text{Pl}}}{2\pi^3 \sqrt{g_{*}(x_F)} } \frac{ a+6b/x_F}{\sqrt{x_F}} \rb],  \label{eq:x_F}
\end{align}
\end{subequations}
where $x_F=M/T_F \sim \mathcal{O}(20)$ with $T_F$ the freeze-out temperature \cite{EarlyUniverse}, $M$ is the WIMP mass,
$M_{\text{Pl}} = 1.22\times10^{19}$ GeV is the Planck mass, the
numerical parameter $c$ is chosen such that $c(c+2) = 1$, and the
number of relativistic degrees of freedom at freeze-out is taken to be
$g_{*}(x_F) = 106$ \cite{CKW,EarlyUniverse}. For real or complex
scalar DM, $g=1$, and for both Majorana and Dirac fermion DM,
$g=2$. For the results in this paper we will assume non-self-conjugate
DM, since some of the operators under consideration vanish for
self-conjugate DM.

Working in an EFT framework and assuming that the DM annihilation proceeds through only one EFT operator $\mathcal{O}(x)$ of mass-dimension $d$, the term in the Lagrangian responsible for the annihilation can be written $\mathcal{L}(x) \supset \Lambda^{4-d} \mathcal{O}(x)$. It then follows that $\langle \sigma v\rangle$ is proportional to $\Lambda^{2(4-d)}$.   Thus, for a given operator, $\Omega_{DM} h^2$ is a function of $M$ and $\Lambda$. Since we assume the particle is a WIMP accounting for 100\% of the observed DM in the universe, for a given mass $M$ we shall numerically solve Eq.\ \eqref{eq:relic_density} for the EFT scale parameter $\Lambda$ to satisfy the relation $\Omega_{DM} h^2 = 0.12$.  This requirement implies that for each operator, for a given mass $M$ there are no free parameters in the low-energy annihilation cross section.

\subsection{DM Density Profiles}
\label{sect:profiles}
Although thermal freeze-out is predicated on the effective shutoff of annihilation in the early universe, this does not imply that WIMP annihilation remains negligible for the remainder of the evolution of the universe. Structure formation  leads to large local densities of both baryonic and dark matter, and this obviously implies that the annihilation rate $\Gamma \sim n  \sigma v$ may again become large enough for detectable levels of annihilation to occur in regions such as the Milky Way Galactic Center (GC), or in the dark-matter-rich Milky Way dwarf companion galaxies.

Understanding the spatial distribution $\rho(r)=Mn(r)$ of the DM mass density in the relevant regions is important in the interpretation of any putative annihilation signal. The $N$-body simulation community is actively studying this, but there is as yet no consensus in the literature for the exact form this so-called halo profile takes in galaxies such as our own. Consequently, it is conventional to quote all results assuming a number of different profiles.  For constraints from Milky Way GC observations, we will utilize the (generalized) Navarro-Frenk-White, the Einasto, and the Isothermal profiles \cite{Iocco1107,FERMI1305,FERMI1205,HESS1301,Hooper1209}.

\renewcommand*\arraystretch{1.5}
\begin{table}
\caption{\label{tab:norms} Milky-Way DM halo profile normalizations. We utilize a band of normalizations corresponding to the central value of $\rho_\odot \pm 2\sigma$ limits consistent with microlensing and galactic rotational velocities given in \citet{Iocco1107}. For generalized NFW and Einasto profiles, we assume $r_s = 20$ kpc, and $r_\odot = 8$ kpc, and take the $\rho_\odot$ values from \citet{Iocco1107}; for the Isothermal profile we assume $r_s = 5$ kpc, and $r_\odot = 8$ kpc, and use representative values for $\rho_\odot$.} 
\begin{ruledtabular}
\begin{tabular}{ddcc}
&  \multicolumn{1}{c}{$2\sigma$ lower limit on $\rho_\odot$ [GeV/cm$^{3}$]} & \multicolumn{1}{c}{Central value of $\rho_\odot$ [GeV/cm$^{3}$]}  & $2\sigma$ upper limit on $\rho_\odot$ [GeV/cm$^{3}$] 					\\ \hline \hline
 & \multicolumn{3}{c}{\textbf{generalized NFW}} 						\\ \hline
\gamma=1.0		&		0.28		&	0.40		&		0.54		\\
\gamma=1.2		&		0.25		&	0.36		&		0.48		\\
\gamma=1.3		&		0.235	&	0.34		&		0.45		\\ \hline \hline
 & \multicolumn{3}{c}{\textbf{Einasto}} 								\\ \hline
\alpha=0.17		&		0.245	&	0.36		&		0.57		\\ \hline \hline
 & \multicolumn{3}{c}{\textbf{Isothermal}} 								\\ \hline
				&		0.30		&	0.40		&		0.50
\end{tabular}
\end{ruledtabular}
\end{table}

\renewcommand*\arraystretch{1.5}
\begin{table}
\caption{\label{tab:paper_norms} Summary of the main sources used to compute cross section limits, with corresponding choices of DM halo profile normalizations and $J$ factors where applicable. [The $J$ factor is defined in Eq.\ \eqref{eq:J}.] If no value for a parameter was explicitly stated in a given reference, we indicate why we have assumed the indicated value. The region of interest (ROI) used in the analysis is indicated in parentheses following the notation of the source referenced. Further details in each case can be found in the referenced section number. In the final column we give the rescaling factor $K$ applied in order to map published limits onto the halo profiles and normalizations given in \tabref{norms}; further details on the rescaling are given in \sectref{rescaling}. For the most part the rescaling comes from scaling $J$ factors with $\rho_\odot^2$. The only exception is the isothermal case used by Weniger, where there is a significant difference in $r_s$ that affects the $J$ factor by a factor of 2.} 
\begin{ruledtabular}
\begin{tabular}{llddddd}
Profile (ROI) & $\gamma$ or $\alpha$ & \multicolumn{1}{c}{$r_s$ [kpc]} & \multicolumn{1}{c}{$r_\odot$ [kpc]}
 & \multicolumn{1}{c}{$\rho_\odot$ [GeV/cm$^3$]} & \multicolumn{1}{c}{$J$ factor [$10^{21}$GeV$^{2}$/cm$^{5}$]}  & \multicolumn{1}{c}{$K$} \\    \hline
\multicolumn{7}{c}{ Fermi line search, \citet{FERMI1305} --- \sectref{Fermi_line_limits}} \\ \hline
NFWc (R3)	&	$\gamma = 1.3$	&	20.0	&		8.2\footnote{\label{ftnote1}Not explicitly stated in the reference, but using this value correctly reproduces their $J_{\text{R3}}$, for which no point-source masking was applied. The reference also cites \citet{Catena0907} where $r_\odot = 8.2$ kpc is mentioned.}	&	0.4   &    13.9   &   0.689	\\
NFW  (R41)	 	&	$\gamma = 1.0$	&	20.0	&		8.2\up{\ref{ftnote1}}	&	0.4  &   8.48  &  0.960	\\
Einasto  (R16)  	&	$\alpha = 0.17$		&	20.0	&		8.2\up{\ref{ftnote1}}	&	0.4  &  8.53   &  0.770	\\
Isothermal (R90)	&	\multicolumn{1}{c}{---}			&	5.0	&		8.2\up{\ref{ftnote1}}	&	0.4   &    6.94  & 0.962	\\ \hline
\multicolumn{7}{c}{ H.E.S.S.\ line search, \citet{HESS1301} --- \sectref{HESS_line_limits} } \\ \hline
NFWc\footnote{The ROI for all cases is a $1^\circ$ radius circle centered on the GC, with $|b| < 0.3^\circ$ excluded and no point-source masking applied.}    	&	$\gamma = 1.3$	&	20.0	&		8.0	&	0.42   &   22.1\footnote{\label{ftnote3}$J$ factors are not given explicitly in the reference, but we have verified these reproduce the limits.}    &   0.655	\\
NFW 	 	&	$\gamma = 1.0$	&	20.0	&		8.0	&	0.42   &    4.37\up{\ref{ftnote3}}    &   0.907	\\
Einasto  	&	$\alpha = 0.17$		&	20.0	&		8.0	&	0.42   &  2.43\up{\ref{ftnote3}}    &  0.735	\\
Isothermal	&	\multicolumn{1}{c}{---}				&	5.0	&		8.0	&	0.42   &   0.312\up{\ref{ftnote3}}   &  0.907	\\ \hline
\multicolumn{7}{c}{Template-based analysis, \citet{Hooper1209} --- \sectref{Hooperlimits}} \\ \hline
NFWc 	&	$\gamma = 1.2$	&	20.0	&		8.0\footnote{\label{ftnote2}Not explicitly stated in the reference, but assumed since this is the value used in \citet{Iocco1107} from which the reference took all its normalizations.}	&	0.25  &  \multicolumn{1}{c}{---}  &  2.07	\\
NFW	 	&	$\gamma = 1.0$	&	20.0	&		8.0\up{\ref{ftnote2}}	&	0.28   & \multicolumn{1}{c}{---}    &  2.04	\\
Einasto 	&	$\alpha = 0.17$		&	20.0	&		8.0\up{\ref{ftnote2}}	&	0.25    &   \multicolumn{1}{c}{---}  &   2.07	\\ \hline
\multicolumn{7}{c}{ Line search based on Fermi data, \citet{Weniger1204} --- \sectref{Wenigerlimits} } \\ \hline
NFWc  (Reg4)	       &	$\gamma = 1.3$	&	20.0	&		8.5	&	0.4   & \multicolumn{1}{c}{---}  &   0.644	\\
NFW   (Reg4)    	&	$\gamma = 1.0$	&	20.0	&		8.5	&	0.4   &  \multicolumn{1}{c}{---}  &  0.890	\\
Einasto   (Reg4)	&	$\alpha = 0.17$		&	20.0	&		8.5	&	0.4   &  \multicolumn{1}{c}{---}    &	0.715 \\
Isothermal   (Reg4)	&	\multicolumn{1}{c}{---}				&	3.5	&		8.5	&	0.4   &  \multicolumn{1}{c}{---}   &   0.443	\\ 
\end{tabular}
\end{ruledtabular}
\end{table}

The generalized Navarro-Frenk-White (NFW) DM density profile is given by \cite{Iocco1107}
\begin{align}
\rho(r;\gamma) & = \frac{\rho_0}{\lb(r/r_s\rb)^\gamma \lb( 1 +r/r_s \rb)^{3-\gamma} },
\end{align}
where the scale parameter is chosen to be $r_s = 20$ kpc. The parameter $\gamma$ controls the central slope of the profile: $\gamma = 1$ for the canonical NFW profile, while $\gamma > 1$ defines a profile with a steeper central region (where $\rho \sim r^{-\gamma}$), known as a contracted profile (NFWc) which may arise due to gravitational interactions between the dark and baryonic matter as the latter cools during galaxy formation \cite{Gnedin:fk}.  In either case, the profile is peaked toward $r = 0$ and ``cuspy.'' The overall scale $\rho_0$ is chosen for galactic measurements such that the local (in the vicinity of the Sun) DM density $\rho(r_\odot) \equiv \rho_\odot$ takes some desired value (see \tabref{norms}), while for extragalactic (dwarf) measurements, $\rho_0$ is chosen to obtain the correct observed stellar rotational velocities \cite{FERMI1108}. 

The Einasto profile is given by \cite{Iocco1107}
\begin{align}
\rho(r) & = \rho_0 \exp \lb\{ -\frac{2}{\alpha} \lb[ \lb(\frac{r}{r_s}\rb)^\alpha - 1 \rb] \rb\}
\end{align}
where the scale parameter is chosen to be $r_s = 20$ kpc. In line with large-scale numerical simulations, the parameter $\alpha = 0.17$ is used.  This profile is cuspy. Overall normalizations are chosen similarly to the NFW profile; see \tabref{norms}. 

The Isothermal profile is given by \cite{Iocco1107}
\begin{align}
\rho(r) & = \frac{\rho_0}{ 1 + (r/r_s)^2 }
\end{align}
where the scale-parameter $r_s = 5$ kpc is chosen. This profile is known as ``cored'' since $\rho(r)$ flattens off to a constant value at small $r$. Overall normalizations are again chosen similarly to the NFW profile; see \tabref{norms}. 

We have scaled all the limits presented in this paper to reflect the halo profile parameters presented in \tabref{norms} (original parameters assumed in the literature are shown in \tabref{paper_norms}; see \sectref{rescaling} for more details on the rescaling).

\subsection{Annihilation Photon Flux}
\label{sect:flux}
For all the EFT operators under consideration, the detectable annihilation signal in $\gamma$-rays includes a) one or more monochromatic prompt photon ``lines'' arising from the direct annihilation DM DM $\rightarrow \gamma X$ where $X$ can be any SM boson consistent with gauge symmetries ($X = \gamma,Z,h$), and/or b) a prompt diffuse continuum photon flux arising from final-state radiation or radiative hadron decay arising when considering various non-photon primary DM annihilation products: DM DM $\rightarrow X Y \rightarrow \gamma$'s where $X$ and/or $Y$ is not a photon.

For each particular operator there will be a characteristic per-annihilation differential prompt photon spectrum 
\begin{equation}
\frac{dN_\gamma(E)}{dE} \equiv \sum_f \text{BR}_f \lb( \frac{dN^f_\gamma(E)}{dE} \rb)
\label{eq:spectrum}
\end{equation}
where $\text{BR}_f \equiv \sigma\lb(\text{DM DM} \rightarrow f\rb) / \ \sigma_\text{total} $ is the branching ratio for the annihilation mode to final-state $f$, $dN^f_\gamma(E)/dE$ is the per-annihilation differential prompt photon spectrum for that mode, and the sum runs over all annihilation modes for the operator. Given this spectrum, the differential DM-annihilation prompt photon flux which an experiment would observe at photon energy $E$ is given by \cite{FERMI1205} 
\begin{subequations}
\begin{align}
\frac{d\Phi(E)}{dE} &= \frac{[ \sigma v]_\text{total}}{16\pi M^2} \ J \  \frac{dN_\gamma(E)}{dE} \quad \times \lb\{ \begin{array}{ll} 2 & \text{for self-conjugate DM} \\  1 & \text{for non-self-conjugate DM} \end{array}\rb. \label{eq:flux} \\
\text{where}~~~~   J &= \int_{\text{LOS-ROI}} \rho^2[r(s,l,b)]\ ds \ \cos b\ db \ dl. \label{eq:J}
\end{align}
\end{subequations} 
Hereafter we will assume non-self-conjugate DM, \textit{i.e.,} complex scalars
and Dirac fermions.  In the equation above, $\rho(r)$ is the DM density
profile at Galactocentric distance $r$. The astrophysical ``$J$
factor'' [Eq.\ \eqref{eq:J}] is computed by integrating the squared
DM density along the line of sight (LOS) from the Earth, and over the
relevant Galactic lat/long coordinates.\footnote{The Galactic Center (GC) is defined to be $l=b=0$, where $l$   is Galactic longitude and $b$ Galactic latitude. We work in a convention where $b\in[-90^\circ,90^\circ]$ and   $l\in[-180^\circ,180^\circ)$.}
The integral over $b$ and $l$ defines a search region of interest
(ROI). A point observed at LOS distance $s$ at lat/long
coordinates $(l,b)$ has Galactocentric distance $r = \left( s^2 +
r_\odot^2 - 2sr_\odot \cos l \cos b\right)^{1/2}$ where $r_\odot$ is
the Sun--GC distance. We will always take $r_\odot = 8$ kpc
\cite{Iocco1107}. Finally note that the polar angle $\theta$ from the
GC satisfies $\cos\theta = \cos l \cos b$.

It is our aim to place upper limits on the non-relativistic cross sections $\sigma v$ using experimentally determined upper limits on such photon fluxes.

\subsubsection{Profile rescaling}
\label{sect:rescaling}

The constraints derived in this paper draw on published gamma-ray line
searches and limits on diffuse emission from several sources, each of
which employ different assumptions for Milky Way halo profiles. We
give the parameters used by these references in
\tabref{paper_norms}. For consistency, however, we would like to derive limits
for a single set of halo profile parameters (given in \tabref{norms}).

The difference in halo profiles is encapsulated in the $J$ factor;
typically we wish to take some given $J$ factor from the literature
(using some values of $r_\odot$, $\rho_\odot$ and/or $r_s$) and
rescale it to values of $r_\odot'$, $\rho_\odot'$ and/or $r_s'$ that
are our choices.  All other things being equal, the limit for $\sigma v \propto 1/J$, so the cross section limit that would be
set with the new set of halo parameters is
 \begin{align} 
[\sigma v]'  = \frac{J}{J'}  \ [\sigma v] \equiv 
\ \frac{[ \sigma v]}{K} .
 \end{align}
Here $J'$ is the $J$ factor computed with the set of parameters we
wish to use and $J$ is the $J$ factor we compute using the parameters
in the literature. We thus rescale the published limits by the factor
$K$ (see \tabref{paper_norms}).

We find that scaling the $J$ factors with $\rho_\odot^2$
accounts for most of the difference in the halo profile
normalizations. This is primarily because the variation in
$\rho_\odot$ values used in the literature is large (up to 40$\%$),
while the variation in the $r_\odot$ values is less than 10$\%$, and the $r_s$
values are almost uniform.  There is only one case where there is a
large difference in $r_s$; for the isothermal profile for
\texttt{Reg4} of Ref.\ \cite{Weniger1204} where $r_s = 3.5\ \text{kpc}
\rightarrow 5.0\ \text{kpc}$. In this case the rescaling factor was
roughly a factor of 2 smaller than one would have obtained from simply
scaling $J$ with $\rho_\odot^2$, accounting for the fact that the core
is less dense when $r_s$ is increased.

Note that for the Fermi ROIs, we have computed these rescaling factors
without considering point source masking effects which are estimated
to impact the $J$ factor by less than about $10\%$. Our justification for this
simplification is that the halo profiles are smooth away from $r=0$,
so that the {\it ratios} $J'/J$ computed with or without point-source
masking being considered should be very nearly equal (certainly much
closer than 10$\%$).

We also point out that the rescaling of the \texttt{Reg4} values from
Ref.\ \cite{Weniger1204} is appropriate even though the ROIs in that
reference were chosen based on an expected signal/background analysis
with a specific profile in mind. \citet{Weniger1204} reports
cross section results from each such ROI choice assuming a number of
different profiles \emph{in the conversion from flux to cross section
  while holding the ROI fixed}, and it is only these
flux-to-cross-section conversions we are rescaling.  Similarly, the
ROIs in Ref.\ \cite{FERMI1305} were chosen based on an expected
signal/background analysis; but again, we are simply holding their
chosen ROI, and hence fluxes, fixed and are only rescaling the
flux-to-cross-section conversion.

\subsection{Photon Spectra}
\label{sect:spectra}

In the models we study, dark matter annihilation produces primary electroweak gauge bosons ($\gamma$, $Z$, $W^\pm$), higgs bosons, or fermions. In addition to primary photons, the subsequent shower and hadronization of final states will produce abundant hadrons, giving rise to additional prompt photons dominantly through decay of neutral pions $\pi^0 \rightarrow \gamma \gamma$. To extract cross section upper limits from the measured flux upper limits requires knowledge of this per-annihilation prompt photon spectrum, see Eq.\ \eqref{eq:flux}. Here we summarize the physics of prompt photon production from DM annihilation.

\begin{figure}[t]
\begin{center}
\includegraphics[width = 0.8\textwidth]{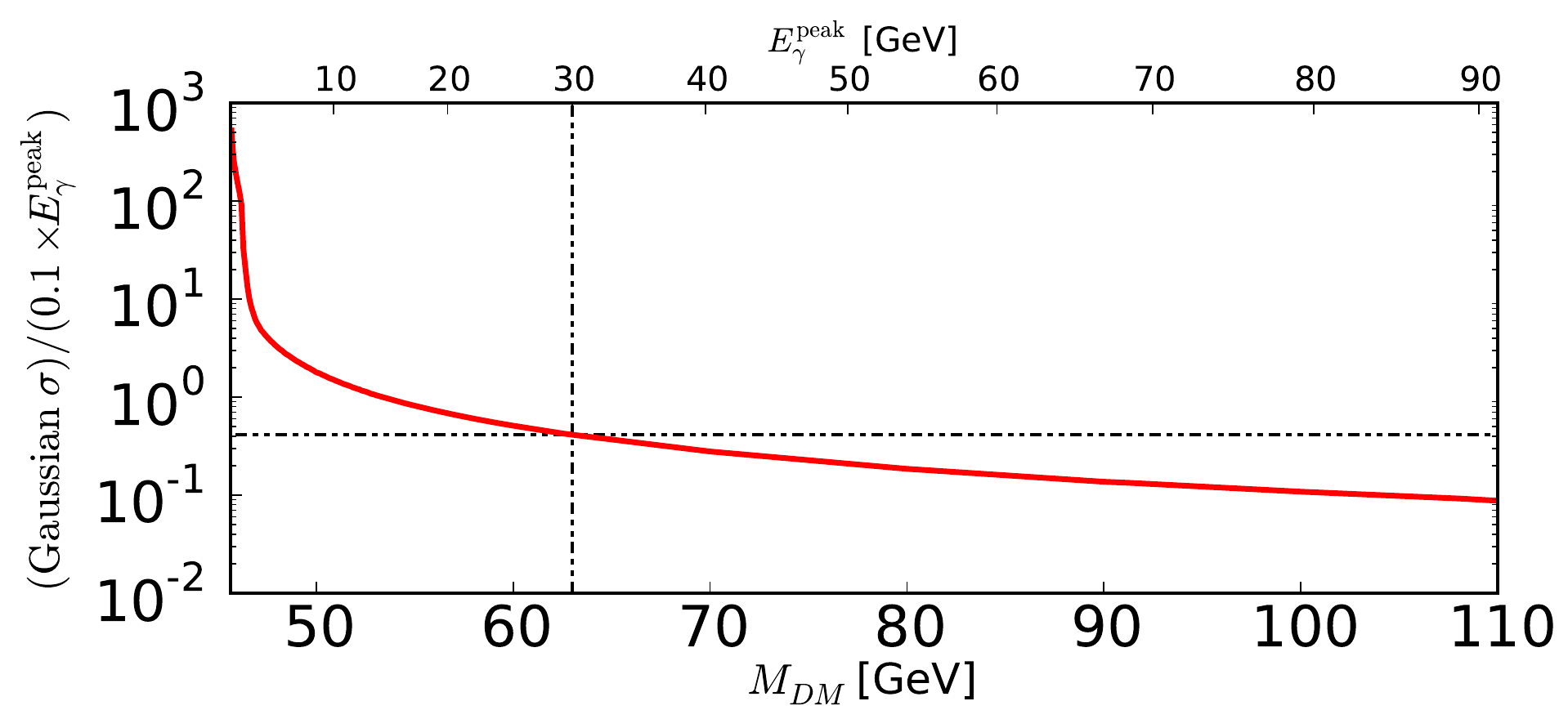}
\caption{\label{fig:Zlineshape} The Gaussian standard deviation of the
  photon line arising from the $\gamma Z$ primary final state as a
  function of DM mass, expressed in units of $0.1 \times E_\gamma$,
  approximately the Fermi-LAT energy resolution (68\% containment
  half-width). The line-width is large with respect to the resolution
  for DM masses less than about $60$ GeV; we only set limits on photon
  lines from $\gamma Z$ for $M > 63$ GeV ($E_\gamma >30$ GeV), as discussed in the
  text. The standard deviation here has been extracted by fitting a Gaussian to photon
  line shapes computed with Pythia
  (this provides a good fit to the central region of the peak but
  mismodels the tails). }
\end{center}
\end{figure}

We begin our discussion with monochromatic photons. The $\gamma\gamma$ and $\gamma h$ final states give rise to photon lines.\footnote{We assume $m_h = 125$ GeV in this paper. For a 125 GeV SM Higgs, $\Gamma_h \approx 4$  MeV \cite{Barger1203}, so the width of the photon line is negligible.} The contribution to the photon spectrum is a delta function: $dN(E)/dE = N_\gamma \delta( E - \widetilde{E}_\gamma )$, where simple kinematics gives $\widetilde{E}_\gamma = M - m_X^2/4M$ for a $\gamma X$ final state, and $N_\gamma$ counts the number of monochromatic photons in the decay mode ($N_\gamma = 2$ for the $\gamma \gamma$ annihilation modes and $N_\gamma = 1$ otherwise). The photon line from a $\gamma Z$ annihilation mode is similarly peaked at $E_\gamma = M - m_Z^2/4M$, but obviously has a finite width due to the finite width of the $Z$. The width of the line will be relevant to our discussion of experimental constraints later and is shown in \figref{Zlineshape}. 

For annihilation modes $W^+W^-,\ ZZ,\ Zh,\ hh, \ f\bar{f},\ \gamma
h,\ \mathrm{and}\ \gamma Z$, there is also a continuum prompt photon
spectrum. We employ Pythia 8.176 \cite{PYTHIA} to perform the shower
and hadronization (and hadron decay) to obtain the photon
spectra. Following Ref.\ \cite{PPPC4DMID}, we create a fictitious
spin-1 resonance at $m = 2 M$ with a width of 1 GeV, which
we populate using fictitious $e^+e^-$ colliding beams free of
initial-state radiation at $\sqrt{s} = 2M$. We set by hand the
resonance decay channels to be, in turn, 100\% to each of the desired
primary final states, and the per-annihilation differential photon
spectra are computed by correctly normalizing\footnote{$dN/dE$ =
counts per bin / bin width / total number of annihilations.}
the histogrammed photon energies from 10\up{5}
simulated annihilations per mass point per primary final state. 

Good agreement is found between these computed Pythia spectra and a
set of spectra from \citet{PPPC4DMID} for DM masses below 300 GeV. For
larger DM masses, the effects of electroweak (EW) final-state
radiation (FSR) of $W$s and $Z$s (which are not included in the Pythia
shower) can significantly alter the spectra at low $E_\gamma$
\cite{Ciafaloni1009,PPPC4DMID}; these effects have been included in
the spectra in \citet{PPPC4DMID}. However, we shall see below that we
will be primarily interested in the spectrum for $E_\gamma \in
[10,100]$ GeV at large DM mass.  In this photon energy range, the EW
corrections are unimportant for $M \lesssim 1$ TeV except for primary
final states containing charged leptons, which in any event give rise
to about an order of magnitude fewer photons per annihilation compared
to other channels. Furthermore, the branching fraction for
annihilation to leptonic final states in particular, and fermionic
final states more generally, is subdominant at large $M$ (see the
result plots in \sectref{results}), with the exception of operator
XX-1 where all annihilations are mediated through an $s$-channel $Z$
or $\gamma$.  Therefore, since the error of neglecting the EW FSR is
small, and since the necessary spectra are not all available in
\citet{PPPC4DMID}, we have chosen for the sake of uniformity to use
the Pythia spectra which we have computed rather than attempting to
use the corrected spectra from \citet{PPPC4DMID}.

\subsubsection{Qualitative understanding of the shape of the photon spectra}

The shape, and in particular the DM-mass dependence, of the continuum
photon spectrum differs qualitatively depending on whether the primary
final states are leptonic, non-leptonic and color neutral, or
colored. We consider illustrative cases for the latter two: annihilation
to $W^+W^-$ and annihilation to $\bar b b$ (light $q\bar{q}$ modes are qualitatively similar), and comment on the leptonic modes.

\begin{figure}
\includegraphics[width=\textwidth]{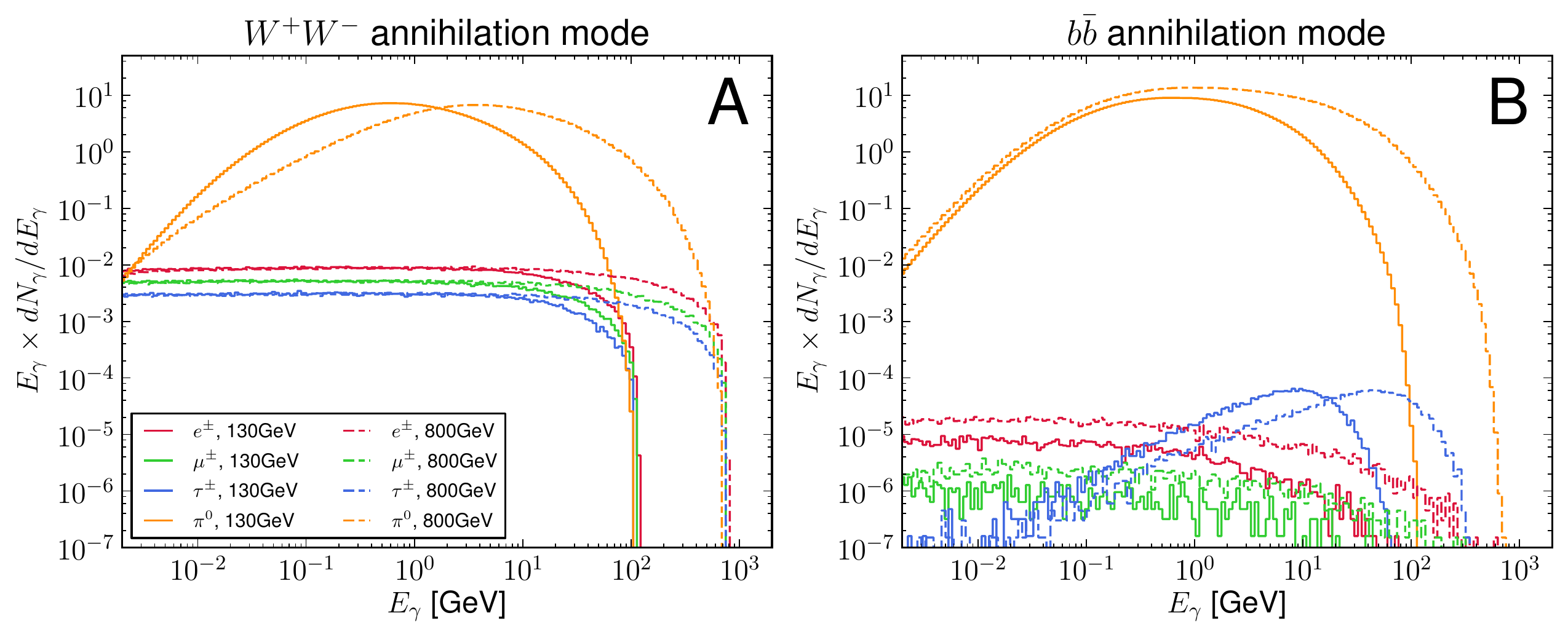}
\caption{\label{fig:WW_origins} The photon spectrum arising from DM
  annihilation, decomposed according to the production mechanism of
  either $\pi^0$ decay or final state radiation of charged leptons. We
  show annihilation to $WW$ (Panel A) and to $b\bar b$ (Panel B), for
  DM mass of 130 GeV and 800 GeV. It is clear that for $E_\gamma
  \gtrsim 1$ MeV, $\pi^0$ decay dominates the spectrum. In the case of
  the $WW$ final state, the shift of the pions to higher energies
  reflects the increasing boost of each $W$ (and thus of its decay
  products). For the $b\bar b$ final state, the yield of pions
  increases over the entire energy range of interest. We have 
  omitted the spectra from other hadronic decay modes for clarity.}
\end{figure}

Consider first the $WW$ case. In \figref{WW_origins} we plot the
photon spectrum decomposed into photons from $\pi^0$ decay and those
from final state radiation (FSR) of charged leptons (omitting other 
hadronic-decay photons for clarity).  For $E_\gamma
\gtrsim 1$ MeV, the production mechanism is dominantly due to pion
decay. The strongest continuum $\gamma$-ray constraints are for energies
from 0.1 GeV to 100 GeV, so we will always be in the regime where the
behavior of the photon spectrum from diboson annihilation channels is
dominated by photons from $\pi^0\rightarrow \gamma\gamma$.  Meanwhile,
the FSR from leptons has a spectrum that goes like
$dN_\gamma/dE_\gamma \sim E_\gamma^{-1}$ and is subdominant for
$E_\gamma \gtrsim 1$ MeV.

Since $W$ is color neutral, the spectrum and multiplicity of photons
is determined by the independent decay and shower of each $W$, boosted into
the center of mass frame of the annihilation. As $M$ increases, the
$W$s are increasingly boosted and the decay products also become more
energetic while the multiplicity (namely the pion yield) remains
roughly constant. We can observe this in the shifting of the photon
spectrum in \figref{WW_origins} to higher energies. Furthermore, this
migration to higher energy decay products, while the yield remains
constant, leads to the effect that as $M$ increases there is a
decrease in the number of photons produced at low energy. This can
also be observed in \textit{e.g.,} Panel B of \figref{TabVI_OP3}, where for $M >
100$ GeV where the $WW$ final state dominates, there is a decrease in
the photon spectrum around $E_\gamma \sim 0.1$ GeV as $M$ increases.

Since most of the operators considered here have annihilation modes
into gauge bosons or higgs, the mass dependence of the $\gamma$-ray spectra
has similar behavior to the $WW$ case, modulo the effect of changing
branching fractions.  Some operators which have only diboson
annihilation modes also show first an increase in the low-energy
photon spectrum with increasing $M$ below or around a hundred GeV
before the decrease discussed above sets in. This has less to do with
the intrinsic photon spectrum from each annihilation mode and more to
do with the branching fractions for each mode.  If the $\gamma\gamma$
mode is present, then the branching fraction to $\gamma Z$ starts off
small around $M \sim m_Z/2$ and rapidly increases with $M$ giving
the observed increase in the continuum spectrum (\textit{e.g.},
Operator VI-1 in \figref{TabVI_OP1} of \sectref{results}). In
this case, there may be a $WW$ or $ZZ$ mode entering once $M$ gets to
$m_{W,Z}$, which results in a further increase since the spectra are
normalized per-annihilation and twice as many $W/Z$s per annihilation
gives twice as many photons per annihilation (the $W$ and $Z$
intrinsic photon spectra are almost identical, modulo kinematic
effects). For $M$ above a (few) hundred GeV, the low-energy spectrum begins to decrease
again as per the discussion in the previous paragraph. Explanations of
a similar nature are possible for cases where other mixtures of
diboson annihilation modes are present.

For annihilation to $b\bar{b}$, shown in \figref{WW_origins}, the dominant
photon-production mechanism is still neutral pion decay, but the same
qualitative picture does not obtain: there is a significant increase
in the pion yield with increasing $M$ because in this case the
initial hard scale for the parton shower is set by $M$ and so the
number of FSR gluons and other strongly interacting partons produced
in the shower increases significantly as the DM mass
increases. Therefore, although the average pion energy does increase,
the increase in the pion yield at the same time means that the pion
yield at low energies actually slowly \emph{increases} as $M$
increases; the photon yield thus shows a similar increasing trend.
This is evident in the slow overall monotonic increase in the spectrum
as $M$ increases for Operator XX-1 shown in
\figref{TabXX_OP1}. (For the other operators considered here, this
behavior gets cut off near $M \sim 100$ GeV since the branching
fraction to quarks typically becomes subdominant once diboson
annihilation modes become kinematically allowed.)

Finally, for annihilation to charged leptons, the dominant photon production mechanism is direct final-state photon radiation from the leptons. This spectrum falls roughly as $E_\gamma^{-1}$ over most of the energy range with a high-energy kinematic cutoff for the photon spectrum at $M$, while the normalization of the photon spectrum grows slowly as $M$ increases. 
Again however, with the exception of operator XX-1 already noted above, fermionic final states tend to have rapidly falling branching fractions once on-shell diboson final states become kinematically allowed as $M$ increases, so the increase in the spectrum at low $E_\gamma \sim 0.1$ GeV is cut off around $M \sim 100$ GeV.

\section{Limits}
\label{sect:limits}

In this section we discuss the experimental limits used in
constraining the effective operators. We consider line searches in
gamma-ray data from Fermi-LAT and H.E.S.S., as well as constraints on
diffuse gamma-ray emission derived from Fermi-LAT observations of the
GC and dwarf galaxies, for which the continuum photon emission is
relevant. We describe our methods of extracting limits for the
operators, re-interpreting published results for the specific final
states and branching ratios of each operator. Our final results are
shown in the figures in \sectref{results}, along with a universal
caption and summary of the constraints on the operators.

\subsection{Low-energy line limits: Fermi-LAT}
\label{sect:Fermi_line_limits}

The Fermi-LAT Collaboration \cite{FERMI1305} presents 95\% confidence level (CL) upper limits
(UL) on the flux $\Phi$ of photons from regions on the sky centered on the
Galactic Center due to a strictly monochromatic underlying photon line
in the photon energy range 5 to 300 GeV. The underlying photon energy
$E_\gamma$ was scanned across this range, and for each energy
they obtained a flux limit by performing a maximum likelihood analysis in which the Fermi-LAT line shape plus a
power-law background was fitted\footnote{The position of the line in the fit was fixed; the line and background normalizations as well as the background spectral index were allowed to float.} to their
3.7-year data over sliding energy ranges $E_\gamma \pm 6 \sigma_E$
where $\sigma_E$ is the Fermi-LAT energy resolution (68\% containment
half-width). The flux limit $\Phi$ they report is the 95\% CL UL of
the normalization of the line in this fit, multiplied by a factor
giving the effective exposure time of the experiment. 

The Fermi-LAT search methodology selected a different signal region of
interest (ROI) for each of four DM halo profiles (NFW, Einasto,
Isothermal, NFWc$_{\gamma = 1.3}$) they considered in order to
maximize the expected signal-to-background ratio\footnote{This is the
  approach advocated in \citet{Weniger1204}. The methodology of the
  previous Fermi-LAT analysis \cite{FERMI1205} was to look at a single
  ROI for all profiles, which obviously leads to a single flux limit,
  which is then interpreted differently for each profile.} as
estimated through Monte Carlo simulations. Thus, they present four
distinct flux limits, one for each ROI.
These ROIs are denoted R$\alpha$ where $\alpha$ is the angular size of a
circle centered on the GC, with the region $|l|>6^\circ$ and $|b| <
5^\circ$ masked; regions around known $\gamma$-ray point sources are also masked
for all ROIs except R3 \cite{FERMI1305}.

We derive constraints on $\sigma v$ from the flux limits presented in
Ref.\ \cite{FERMI1305}, which are given at discrete $E_\gamma$ in the
energy range 5 to 300 GeV; we have interpolated (linearly in $\log
\Phi$ vs.\ $E_\gamma$) these limits to
intermediate energies where necessary. If the underlying photon line
arising from a $\gamma X$ final-state mode is monochromatic or
monochromatic to an excellent approximation, we could merely use the
flux upper limit at $\widetilde{E}_\gamma = M - M_X^2/4M$ combined
with Eq.\ \eqref{eq:J} and the $J$ factors in Ref.\ \cite{FERMI1305}
(rescaled per \tabref{paper_norms}) to derive limits on $\sigma v$:
\begin{equation}
\lb[ \sigma v \rb]^{95\% \text{ CL UL}} =  \Phi^{95\% \text{ CL UL}}(\widetilde{E}_\gamma) \times \frac{16\pi M^2}{ J N_\gamma },
\label{eq:singleline_limit}
\end{equation}
where $N_\gamma$ is the number of photons in the line per
annihilation: 2 for $\gamma\gamma$ and 1 otherwise. This is the case
for the $\gamma h$ annihilation mode; complications however arise for
the $\gamma Z$ and $\gamma \gamma$ modes for the operators we consider.

The photon line arising from $\gamma Z$ annihilation modes has an intrinsic width of order $1$ GeV, which can be significant with respect to the \textit{ca.}\ 5-10\% Fermi-LAT energy resolution at low energies \cite{FERMI1206,FERMI1305} (see \figref{Zlineshape}).  However, even for a line with intrinsic width of 50\% of the Fermi-LAT energy resolution, the resulting observed spectral feature is only expected to be broadened by around 11\% after convolution with the Fermi-LAT energy dispersion function.
If such a feature were fitted with a line shape derived from the assumption of a purely monochromatic intrinsic line,  the number of photons would be \emph{underestimated} by about 7\% (Ref.\ \cite{FERMI1305}, App.\ D2--3) which would set the flux limit correspondingly more stringent than it should. However, this effect is much smaller than both the expected statistical fluctuation (\textit{ca.}\ 50\%) in the limits set by Ref.\ \cite{FERMI1305} and the uncertainty in the halo profile normalization (a factor of \textit{ca.}\ 10; see results) and it is roughly on the same order as other systematic effects (\textit{ca.}\ 10\%) which are present in the Fermi-LAT analysis and ignored in their flux upper limits \cite{FERMI1305}. Provided the linewidth is expected to be less than about half the energy resolution, we therefore make no correction for the finite $Z$ width, and merely use the Fermi-LAT flux limits derived under the assumption of a monochromatic intrinsic line in deriving limits on $\lb[ \sigma v \rb]_{\gamma Z}$.
To satisfy this linewidth requirement, we follow the older Fermi-LAT line search analysis \cite{FERMI1205} and only present limits on $\sigma v$ from the $\gamma Z$ annihilation mode for $E_\gamma \geq 30$ GeV, which corresponds to $M \gtrsim 63$ GeV (see also \figref{Zlineshape}).

There is a significant complication \cite{Fan1307} in interpreting the flux limits for all the operators with $\gamma \gamma$ annihilation modes: due to their gauge structure, all such operators under consideration also have a $\gamma Z$ annihilation mode which yields a second line at an energy $\delta E = m_Z^2/4M$ lower than the $\gamma\gamma$ line. Since the unbinned likelihood analysis in Ref.\ \cite{FERMI1305} is sensitive to both shape and normalization of the line spectral feature, this makes the interpretation of the single-line flux limits difficult in the region of DM masses where the two photon lines are separated by an amount of the order of the Fermi-LAT energy resolution. Depending on the WIMP mass, we have adopted a few different techniques to interpret the single-line flux limits for these operators. In all cases, we choose to use the flux limits to set a single limit on the quantity $\lb(2\sigma_{\gamma\gamma} + \sigma_{\gamma Z}\rb)v$.

\subsubsection{$M <$ \rm{63 GeV} }
This mass range corresponds to $E_\gamma \leq 30$ GeV for the $\gamma Z$ channel (when kinematically allowed), so we do not set a $\gamma Z$ line limit in this mass range. We find the 95\% CL UL on $\sigma_{\gamma\gamma}v$ using Eq.\ \eqref{eq:singleline_limit} with the flux at $E_\gamma = M$, and use this to set, at each WIMP mass, the limit
\begin{align}
\lb[ \lb(2\sigma_{\gamma\gamma} + \sigma_{\gamma Z}\rb)v \rb]^{95\% \text{ CL UL}} = 2 (\sigma_{\gamma\gamma}v)^{95\% \text{ CL UL}} \lb[ 1 + \frac{\sigma_{\gamma Z}v(M)}{2 \sigma_{\gamma\gamma}v(M) } \rb]
\end{align}
where the cross section ratio in the $[\ \cdots\ ]$ brackets is computed using the analytic expressions in CKW and rescales the $\sigma_{\gamma \gamma}v$ limit to the quantity $\lb(2\sigma_{\gamma\gamma} + \sigma_{\gamma Z}\rb)v$.

\subsubsection{\rm{63 GeV} $< M <$ \rm{80 GeV} \label{method:low_inter}}
In this mass region, we have two lines which are relatively well separated\footnote{To be strictly correct, to use the limits independently, the two lines must be separated enough that they do not lie in the same sliding-fit energy range used in the Fermi-LAT line search, which for a peak at $E_\gamma$ was $E_\gamma \pm 6 \sigma_E$ where $\sigma_E$ is the 68\% containment half-width 
for the Fermi-LAT energy dispersion \cite{FERMI1305}. We have been a little more permissive in setting limits up to $M = 80$ GeV, where the two lines are only about $3\sigma_E$ to $6\sigma_E$ apart (depending on whether one assumes a 10- or 5-percent energy resolution, respectively), but we do not expect this to be much of an issue.} assuming a 5 to 10\% energy resolution for Fermi-LAT \cite{FERMI1206}. We thus use Eq.\ \eqref{eq:singleline_limit} to set independent 95\% CL UL on $\sigma_{\gamma\gamma}v$ using the flux at $E_\gamma = M$, and on $\sigma_{\gamma Z}v$ using the flux at $E_\gamma = M - m_Z^2/4M$. We use these two quantities to set, at each WIMP mass, the limit
\begin{align}
\lb[ \lb(2\sigma_{\gamma\gamma} + \sigma_{\gamma Z}\rb)v \rb]^{95\% \text{ CL UL}} = \text{min} \lb\{ 2 (\sigma_{\gamma\gamma}v)^{95\% \text{ CL UL}} \lb[ 1 + \frac{\sigma_{\gamma Z}v(M)}{2 \sigma_{\gamma\gamma}v(M) } \rb] \quad , \quad (\sigma_{\gamma Z}v)^{95\% \text{ CL UL}} \lb[ 1 + \frac{2 \sigma_{\gamma \gamma}v(M)}{\sigma_{\gamma Z}v(M) } \rb] \rb\}
\end{align}
where the cross section ratios in the $[\ \cdots \ ]$ brackets are computed using the analytic expressions in CKW.

\subsubsection{\rm{80 GeV} $< M <$ \rm{160 GeV}}
In this intermediate mass region, the spectral feature expected in the Fermi-LAT data corresponding to the two lines will be very broad with respect to the energy resolution, and may be a double-peaked structure depending on the relative strengths of the lines. Such a feature looks very different from the line shape used in the Fermi-LAT line search \cite{FERMI1305} and since, as noted above, that analysis was sensitive to the shape of the fitted feature as well as its normalization, we exercise an abundance of caution and do not set any limits on $\sigma v$ in this region. This caution notwithstanding, we do not expect the limits to differ from those at nearby energies by more than a factor of a few. Since a reproduction of the analysis in Ref.\ \cite{FERMI1305} with a different assumed line shape is a task best suited for the experimental collaboration, we would encourage the Fermi-LAT Collaboration to look into setting flux limits for the case where there may be two partially resolved lines present, with some variable ratio of strengths.

\subsubsection{\rm{160 GeV} $<M $ \label{method:high_inter}}
In this mass region, the two lines are sufficiently close that the spectral feature expected in the Fermi-LAT is merely a slightly broadened (width increased by $\lesssim 10$\%) line and a single limit can be set on the combined quantity $\lb(2\sigma_{\gamma\gamma} + \sigma_{\gamma Z}\rb) v$. We do this by using the flux upper limit from Ref.\ \cite{FERMI1305} evaluated at the flux-weighted average energy
\begin{align}
\bar{E}_\gamma = \frac{ 2 \sigma_{\gamma\gamma}v(M) \times M +  \sigma_{\gamma Z}v(M) \times \lb( M - m_Z^2/4M \rb)}{2 \sigma_{\gamma\gamma}v(M) +  \sigma_{\gamma Z}v(M)},
\end{align}
where the cross sections here are computed using the analytic expressions in CKW. The limit is set as 
\begin{align}
\lb[ \lb(2\sigma_{\gamma\gamma} + \sigma_{\gamma Z}\rb)v \rb]^{95\% \text{ CL UL}} =  \Phi^{95\% \text{ CL UL}}(\bar{E}_\gamma) \times \frac{16\pi M^2}{ J } \ .
\end{align}

\subsection{High energy line limits: H.E.S.S.}
\label{sect:HESS_line_limits}
The H.E.S.S. Collaboration \cite{HESS1301} presents 95\% confidence level (CL) upper limits (UL)
on the photon flux $\Phi$ from a strictly monochromatic photon line in
the energy range 500 GeV to 20 TeV. The ROI is a $1^\circ$ radius circle
centered on the GC, with $|b| < 0.3^\circ$ excluded. No masking of
point sources was performed. Their flux upper limits were extracted by
performing a maximum likelihood analysis in which a Gaussian peak
(photon line) was fitted along with a parametrized background to 112 hours
of data from the H.E.S.S.\ VHE $\gamma$-ray experiment. The position of the
Gaussian peak was scanned over the search energy range (with the
standard deviation constrained to be equal to the H.E.S.S.\ energy
resolution; see below), and the Gaussian normalization and background
parameters were fitted at each search energy. The line flux limits
reported are the 95\% CL UL on the fitted Gaussian normalizations.

Flux limits are only presented at discrete $E_\gamma$ on the energy
range 500 GeV to 20 TeV in Ref.\ \cite{HESS1301}; we have interpolated
(linearly in $\log \Phi$ vs.\ $\log E_\gamma$) these limits to intermediate
energies where necessary.  Isothermal limits here are very weak as a
consequence of the cored profile and the ROI which is restricted to a
very small region near the GC.  It was necessary for interpreting the
H.E.S.S.\ limits to compute the $J$ factors defined in
Eq.\ \eqref{eq:J} for the halo profiles of our choice, given in
\tabref{norms}. We did this for by computing $J$ factors for the
parameters used in the publication (see \tabref{paper_norms}) and then
rescaling the resulting $J$ factors as necessary, as discussed in
\sectref{rescaling}.

Additionally, in computing cross section limits from the fluxes, we
neglected the small shift in the photon line position for final states
involving a massive particle (the largest this shift gets is  about 7
GeV for the $\gamma h$ line at $M = 500$ GeV) and have simply
evaluated all limits assuming $E_\gamma = M$. Furthermore, since the
H.E.S.S.\ energy resolution is 17\% at 500 GeV, dropping to 11\% at 10 TeV
\cite{HESS1301}, it is always a good approximation to combine the
unresolved $\gamma\gamma$ and $\gamma Z$ lines into a single spectral
feature to derive combined limits on $(2\sigma_{\gamma\gamma} +
\sigma_{\gamma Z})v$ for all the operators with both $\gamma\gamma$
and $\gamma Z$ annihilation modes. The width of the photon line from
$\gamma Z$ has no significant effect here.

To summarize, we derived cross section limits from the H.E.S.S.\ flux upper limits using, as appropriate,
\begin{align}
\lb. \begin{array}{rrr} \lb[ \lb(2\sigma_{\gamma\gamma} + \sigma_{\gamma Z}\rb)v \rb]^{95\% \text{ CL UL}}       \\
			        \lb[ \sigma_{\gamma Z} v \rb]^{95\% \text{ CL UL}}  								\\ 
			        \lb[ \sigma_{\gamma h} v \rb]^{95\% \text{ CL UL}}   \end{array} \rb\} =  \Phi^{95\% \text{ CL UL}}(E_\gamma = M) \times \frac{16\pi M^2}{ J } \ .
\end{align}

\subsection{Inclusive spectrum limits: Hooper \textit{et al.}\ analysis}
\label{sect:Hooperlimits}
\citet{Hooper1209} presents 95\% CL UL on the inclusive photon spectrum from the Galactic Center region based on the 3.7-year Fermi-LAT data in a model-independent form as 95\% CL UL limits on the quantity $\mathcal{F}_{\text{bin }j} \equiv (\sigma v/M^2) \int_{\text{bin }j} dN^{\text{total}}_\gamma/dE_\gamma\ dE_\gamma$ for the four photon energy bins of 0.1 to 1 GeV, 1 to 3 GeV, 3 to 10 GeV and 10 to 100 GeV.
These limits were derived using a signal-template-based subtraction methodology with a photon-flux template proportional to $\int_{\text{LOS}} \rho^2(r[s,l,b])\ ds$, rather than the ROI-integrated approach used in the other references we have used.

We employed these model-independent limits to set limits on $\lb[\sigma v\rb]_{\text{total}}$, taking the most stringent of the limits from the four bins listed above. For each $M$, the prompt photon spectrum $dN_\gamma / dE_\gamma$ (including all annihilation modes) was numerically integrated over each of the four energy bins defined above. 
We have included the photons arising from lines where present. For each bin, the limit 
\begin{align}
\lb[ \sigma v \rb]^{95\% \text{ CL UL}}_{\text{bin }j} = 2 \times \mathcal{F}^{\,95\% \text{ CL UL}}_{\text{bin }j} \times M^2 \times \lb[ \int_{\text{bin }j} \frac{dN_\gamma}{dE_\gamma} dE_\gamma \rb]^{-1}
\end{align}
is formed (the factor of 2 is a re-interpretation of a limit set for self-conjugate DM in \cite{Hooper1209} to our choice of fermion or complex scalar DM [see Eq.\ \eqref{eq:flux}]), with the final limit on the total cross section taken as 
\begin{align}
\lb[ \sigma v \rb]^{95\% \text{ CL UL}}_{\text{total}} = \underset{\text{bin }j}{\text{min}}  \lb\{ \lb[ \sigma v \rb]^{95\% \text{ CL UL}}_{\text{bin }j} \rb\}.
\end{align}
For large $M$, bin 4 (10 to 100 GeV) typically gives the most stringent limit (\textit{cf.}\ the comments in \sectref{spectra} on the regions of the continuum photon spectrum in which we are interested). Note that the inclusion of line photons makes the resulting limits much more aggressive, compared to the case if only continuum photons were present, for $M \lesssim 100$ GeV (the exact cutoff here depends on the identity of $X$ for a $\gamma X$ annihilation mode). We have made no attempt to convolve any photon lines here with the Fermi-LAT experimental resolution; doing so would smooth some of the sharp transitions evident in our result plots at values of $M$ where a line crosses one of the bin edges (see \sectref{results}), but would make no other qualitative changes.

Ref.\ \cite{Hooper1209} also presents limits in a model-dependent form
assuming 100\% branching fraction to, in turn, $WW$, $ZZ$, $b\bar{b}$,
$c\bar{c}$, $\mu^+\mu^-$, and $\tau^+\tau^-$ final states.  We find
that there is a multiplicative factor of \textit{ca.}\ 2.8 discrepancy
between the model-independent (more stringent) and the more
model-dependent (more conservative) limits once the continuum spectra
are used to convert the model-independent limits to limits on the
specific final states listed above. Following
Hooper (private communication), we resolve this discrepancy
in favor of the more conservative limits. We do this by multiplying
all $\sigma v$ limits set for our operators using the
model-independent $\mathcal{F}$, as described in the paragraph above,
by a factor of 2.8.

\subsection{Continuum limits: Fermi-LAT}
\label{sect:dwarf_limits}
Assuming particular annihilation channels, the Fermi-LAT Collaboration \cite{FERMI1108} also presents 95\% CL UL on the inclusive photon spectrum using observations of 10 Milky Way dwarf companions (the ``stacked dwarf'' limits). Their limits were obtained by performing a simultaneous maximum likelihood analysis on all 10 dwarfs in which the expected DM annihilation signal was fitted along with a diffuse galactic background model to 2 years of Fermi-LAT data for photons in the energy range 0.2 to 100 GeV. The normalizations of the two components were floated in the fit and cross section limits were extracted from the normalization of the DM signal component assuming in turn, that the annihilation was 100\% via each of the modes $b\bar{b}$, $WW$ (equivalent to $ZZ$ for these purposes), $\mu^+\mu^-$, and $\tau^+\tau^-$. 

Unfortunately, since this analysis is sensitive to spectral shape, and since the continuum photon spectrum for annihilation modes have different shapes in general, it is not possible to simply reverse-engineer these limits to set the exact 95\% CL UL considering all the continuum contributions from a combination of annihilation modes (which is the case for the operators we consider). We therefore extract approximate and conservative limits on $\lb[ \sigma v \rb]_{\text{total}}$ 
by requiring that the individual $WW/ZZ$, $b\bar{b}$, $\mu^+\mu^-$, and $\tau^+\tau^-$ contributions to $\sigma v$ for any operator do not violate the individual experimental upper bounds. For a given final state, we take the limit
\begin{align}
\lb[ \sigma v \rb]^{95\% \text{ CL UL}}_{\text{total, from } f} = 2 \times \frac{ \lb[ \sigma v \rb]^{95\% \text{ CL UL}}_{f} }{ \text{BR}_f },
\end{align}
where $\lb[ \sigma v \rb]^{95\% \text{ CL UL}}_{f}$ is the limit taken directly from Ref.\ \cite{FERMI1108} and BR$_f$ is the branching fraction for the annihilation mode $f \in \left\{ WW/ZZ\right.$, $b\bar{b}$, $\mu^+\mu^-$, $\left.\tau^+\tau^-\right\}$ computed using the analytic expressions in CKW. The factor of 2 is a re-interpretation of a limit set for self-conjugate DM in Ref.\ \cite{FERMI1108} to our choice of fermion or complex scalar DM.

These constraints from dwarf galaxies as we have implemented them are conservative and approximate and are not necessarily indicative of the full ability of such measurements to constrain these operators. Based on Fig. 14 of Ref.~\cite{Hooper1209}, we expect the exclusion reach for an analysis which took into account the entire continuum spectrum at once rather than only looking independently at each component of the spectrum from each final state and comparing to the relevant final state limit from Ref.~\cite{FERMI1108} (thereby paying a cost of BR$_f$) would be similar to the inclusive galactic center limits from Ref.~\cite{Hooper1209} in cases where there are no lines.

Finally, we note that as
this paper was being finalized, the Fermi-LAT collaboration released
updated constraints stacking 25 dwarf satellites based on 4 years of data
\cite{::2013yva}. These limits are {\it weaker} than expected, and in
particular weaker than the limits we take from Ref.\ \cite{FERMI1108}. However,
since these limits are not constraining (as implemented), the updated
analysis in Ref.\ \cite{::2013yva} ultimately does not affect our results.

\subsection{Continuum-to-line ratio limits: Cohen \textit{et al.}\ analysis}
\label{sect:Cohen_supersaturation}

\citet{Wacker1207} finds $5.5\sigma$ (local) evidence for a photon
line in the 3.7-year Fermi-LAT data at roughly 130 GeV, and based on
the same data, presents limits on the ratio $R$ of the annihilation
cross section via modes which give rise to continuum photons, to the
annihilation cross section giving lines ($\gamma \gamma$ and/or
$\gamma Z$). These limits are based on observations of photon fluxes
from Fermi-LAT in the energy range 5 to 200 GeV in an annulus around
the GC with inner and outer radii $1^\circ$ and $3^\circ$,
respectively.  We compare the two classes of $R$ limits in
Ref.\ \cite{Wacker1207} -- shape and supersaturation -- to the ratio
$R_{\text{th}} = \left(\sigma_{WW}v + \sigma_{ZZ}v +
\sigma_{b\bar{b}}v\right)/\left(2\sigma_{\gamma\gamma}v + \sigma_{\gamma
  Z}v \right)$, computed using the analytical expressions in CKW.

The supersaturation limits on $R$ are derived without assuming
a background model, requiring only that the photons from DM
annihilation do not supersaturate the observed photon spectrum. These
limits are only presented for the case where 100\% of the continuum
photon flux comes from annihilations going to a $W^+W^-$ primary final
state, but due to the similarity of the spectra, these limits are also
applicable if the annihilation is to $ZZ$ or $b\bar{b}$. Although the
limits are very conservative, they are more robust to details of
astrophysical backgrounds or the DM model than the shape constraints.

We briefly summarize the procedure they followed to obtain the
supersaturation limits:  for each $M$, the number of line photons was
found by performing a maximum likelihood analysis in which the photon
lines for $\gamma \gamma$ and $\gamma Z$ were fitted to binned
photon-count data, marginalizing over the relative normalization of
the lines.  The sum of the signal from the two lines was taken as the
number of line photons. To find the upper limit on the continuum
normalization, the energy bin where the continuum photon spectrum is
expected to peak relative to an assumed power-law astrophysical
background with a spectral index of 2.8 was first found (this is the
only point where spectral information was used); for a pure $WW$
annihilation mode, this was determined to be the bin 10 to 20 GeV. The
normalization for the continuum annihilation was then scaled up until
the number of photons in the 10 to 20 GeV bin violated the 95\% CL UL
from data. Correcting for the effective area in Fermi-LAT, this
results in a limit on the continuum-to-line cross section ratio.

The second (much stronger) class of limits presented in
Ref.\ \cite{Wacker1207} are shape limits, derived from a maximum
likelihood analysis including the two photon lines, the continuum
photons, and a single power-law background over the energy range 5 to
200 GeV. At each $M$, the fit was performed for the ratio of
continuum-to-line photons in that energy bin, marginalizing over all
other parameters including the relative strengths of the two
lines. This procedure effectively resulted in a 2-dimensional likelihood profile in the
$M-R$ plane; the 2$\sigma$ limit line in this $M-R$ plane is taken as
the 95\% CL UL on $R$. We consider their results for each
of the continuum-photon-producing final states in turn, 100\% $WW/ZZ$
and $b\bar{b}$; the $\mu^+\mu^-$ and $\tau^+\tau^-$ modes
were also considered there, but for our operators annihilation to
charged leptons is typically subdominant. 

It should be borne in mind that the analysis of
Ref.\ \cite{Wacker1207} marginalized over the $\sigma_{\gamma
  \gamma}v/\sigma_{\gamma Z}v$ ratio, while the operators here have
fixed line ratio for a given $M$. Furthermore, the limits are most applicable 
if there is some region of $M$ where a good
fit to the photon line is possible; note that the line fits are
discussed further in \sectref{Cohen_line} and also indicated in panel
C of our results figures. With more recent data \cite{FERMI1305}, evidence for the line has weakened, which requires fewer line photons and would lead to correspondingly weaker limits on the continuum-to-line ratio if this analysis were to be updated. Furthermore, Ref.\ \cite{Wacker1207}
assumes 100\% annihilation to each of the various
continuum-photon-producing modes, while the operators we consider
invariably have some combination of multiple
continuum-photon-producing annihilation modes. Note also that no
inverse Compton scattering (ICS) contribution was included in the
continuum spectrum, as it should be sub-dominant for $WW/ZZ$ and
$b\bar{b}$, and the contribution to the continuum spectrum of the $Z$
in the $\gamma Z$ channel is not included.  Finally, none of the
limits presented is applicable to operators with $\gamma h$ final
states as this annihilation mode introduces a new line which would
dramatically improve the goodness-of-fit in the $M\sim155$ GeV region.

Nevertheless, even given all these issues, the limits remain
indicative for the case where a least one
of the annihilation modes $\gamma\gamma$ or $\gamma Z$ is present, and
the operator's continuum photon spectrum is dominated by one of
$WW/ZZ$ or $b\bar{b}$. Overall, one should interpret the limits
qualitatively, not quantitatively: for operators with
continuum-to-photon ratios near the limit, one would really have to be
careful in re-doing the full likelihood fit with the correct known
branching ratios for the various modes for the specific operator in
question to definitively settle the issue of whether or not the
operator is excluded.

\subsection{Line-like feature: Weniger-like analysis}
\label{sect:Wenigerlimits}
It is interesting to examine whether it is possible for the photon
lines arising from these operators to explain, with the correct
normalization, the line-like feature near 130 GeV\footnote{We shall
  assume here that this line is at exactly 130 GeV, although the most recent best-fit for the line energy is 133 GeV \cite{FERMI1305}.} first reported at
$4.6\sigma$ local significance by \citet{Weniger1204} in the 3.7-year
Fermi-LAT data, assuming for these purposes that we take this feature
seriously as a DM signal.

Assuming the annihilation is to $\gamma \gamma$, \citet{Weniger1204}
reports values for $\sigma v$ required to account for this
feature. For a $\gamma \gamma$ mode, the DM mass must clearly be 130
GeV to explain the line. For the operators we consider here, there may
be annihilation modes to any of the final states $\gamma X$ where $X =
\gamma, Z, h$, where the DM mass must be $M = $ 130, 144, and 155 GeV,
respectively, to explain the observed line. Eq.\ \eqref{eq:flux}
indicates that we must account for this change in mass by performing a
rescaling of the cross section result from
Ref.\ \cite{Weniger1204}. To be explicit, where an operator has
annihilation modes to $\gamma X \ (X = Z,h)$, we rescale the $\gamma
\gamma$ result $\lb[ \sigma v \rb]_{\text{Weniger}}$ from
Ref.\ \cite{Weniger1204} as follows:
\begin{align}
\lb[ \sigma v \rb]_{\gamma X} = 2\times 2 \times \lb( \frac{M}{130 \text{ GeV} } \rb)^2  \lb[ \sigma v \rb]_{\text{Weniger}},
\label{eq:weniger1}
\end{align}
where one factor of 2 is due to the assumption of $\gamma \gamma$ 
and the other factor of 2 is a re-interpretation of a limit set for
self-conjugate DM in \cite{Weniger1204} to our choice of fermion or
complex scalar DM [see Eq.\ \eqref{eq:flux}]. In addition, we perform
the rescaling necessary to account for differing halo profile
normalizations, as discussed in \sectref{rescaling}; the rescaling
factors we use are given in \tabref{paper_norms}.

Owing once again to the only partial resolution in the Fermi-LAT data of the photon lines,
we cannot na\"ively use the required line cross section from Ref.\ \cite{Weniger1204} to give individual required cross sections for the $\gamma \gamma$ and $\gamma Z$ modes when both are present (see discussion in \sectref{Fermi_line_limits}). Assuming either $M = 130$ GeV or $M=144$ GeV, the secondary line at either 114 or 144 GeV, respectively, would give a significant photon contribution at the position of the line at 130 GeV and/or vice versa depending on the relative normalizations of the lines. Similarly, since the line separation here is still on the order of the Fermi-LAT energy resolution (even if assumed to be 10\%), one also cannot assume the lines are completely unresolved to derive a single required combined value for $(2\sigma_{\gamma\gamma} + \sigma_{\gamma Z})v$.  

Nevertheless, while we acknowledge this issue, for the purpose of giving a rough estimate of the annihilation cross section which would be required for our operators with both $\gamma \gamma$ and $\gamma Z$ annihilation modes to explain the line reported at 130 GeV in Ref.\ \cite{Weniger1204}, we have performed the following approximate analysis. We note that it is usually the case that one or other of the lines from either the $\gamma \gamma$ or $\gamma Z$ mode is dominant for $M$ in the approximate range 120 to 150 GeV (in terms of photon flux, for operators VI-1, IV-2, VIII-1, or VIII-2, it is the $\gamma\gamma$ mode which dominates by about a factor of 4, while for operators VI-3, VI-4, VIII-3, or VIII-4, it is the $\gamma Z$ mode which dominates by about a factor of 2.5). Therefore, we shall make the approximation that the line at 130 GeV reported in Ref.\ \cite{Weniger1204} is entirely due to the dominant of the $\gamma X\ (X = \gamma, Z)$ modes. As noted above, there will be a second line in either case, but we ignore this complication in our approximate treatment here (see the next section for a slightly more careful treatment of the same issue making use of results from a different reference). We give results for the quantity $\lb( 2\sigma_{\gamma\gamma} + \sigma_{\gamma Z} \rb) v$, requiring that the normalization of the dominant mode can explain the line at 130 GeV. To be explicit, if the $\gamma X \  (X = \gamma, Z)$ annihilation mode explains the line at 130 GeV, we extend Eq.\ \eqref{eq:weniger1} above to the two-line case:
\begin{align}
\label{eq:weniger2}
\lb[ \lb(2\sigma_{\gamma\gamma} + \sigma_{\gamma Z}\rb) v\rb] = 2 \times 2 \times \lb( \dfrac{M}{130 \text{ GeV} } \rb)^2  \lb[ \sigma v \rb]_{\text{Weniger}} \times \lb\{ \begin{array}{ll} \lb[ 1 + \dfrac{ \sigma_{\gamma Z}v(M)}{ 2\sigma_{\gamma\gamma}v(M)} \rb] & \text{for } M = 130\ \text{GeV, } \ X = \gamma \\[4ex] 
					 \lb[ 1 + \dfrac { 2\sigma_{\gamma\gamma}v(M)}{ \sigma_{\gamma Z}v(M)}\rb] & \text{for } M = 144\ \text{GeV, } X = Z \end{array} \rb. .
\end{align}
The factors in the brackets $[ \ \cdots \ ]$ are computed using the analytical expressions in CKW and merely rescale the individual required $\sigma_{\gamma X} v \ (X = \gamma, Z)$ cross sections for the line at 130 GeV to the quantity $\lb( 2\sigma_{\gamma\gamma} + \sigma_{\gamma Z} \rb) v$.

\subsection{Line-like features: Cohen \textit{et al.}\ analysis 
\label{sect:Cohen_line}}

\begin{figure}
\begin{center}
\includegraphics[width = 0.48\textwidth]{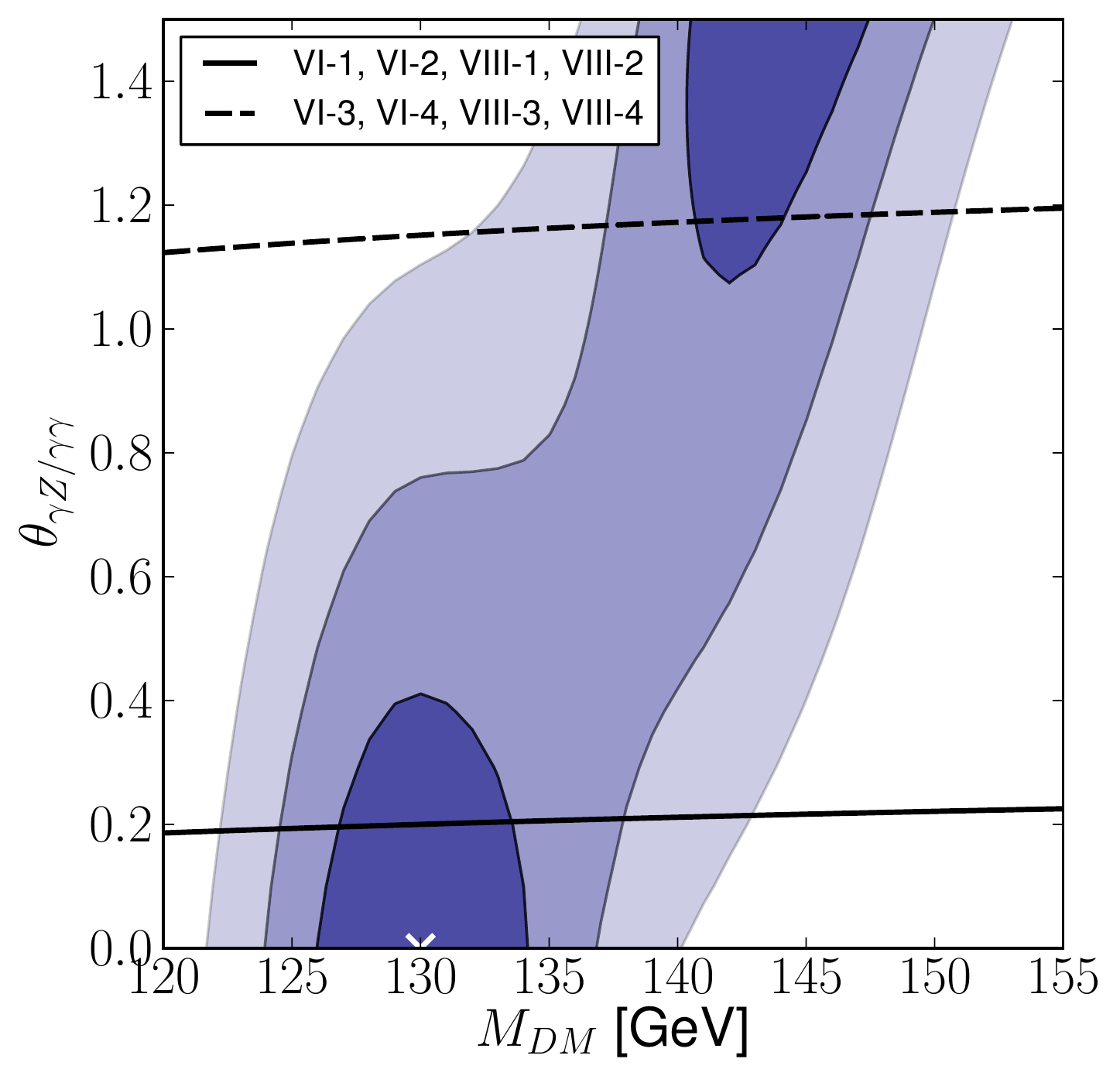}
\caption{\label{fig:Wacker_line} Contours adapted from Fig.\ 2 in \citet{Wacker1207}. The colored bands shown here in order of decreasing darkness are the $1$-, $2$-, and $3$-$\sigma$ delta-log-likelihood contours for the line-existence analysis performed in \citet{Wacker1207}, in which $\gamma\gamma$ and $\gamma Z$ lines plus a power-law background were fitted to the 3.7-year Fermi-LAT data; the white cross at $M = 130$ GeV and $\theta_{\gamma Z/\gamma\gamma} = \arctan\lb( N_{\gamma Z}/N_{\gamma\gamma} \rb) = 0$ is the best fit point corresponding to a $5.5\sigma$ local significance relative to a null hypothesis of no line. The solid and dashed black lines are the theoretical results for the relative normalization of the two lines $\theta_{\gamma Z/\gamma\gamma} = \arctan\lb( \sigma_{\gamma Z}v/2\sigma_{\gamma\gamma}v \rb)$, as computed using the analytical cross section formulae in CKW for the eight operators which have both lines present. Although there are eight operators, the results fall onto only two curves, which are indicated on the figure. There is clearly a region of $M$ parameter space for any of these operators where the \emph{relative} line normalization is such that the double-line provides a good (delta-log-liklihood $< 1\sigma$) fit to the Fermi-LAT data.}
\end{center}
\end{figure}

In addition to presenting limits for the ratio of continuum-to-line cross sections, \citet{Wacker1207} also presents evidence for the existence for a line in the Fermi-LAT 3.7-year data in an annulus centered on the GC with inner and outer radii of $1^\circ$ and $3^\circ$, respectively, with local significance of 5.5$\sigma$ relative to the null hypothesis of no line. To reach this conclusion, they performed a likelihood analysis fitting  binned photon count data in the energy range 5 to 200 GeV  to a single power-law background plus lines at $E_\gamma = M$ (with normalization $N_{\gamma\gamma}$) and at $E_\gamma = M - m_Z^2/4M$ (with normalization $N_{\gamma Z}$), both suitably convolved with the Fermi-LAT energy dispersion. No continuum photon contribution from DM annihilation was included in this analysis. In the fit, $M$ and $\theta_{\gamma Z/\gamma\gamma} = \arctan\lb(N_{\gamma Z}/N_{\gamma\gamma} \rb)$ were scanned over while the background normalization and spectral index, as well as the total line normalization $N_{\gamma\gamma} + N_{\gamma Z}$, were marginalized over. The best fit point for this analysis was found to be $M = 130$ GeV and $\theta_{\gamma Z/\gamma\gamma} = 0$. (An earlier analysis \cite{Rajaraman:2012db} fitting the data with $\gamma \gamma$ and $\gamma Z$ lines found very similar results; we have followed \citet{Wacker1207} since that work also gave limits on the continuum-to-line ratio.)

\citet{Wacker1207} presents $1$-, $2$-, and $3$-$\sigma$ delta-log-likelihood contours for their fitting procedure in the $M - \theta_{\gamma Z/\gamma\gamma}$ plane, which we have reproduced in \figref{Wacker_line}, along with the theoretical values for $\theta_{\gamma Z/\gamma\gamma} = \arctan\lb(\sigma_{\gamma Z}v/2\sigma_{\gamma\gamma}v \rb)$ for the eight operators that have both $\gamma\gamma$ and $\gamma Z$ lines, as computed using the analytical expressions of CKW. We see that the \emph{relative} normalizations of the lines for these operators are such there are regions around $M =130$ GeV or $M=144$ GeV where the sum of the two lines could fit the Fermi-LAT data very well (within $1\sigma$ in delta-log-likelihood). 

Cohen \textit{et al.}\ also gives contours displaying the total normalization (photon count) $N_{\text{total}} \equiv N_{\gamma\gamma} + N_{\gamma Z}$ of the lines in the $M - \theta_{\gamma Z/\gamma\gamma}$ plane which, knowing the ROI-averaged mission-time-integrated exposure-times-effective-area $\mathcal{E}_{\text{ROI}}$ relevant to the Fermi-LAT data they considered, would allow us to extract the cross sections required for the operators with both $\gamma\gamma$ and $\gamma Z$ modes to explain the fitted lines. As $\mathcal{E}_{\text{ROI}}$ was not relevant to the analysis presented in Ref.\ \cite{Wacker1207}, it was not given there. However, we have made use of the Fermi ScienceTools package to reproduce a sufficient portion of the data-extraction and analysis in that reference to enable us to compute it ourselves (see \appref{exposure} for the full details of our analysis and the cross-checks we have performed). We find that the ROI-averaged exposure factor for photon energies from \textit{ca.}\ 120 to 150 GeV (applicable to the line-fits presented in Ref.\ \cite{Wacker1207}) is $\mathcal{E}_{\text{ROI}} = 1.05\eten{11}$cm$^2$s. Knowing this factor, and extracting the fitted total line-normalization $N_{\text{total}}$ from the contours in Fig.\ 2 of \citet{Wacker1207}, we have computed the photon flux corresponding to the fitted line normalization as 
\begin{align}
\Phi = \frac{N_{\text{total}}}{ \mathcal{E}_{\text{ROI}} }.
\end{align}
Finally, making use of Eq.\ \eqref{eq:flux} we have converted this to the annihilation cross section required as a function of $M$ to explain the line for our particular operators having both $\gamma \gamma $ and $\gamma Z$ modes,
\begin{align}
(2\sigma_{\gamma\gamma} + \sigma_{\gamma Z})v = \frac{16\pi M^2}{J} \Phi =  \frac{16\pi M^2}{J} \frac{ N_{\text{total}} }{ \mathcal{E}_{\text{ROI}} },
\label{eq:cohenlinelimit}
\end{align}
for the (operator-specific) range of $M$ within the 3$\sigma$ $|$delta-log-likelihood$|$ contours shown in \figref{Wacker_line}. In performing this conversion, we have also computed the $J$ factors for the ROI in Ref.~\cite{Wacker1207}, which we give in Appendix~\ref{app:exposure}.

\section{Results}
\label{sect:results}

\begin{figure}
\begin{center}
\includegraphics[width=0.95\textwidth]{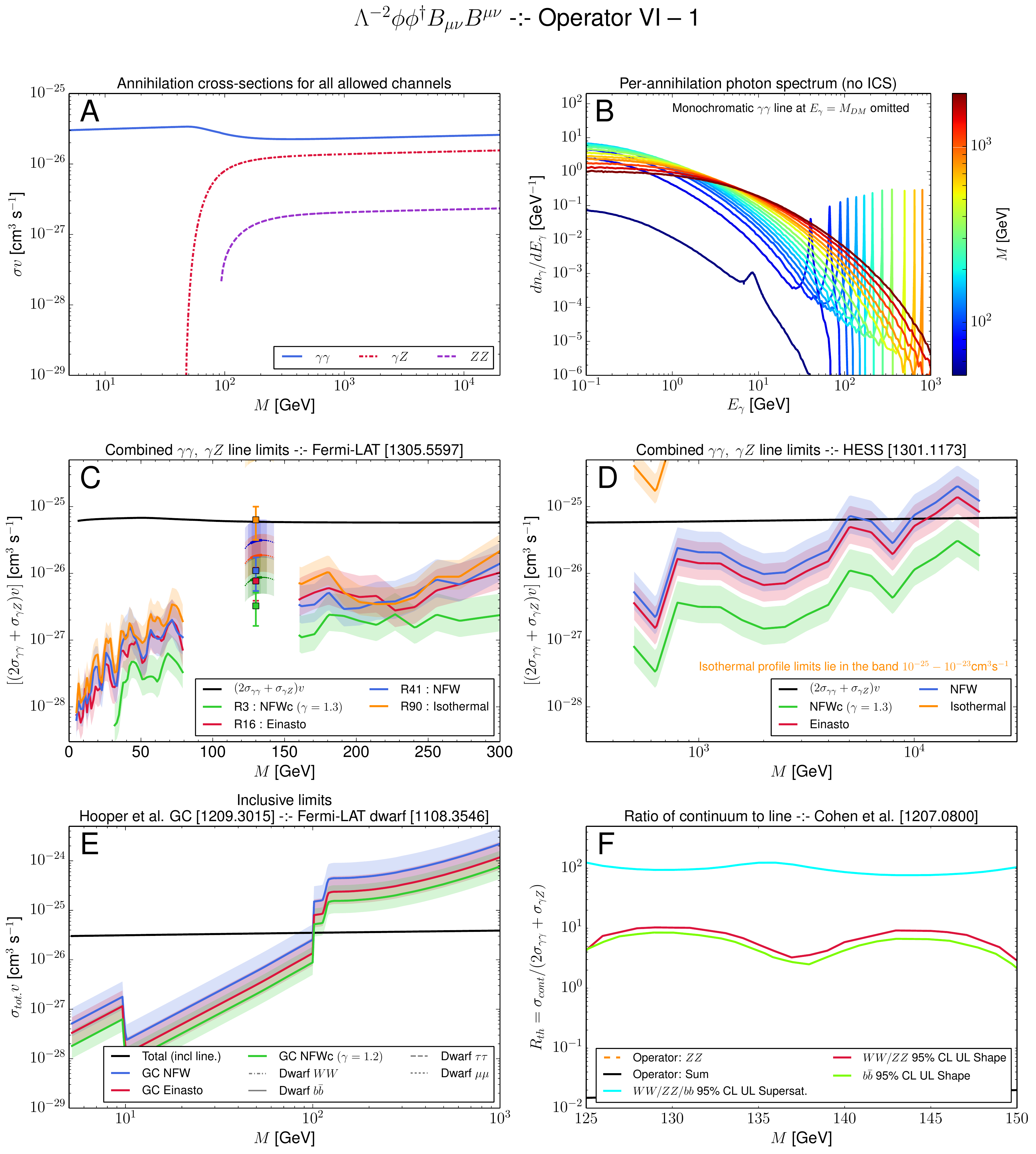}
\caption{\label{fig:TabVI_OP1} Figure captions are provided in \sectref{panela} through \sectref{panelf}.}
\end{center}
\end{figure}

\begin{figure}
\begin{center}
\includegraphics[width=0.95\textwidth]{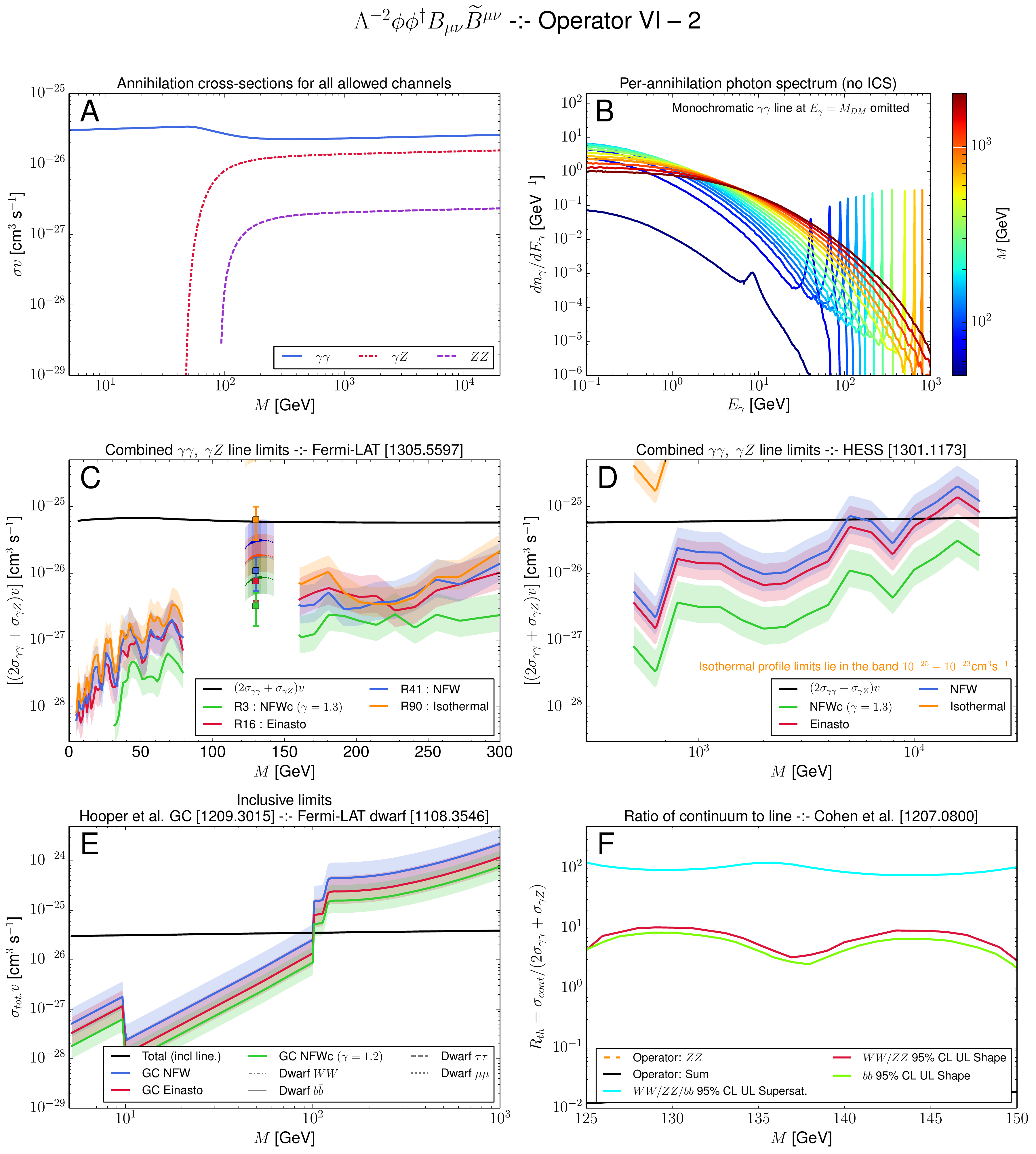}
\caption{\label{fig:TabVI_OP2} Figure captions are provided in \sectref{panela} through \sectref{panelf}.}
\end{center}
\end{figure}

\begin{figure}
\begin{center}
\includegraphics[width=0.95\textwidth]{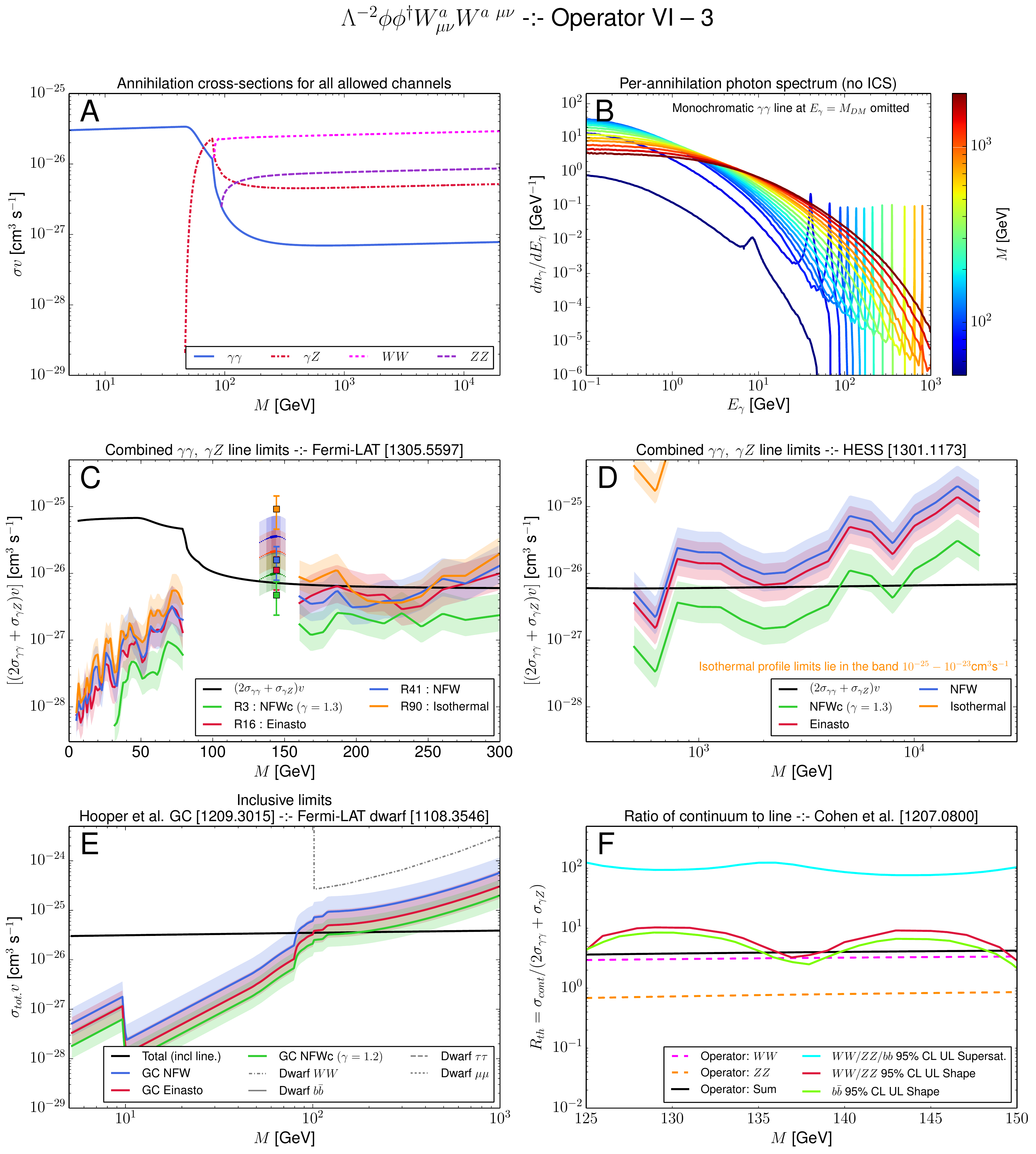}
\caption{\label{fig:TabVI_OP3} Figure captions are provided in \sectref{panela} through \sectref{panelf}. This operator may be compatible with present experimental limits and capable of accounting for the 130 GeV photon line for some profile choices.}
\end{center}
\end{figure}

\begin{figure}
\begin{center}
\includegraphics[width=0.95\textwidth]{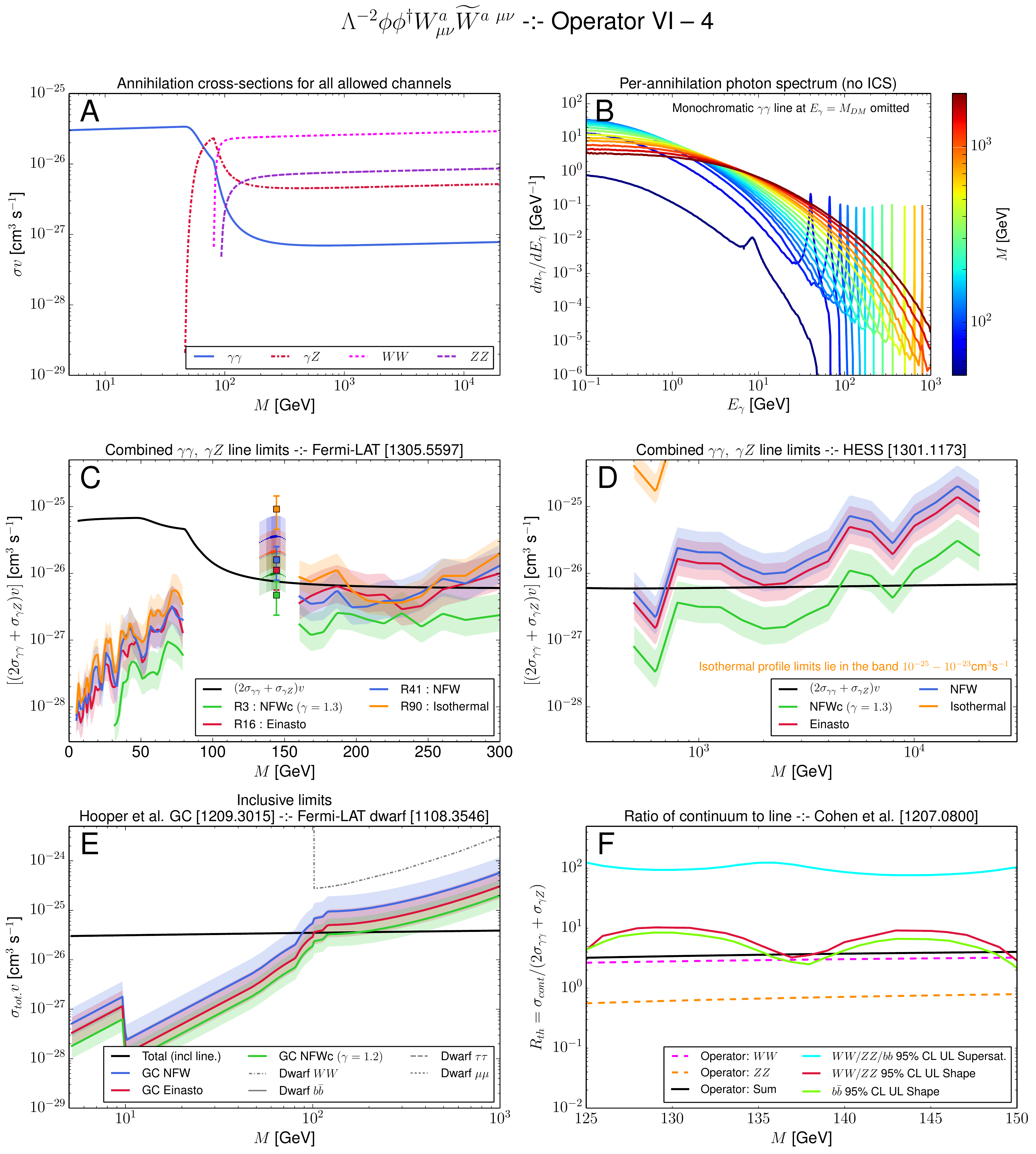}
\caption{\label{fig:TabVI_OP4} Figure captions are provided in \sectref{panela} through \sectref{panelf}.This operator may be compatible with present experimental limits and capable of accounting for the 130 GeV photon line for some profile choices.}
\end{center}
\end{figure}

\begin{figure}
\begin{center}
\includegraphics[width=0.95\textwidth]{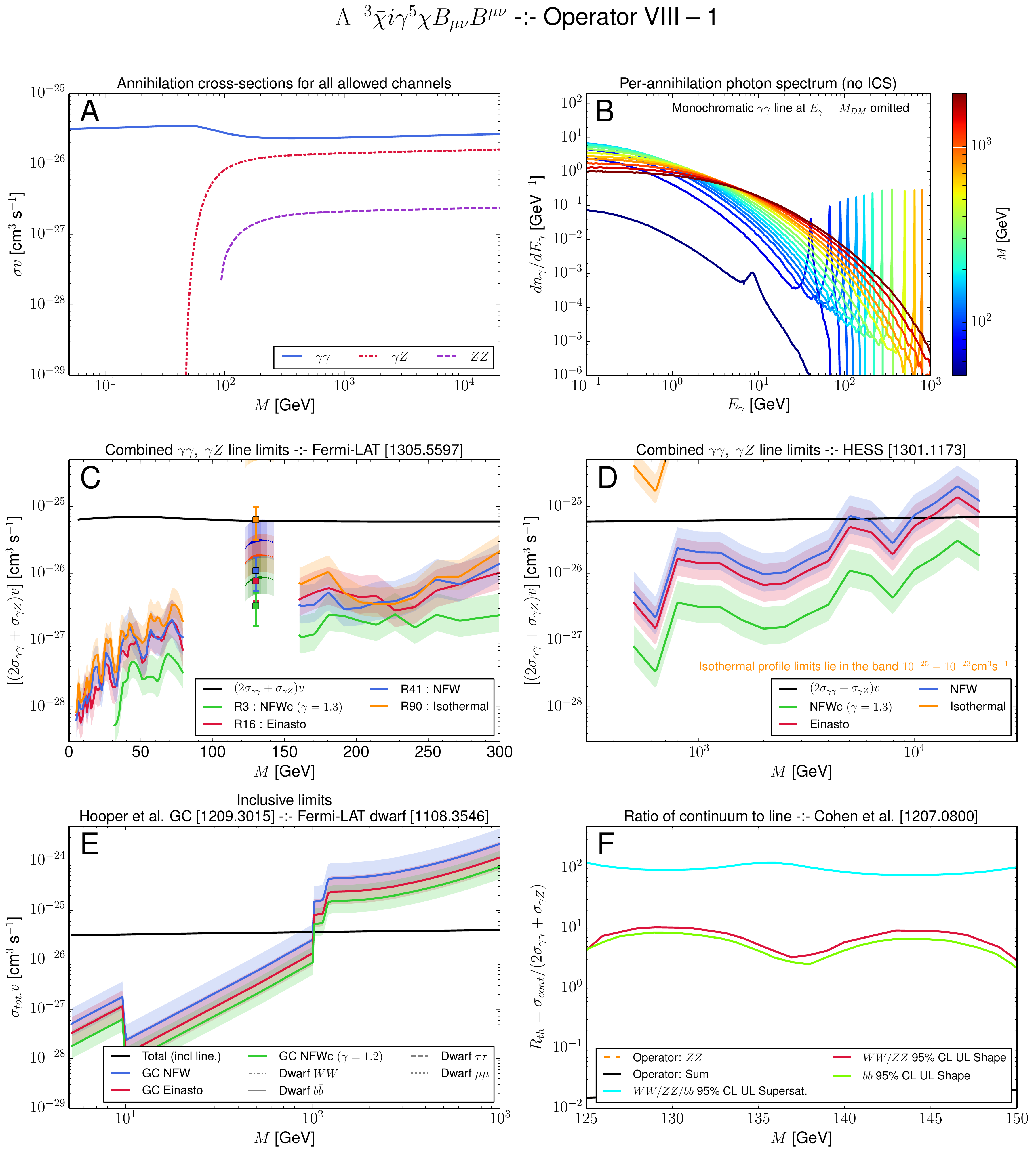}
\caption{\label{fig:TabVIII_OP1} Figure captions are provided in \sectref{panela} through \sectref{panelf}.}
\end{center}
\end{figure}

\begin{figure}
\begin{center}
\includegraphics[width=0.95\textwidth]{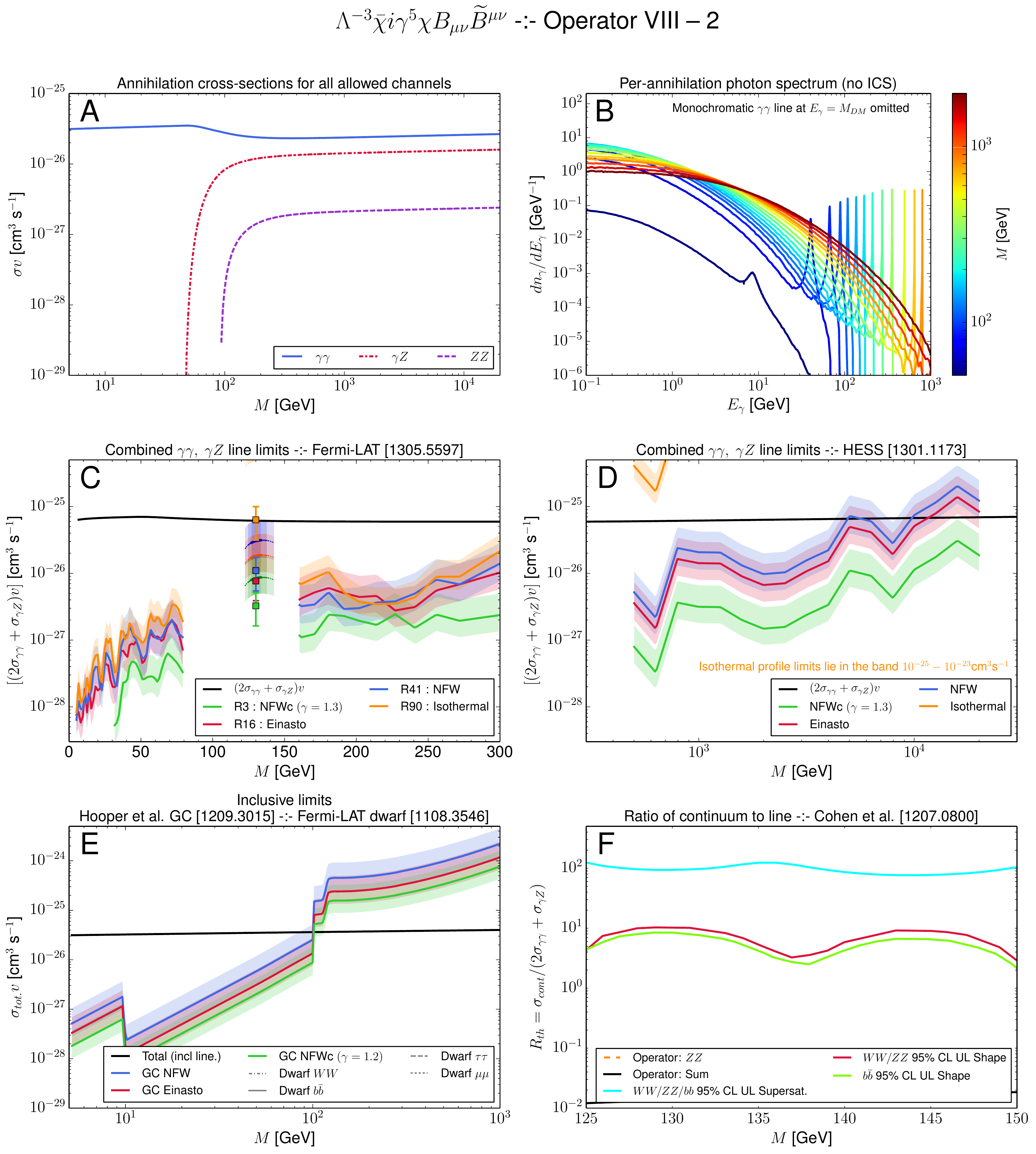}
\caption{\label{fig:TabVIII_OP2} Figure captions are provided in \sectref{panela} through \sectref{panelf}.}
\end{center}
\end{figure}

\begin{figure}
\begin{center}
\includegraphics[width=0.95\textwidth]{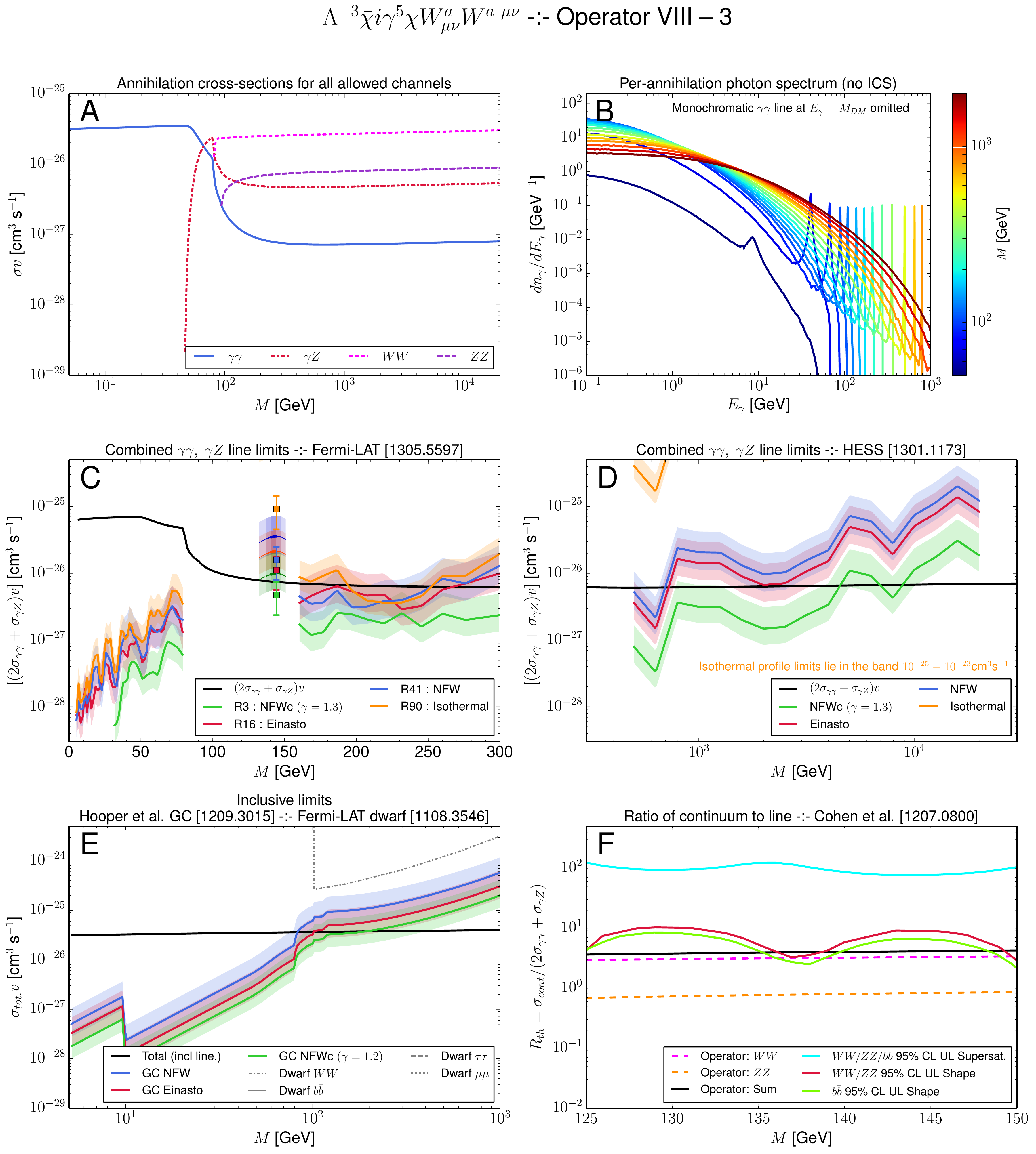}
\caption{\label{fig:TabVIII_OP3} Figure captions are provided in \sectref{panela} through \sectref{panelf}. This operator may be compatible with present experimental limits and capable of accounting for the 130 GeV photon line for some profile choices.}
\end{center}
\end{figure}

\begin{figure}
\begin{center}
\includegraphics[width=0.95\textwidth]{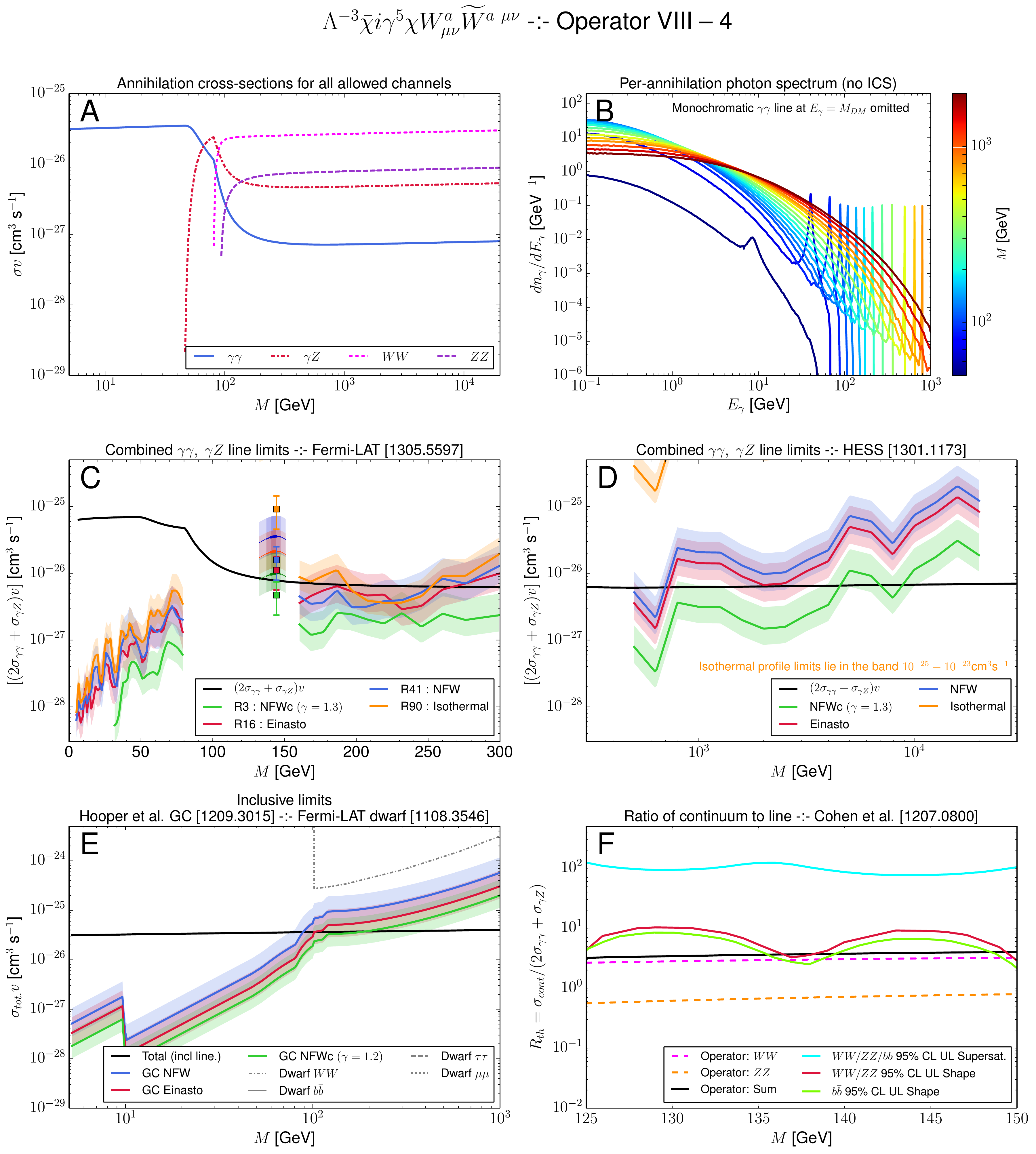}
\caption{\label{fig:TabVIII_OP4} Figure captions are provided in \sectref{panela} through \sectref{panelf}. This operator may be compatible with present experimental limits and capable of accounting for the 130 GeV photon line for some profile choices.}
\end{center}
\end{figure}

\begin{figure}
\begin{center}
\includegraphics[width=0.95\textwidth]{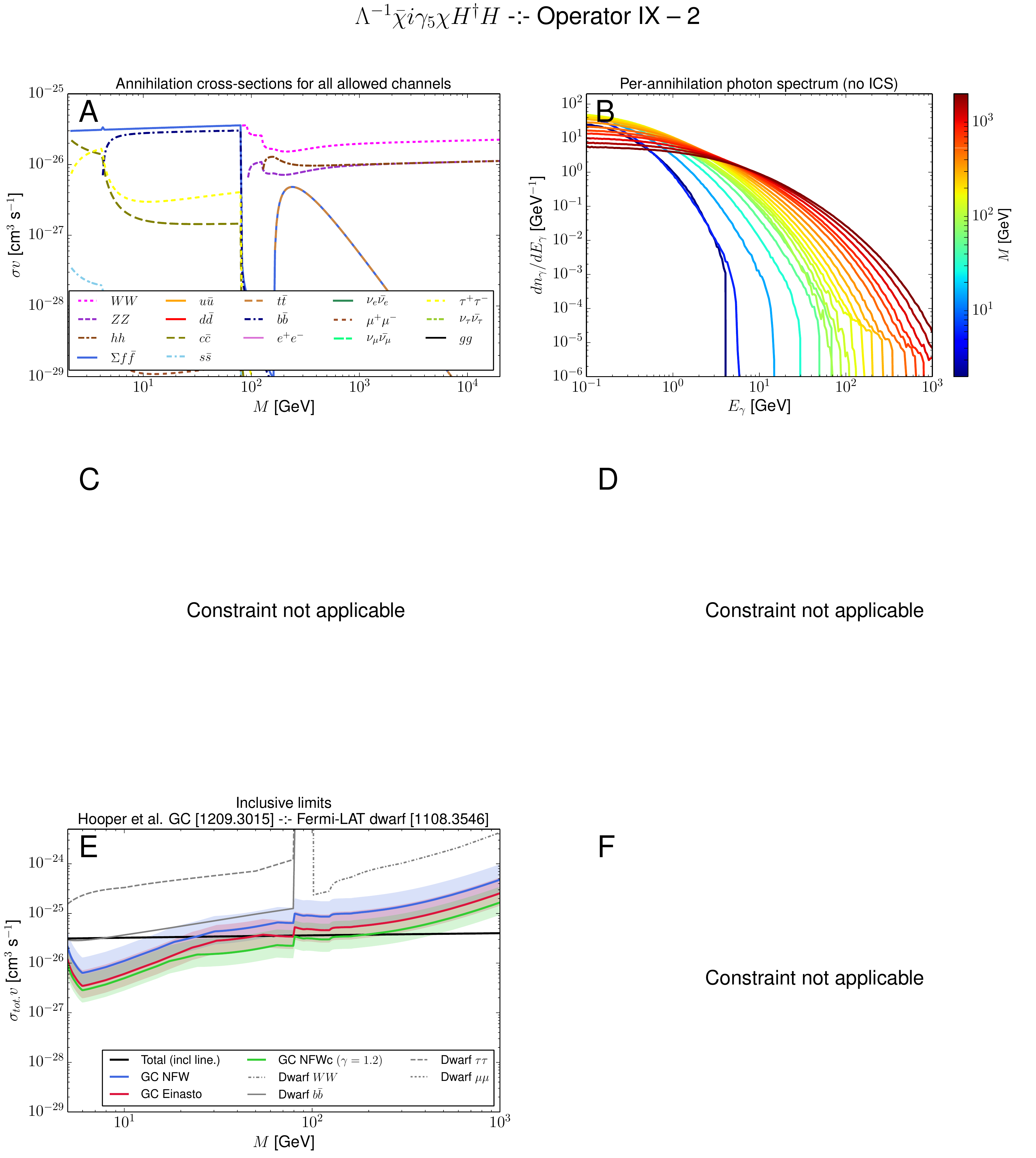}
\caption{\label{fig:TabIX_OP2} Figure captions are provided in \sectref{panela} through \sectref{panelf}. The DM annihilation for this operator proceeds via $s$-channel Higgs exchange (except for the $4$-body contact contribution to the $hh$ final state), and we have only utilized the cross sections for the 2-body on-shell final states that couple at tree-level to the $h$; we have however checked that including the loop-induced couplings to the $gg$ final state makes only a negligible difference to the limits. We have not quantitatively estimated the effect of the $3$- or $4$-body branchings involving intermediate off-shell $W/Z$ bosons, but again we would not expect the limits to change too dramatically.}
\end{center}
\end{figure}

\begin{figure}
\begin{center}
\includegraphics[width=0.95\textwidth]{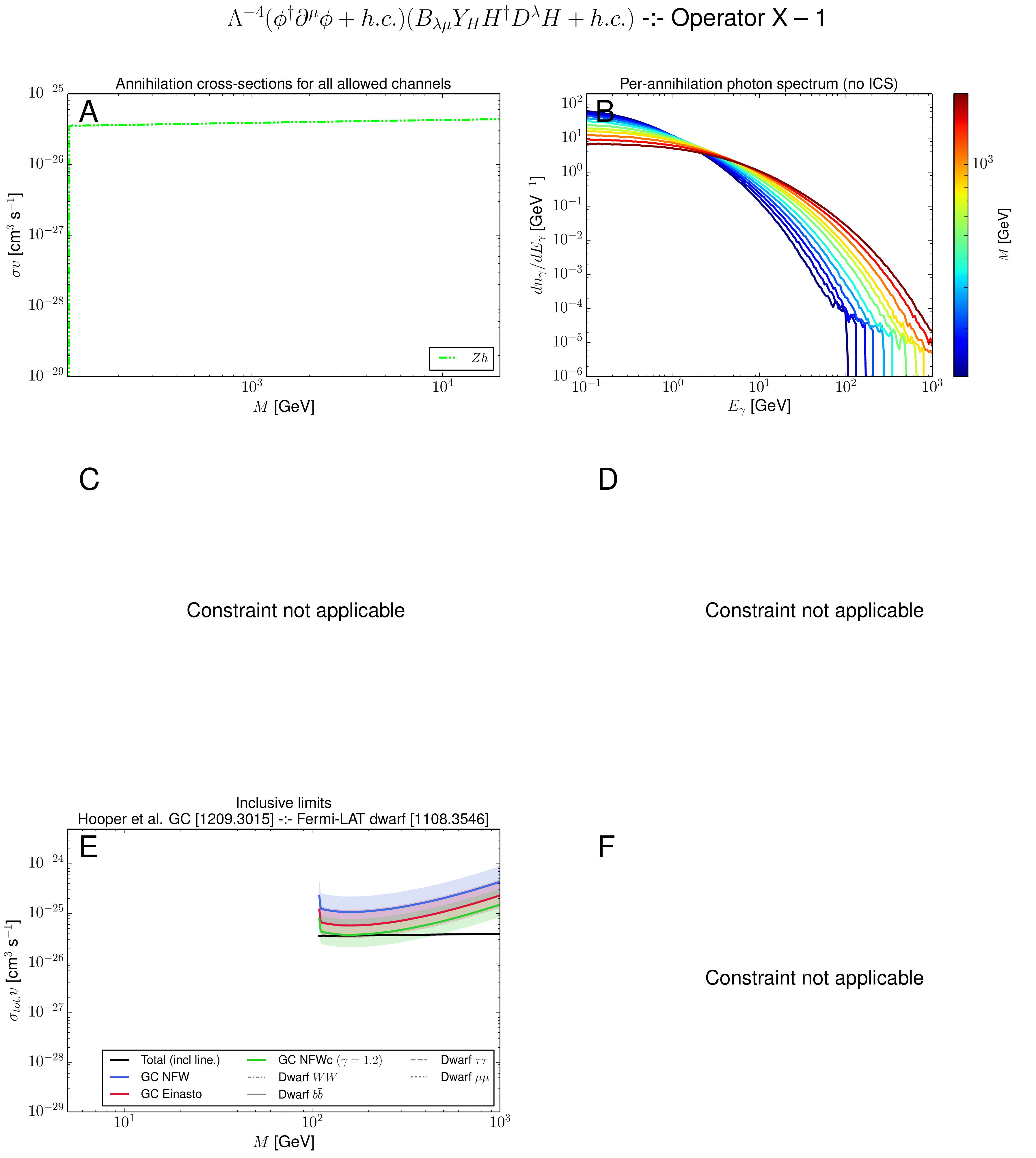}
\caption{\label{fig:TabX_OP1} Figure captions are provided in \sectref{panela} through \sectref{panelf}.}
\end{center}
\end{figure}

\begin{figure}
\begin{center}
\includegraphics[width=0.95\textwidth]{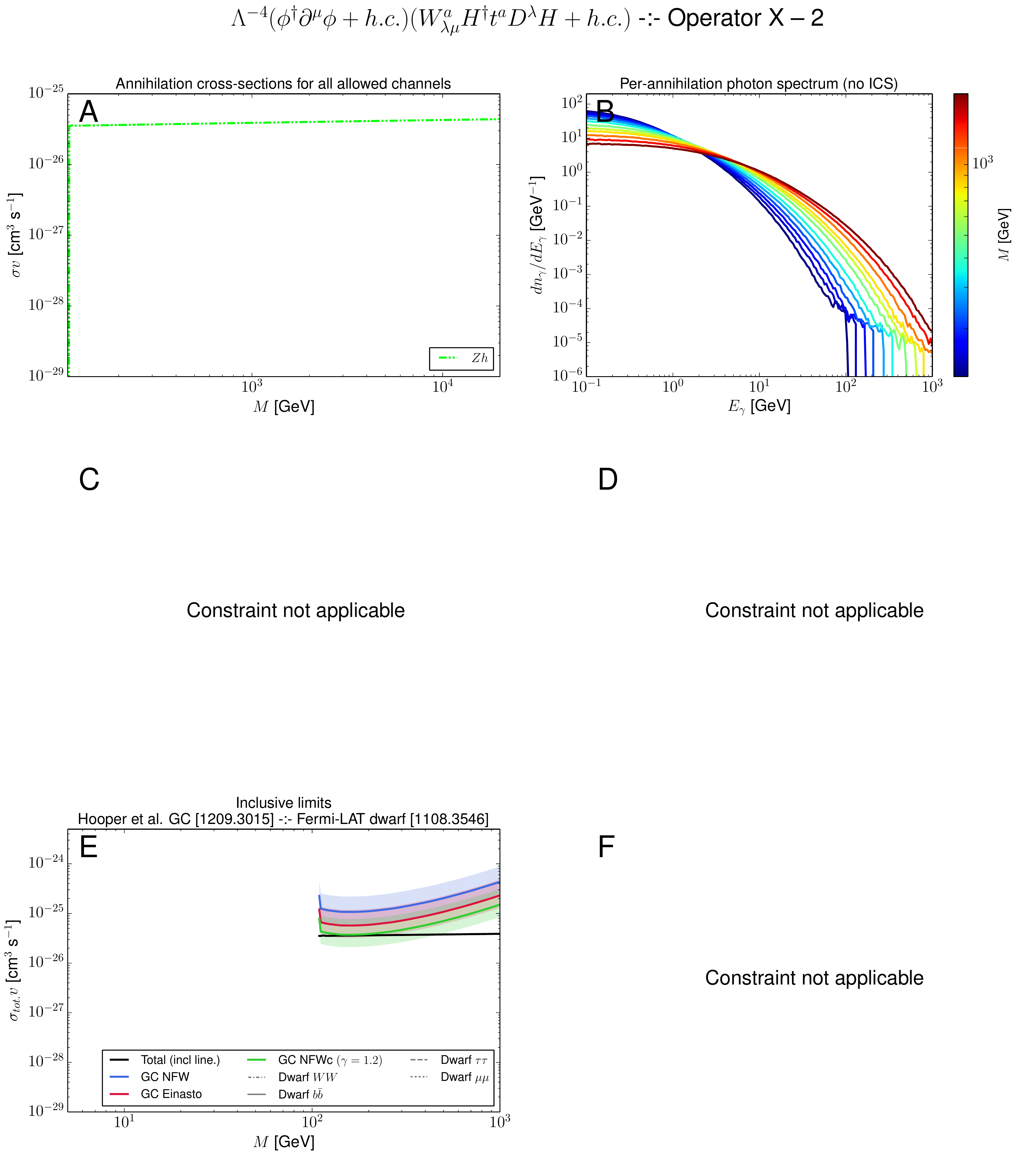}
\caption{\label{fig:TabX_OP2} Figure captions are provided in \sectref{panela} through \sectref{panelf}.}
\end{center}
\end{figure}

\begin{figure}
\begin{center}
\includegraphics[width=0.95\textwidth]{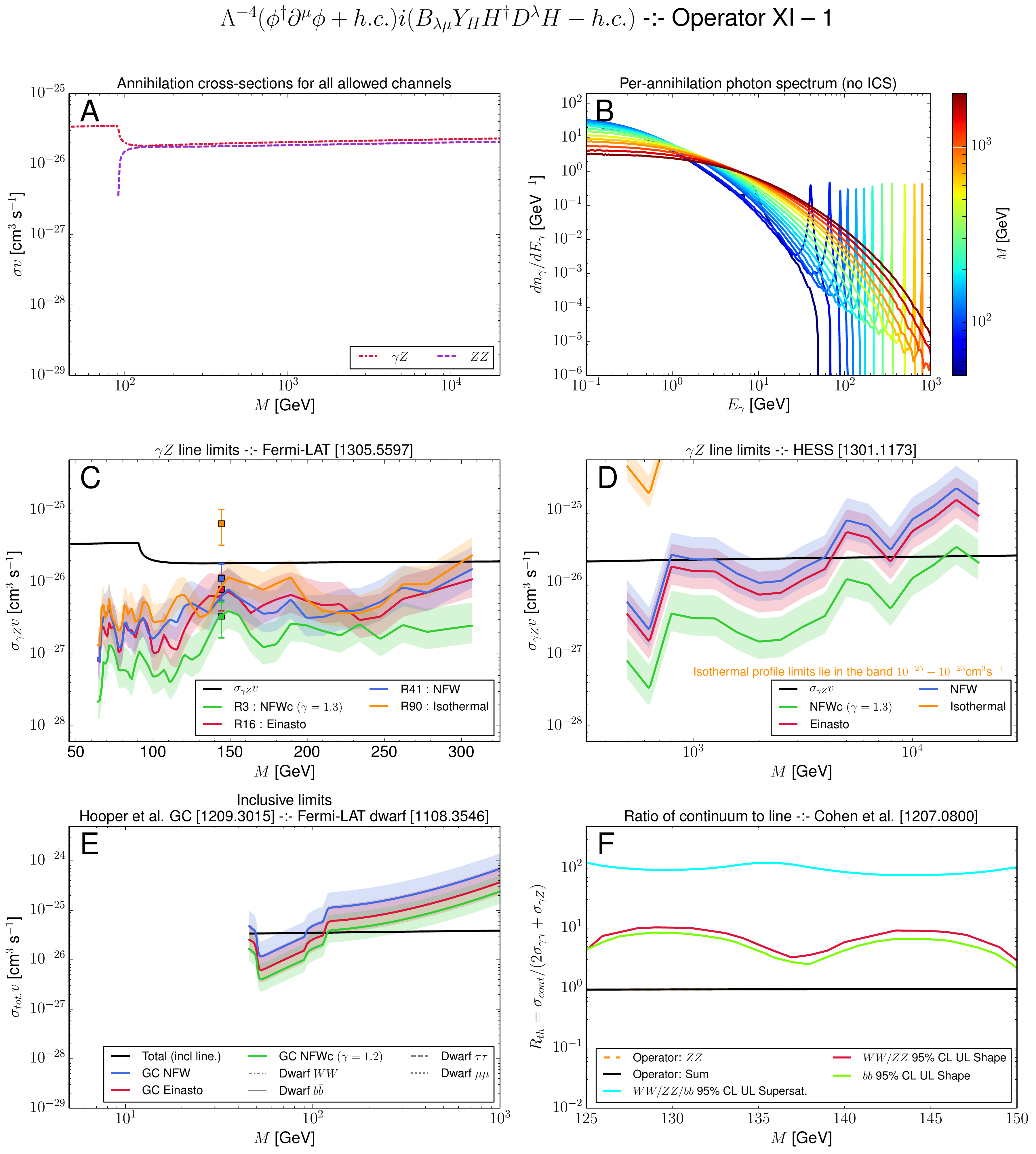}
\caption{\label{fig:TabXI_OP1} Figure captions are provided in \sectref{panela} through \sectref{panelf}.}
\end{center}
\end{figure}

In Figs.\ \ref{fig:TabVI_OP1} through \ref{fig:TabXX_OP1} we present
the indirect detection limits for all operators with $s$-wave
annihilations as studied in CKW.  The cross sections for various
operators are listed in Tables VI-XX of CKW.  We refer to a process
studied in CKW by a table number, and an operator number.  For
instance, operator VI-3 is the third operator of Table VI in CKW, in
this case, $\phi^\dagger\phi W^a_{\ \mu\nu}W^{a\, \mu\nu}$.  Our
notation for field operators will follow the notation in CKW.

There is one figure for each operator consisting of up to six panels.
The captions for each panel are the same for each figure and described
below. In addition, we make operator-specific comments in some
individual figure captions.

\subsection{Panel A: \label{sect:panela}} 
We show the non-relativistic DM annihilation cross sections $\sigma v$ as a
function of DM mass for each of the annihilation modes allowed for
the particular EFT operator, assuming that a) the DM annihilates only
through that operator and b) the DM particle is a thermal relic with
$\Omega_{DM} h^2 = 0.12$.

If all annihilation modes are pure $s-$wave, the total cross section for annihilation attains a value of around $\sigma v \approx 3.6\times 10^{-26}$cm$^{3}$s$^{-1}$ for $M =100$ GeV [showing logarithmic variation with $M$; \textit{cf.}\ Eq.\ \eqref{eq:relic_density}]. However, $p-$wave components to some annihilation channels can cause the the present-day total annihilation cross section to be suppressed due to their non-negligible impact in the early universe when the relic density is set; these contributions naturally drive $\Lambda$ larger and hence the present-day annihilation cross section smaller. This effect is particularly pronounced in operators where the $s-$wave annihilation is essentially independent of $M$ but the $p-$wave annihilation contribution grows with $M$, \textit{e.g.}, the operator XVII-1, $\Lambda^{-4} \bar{\chi}\gamma^{\mu5}\chi i \lb( B_{\lambda\mu}Y_H H^\dagger D^\lambda H - \text{h.c.} \rb)$.

Ignoring kinematic thresholds, for all final states involving fermions except those in Table IX of CKW, the fermion contributions separate into four distinct types: leptons with $T^3 = \pm 1/2$ and quarks with $T^3 = \pm 1/2$ (\textit{cf.}, the results for all operators XVIII-XX).

\subsection{Panel B: \label{sect:panelb}}
The per-annihilation total differential spectrum $dN_\gamma/dE_\gamma
= \sum_f \text{BR}_f \cdot dN^f_\gamma/dE_\gamma$ of prompt gamma rays
as a function of $E_\gamma$ is plotted for different values of DM mass, indicated by the color
scale. The mass dependence of the spectra depends on whether the
dominant final states are fermions or if they are gauge bosons and
Higgs. In the latter case, the spectrum becomes harder with increasing
DM mass because the final states become more boosted. Further details
on these spectra can be found in \sectref{spectra}. No inverse Compton
scattering (ICS) component is included.  An estimate of the error introduced by neglect of ICS is given in \appref{ICS}.

\subsection{Panel C: \label{sect:panelc}}
If applicable, this panel shows the line search limits for the DM mass
region 5 GeV to approximately 300 GeV from the Fermi-LAT \cite{FERMI1305} line
search using observed photon fluxes from the Galactic center. We show
these limits in terms of the total annihilation rate to monochromatic
photons (\textit{i.e.,} the annihilation rate to each final state weighted by
the number of monochromatic photons in that final state). The solid
black line shows the specific $\sigma v$ required for a thermal relic.

The solid colored lines show 95\% CL UL for various halo profiles and
ROIs, assuming the central values for halo normalization $\rho_\odot$
from Ref.\ \cite{Iocco1107} (see \tabref{norms}). The like-colored
bands show the variation in the UL as $\rho_\odot$ is varied through
$2\sigma$ limits. For operators with multiple possible lines
(\textit{i.e.,} $\gamma \gamma$ and $\gamma Z$ annihilation modes) we
do not set limits in the region of DM masses 80 GeV to 160 GeV where
we expect the Fermi-LAT line shape would become a very broad or
double-peaked structure (see \sectref{Fermi_line_limits}).

If present, the colored squares with their error bars
represent $(M,\sigma v)$ values where the operator in question could
additionally supply the line reported in \citet{Weniger1204}, assuming
the central values for $\rho_\odot$ for each halo profile.  As
discussed in \sectref{Wenigerlimits}, for operators where there are
multiple photon lines we make the assumption that the line at 130 GeV
is due to the dominant photon line mode.

For operators with $\gamma \gamma$ and $\gamma Z$ lines, we show colored lines near 130-150 GeV, which indicate the cross sections required to explain a possible double-line feature in the Fermi-LAT data based on the analysis of \citet{Wacker1207} as extended by us in \appref{exposure}. The opacity of the colored band around each line indicates the delta-log-likelihood of the double-line fit compared to the best-fit point, with the darkest band being $1\sigma$, the medium band being $2\sigma$ and the lightest being $3\sigma$. The central line corresponds to the central value of the halo profile normalization, with the shaded band vertically giving the $\pm 2\sigma$ normalization uncertainties on the halo profiles.

\subsection{Panel D: \label{sect:paneld}}
If applicable, this panel shows the line search limits for the WIMP
mass region from 500 GeV to 20 TeV from the H.E.S.S \cite{HESS1301} line search
discussed in \sectref{HESS_line_limits}. The ROI is a circle of radius
$1^\circ$ centered on the GC with Galactic latitudes $|b|<0.3^\circ$
excluded. Note the change from Panel C to a logarithmic scale in $M$.

The black and solid colored lines and similarly colored bands are all
as described in Panel C. Isothermal limits are very weak for this line
search owing to the cored nature of the profile and very small ROI
near the GC.

\subsection{Panel E: \label{sect:panele}}
This panel shows the inclusive (continuum and line) spectrum limits
for DM mass from 5 GeV to 1 TeV based on the
``model-independent'' limits given in \citet{Hooper1209} and described
in \sectref{Hooperlimits}. These are given as the solid colored lines
labelled `GC ...'; the colored bands show the range of halo profile
normalizations, as in Panel C. The solid black line is the result for the total $\sigma v$ for the
operator.

The sharp features in the GC limits that can be seen in some cases are
due to the $\gamma$ lines changing through the bins used in
\cite{Hooper1209}. For operators with lines only from $\gamma Z$ or
$\gamma h$, we also see sharp features in the limits near threshold as
the line energy migrates rapidly across the bins; had we convolved the lines 
with the Fermi-LAT energy dispersion, these cross-over regions would be smoothed.

Also shown are Fermi-LAT \cite{FERMI1108} stacked dwarf galaxy
limits, discussed in \sectref{dwarf_limits}. These are given as a
variety of grey lines and are a conservative estimate of the 95\% CL
UL limits; because the published limits were presented in terms of
specific final states, there is not enough information to fully derive
the limits for the continuum spectra here.

\subsection{Panel F: \label{sect:panelf}}
If applicable, the limits presented here are on the ratio of selected continuum final state annihilation cross sections to the line final states $\gamma\gamma$ and $\gamma Z$ as described in \citet{Wacker1207} and summarized in \sectref{Cohen_supersaturation}; these limits are only applicable if the operator has no final state $\gamma h$, and has dominant annihilation branching ratios to at least one of the final states $WW/ZZ$ or $b\bar{b}$.

The solid black line is the ratio $R_{\text{th}} \equiv \lb( \sigma_{WW}v + \sigma_{ZZ}v + \sigma_{b\bar{b}}v \rb) / \lb( 2\sigma{\gamma\gamma}v + \sigma_{\gamma Z}v \rb)$ for this operator; the dashed colored lines are individual annihilation mode contributions to this summed result. The various solid colored lines give either the supersaturation 95\% CL UL on $R$ for fixed $M$, or the boundary of the `$2\sigma$' confidence region\footnote{That is, where the log-liklihood of the fit lies within 4.86 (4 d.o.f.) of that for the best fit point \cite{Wacker1207}.} in the $R-M$ plane for the shape constraints from \citet{Wacker1207} assuming 100\% annihilation to the indicated final state. Further discussion of the applicability of these limits is given in the text in \sectref{Cohen_supersaturation}, but it should be borne in mind that operators which are only marginally excluded or allowed per the results given here merit further investigation to determine whether or not they are actually excluded. In particular, the continuum radiation from the $Z$ in $\gamma Z$ is not factored in. Furthermore, the ratio of  $\gamma \gamma$ to $\gamma Z$ is not fixed at the correct ratio for each operator here considered, but is rather marginalized over in the analysis of Ref.\ \cite{Wacker1207}.

\section{Discussion and Conclusions}
\label{sect:conclusion}

\renewcommand{\arraystretch}{1.3}
\newcommand{\colwidth}{9.0cm}
\newcommand{\colwidthbuffer}{9.0cm}
\begin{table}
\caption{\label{tab:classification} Classification of operators by limit results.  Starred (*) categories have operators which could fit the Fermi 130 GeV $\gamma$-ray line and also match the correct relic abundance.}
\begin{ruledtabular}
\begin{tabular}{p{0.5cm} p{1.2cm}p{5.5cm}p{\colwidthbuffer}} 
\#&Labels & \multicolumn{1}{c}{Operators} & \multicolumn{1}{c}{Summary of annihilation modes and limits} \\[1ex]\hline
1& VI-1 & $\phi^\dagger\phi\ B_{\mu\nu}B^{\mu\nu}$ &
\multirow{3}{\colwidth}{\justifying{$\gamma\gamma$ is present and
    dominant for all masses, leading to strong line limits for masses
    up to 10 TeV; $\gamma Z$ and $ZZ$ enter above kinematic thresholds
    but are sequentially more parametrically suppressed by powers of
    $(\sin\theta_W/\cos\theta_W)^2$. }} \\			
&VI-2 & $\phi^\dagger\phi\ \widetilde{B}_{\mu\nu}B^{\mu\nu}$ \\
& VIII-1 & $\bar{\chi} i \gamma^5 \chi\ B_{\mu\nu}B^{\mu\nu}$ \\ 
& VIII-2 & $\bar{\chi} i \gamma^5 \chi\ \widetilde{B}_{\mu\nu}B^{\mu\nu}$ \\
[1ex]\hline 
2 & XI-1 & $\lb( \phi^\dagger \partial^\mu \phi + \text{h.c.} \rb) i
\lb(B_{\lambda\mu} Y_H H^\dagger D^\lambda H - \text{h.c.}\rb)$ &
\multirow{3}{\colwidth}{\justifying{$\gamma X$ and
    $ZX$ where $X = Z$  (for ``$-\text{h.c.}$" operators) OR $X=h$
     (for ``$+\text{h.c.}$") are present above thresholds, with the latter in
    each case  typically suppressed by $(\sin\theta_W/\cos\theta_W)^2$. Line limits are constraining up to a TeV. } } \\
& XI-2	& 	$\lb( \phi^\dagger \partial^\mu \phi + \text{h.c.} \rb) i \lb(\widetilde{B}_{\lambda\mu} Y_H H^\dagger D^\lambda H - \text{h.c.}\rb)$ 	\\
& XIV-1	&	$ \bar{\chi}\gamma^\mu \chi \lb(B_{\lambda\mu} Y_H H^\dagger D^\lambda H + \text{h.c.}\rb)$	\\
& XIV-2	&	$ \bar{\chi}\gamma^\mu \chi \lb(\widetilde{B}_{\lambda\mu} Y_H H^\dagger D^\lambda H + \text{h.c.}\rb)$ \\
& XV-1	&	$ \bar{\chi}\gamma^\mu \chi i \lb(B_{\lambda\mu} Y_H H^\dagger D^\lambda H - \text{h.c.}\rb)$	\\
& XV-2	&	$ \bar{\chi}\gamma^\mu \chi i \lb(\widetilde{B}_{\lambda\mu} Y_H H^\dagger D^\lambda H - \text{h.c.}\rb)$ \\
[1ex] \hline	 
3 & XVII-1 & $\bar{\chi} \gamma^{\mu 5} \chi\ i \lb(B_{\lambda\mu} Y_H
H^\dagger D^\lambda H - \text{h.c.}\rb)$ &
\multirow{2}{\colwidth}{\justifying{$\gamma Z$ and
    $ZZ$ are present above thresholds.  Line limits are constraining
    up to a few hundred GeV, however $p$-wave components to the
    annihilation cross section during the freeze out become important 
    at larger DM mass
    leading to suppression of the $s$-wave component and significant
    weakening of constraints. } } \\
& XVII-2	& 	$\bar{\chi} \gamma^{\mu 5} \chi\ i \lb(\widetilde{B}_{\lambda\mu} Y_H H^\dagger D^\lambda H - \text{h.c.}\rb)$	\\
[8ex] \hline
4 & XVIII-1 & $\bar{\chi}\gamma^{\mu\nu}\chi\ B_{\mu\nu} Y_H H^\dagger H $
& Annihilation to fermions is dominant and $s$-wave at low WIMP mass,
with strong inclusive limits. The $\gamma h$ line annihilation mode
becomes dominant above threshold, and the line limits are constraining
for masses of 100 GeV up to a few TeV. \\
[1ex] \hline
5 & XVIII-2	 &  $\bar{\chi}\gamma^{\mu\nu}\chi\ \widetilde{B}_{\mu\nu} Y_H H^\dagger H $	 & Similar to the operator above, but the continuum final states are $p$-wave, so inclusive limits not constraining below 70 GeV. \\				
[1ex] \hline
6* & VI-3 & $\phi^\dagger\phi\ W^a_{\mu\nu}W^{a\ \mu\nu}$	&
\multirow{3}{\colwidth}{\justifying{$\gamma\gamma$ is present for all
    masses; $\gamma Z$, $ZZ$ and $WW$ enter above kinematic
    thresholds. Above the $W^+W^-$ threshold, annihilation modes with
    lines are heavily parametrically suppressed by powers of  $\sin^2
    \theta_W$. Line searches are constraining below the $W^+W^-$
    threshold.}} \\								
& VI-4   & $\phi^\dagger\phi\ \widetilde{W}^a_{\mu\nu}W^{a\ \mu\nu}$ \\
& VIII-3 & $\bar{\chi} i \gamma^5 \chi\ W^a_{\mu\nu}W^{a\ \mu\nu}$ \\
& VIII-4 & $\bar{\chi} i \gamma^5 \chi\ \widetilde{W}^a_{\mu\nu}W^{a\ \mu\nu}$	\\
[1ex] \hline
7* &XI-3 & $\lb( \phi^\dagger \partial^\mu \phi + \text{h.c.} \rb) i
\lb(W_{\lambda\mu} H^\dagger t^a D^\lambda H - \text{h.c.}\rb)$ &
\multirow{3}{\colwidth}{\justifying{$\gamma X$, $ZX$ and $W^+W^-$
    where $X = Z\ (``-\text{h.c.}") \text{ OR } h \ (``+\text{h.c.}")$
    are present above thresholds. Line limits are severely
    constraining for masses less than around 100 GeV; once the
    $W^+W^-$ mode enters, the $\gamma X$ annihilation mode is heavily
    suppressed. Inclusive limits are weakly constraining for masses up
    to a few hundred GeV. Where applicable, the ratio limits are
    either marginal or quite constraining on 125-150 GeV. } } \\
& XI-4	& 	$\lb( \phi^\dagger \partial^\mu \phi + \text{h.c.} \rb) i \lb(\widetilde{W}^a_{\lambda\mu} H^\dagger t^a D^\lambda H - \text{h.c.}\rb)$ 	\\
& XIV-3	&	$ \bar{\chi}\gamma^\mu \chi \lb(W^a_{\lambda\mu} H^\dagger t^a D^\lambda H + \text{h.c.}\rb)$	\\
& XIV-4	&	$ \bar{\chi}\gamma^\mu \chi \lb(\widetilde{W}^a_{\lambda\mu} H^\dagger t^a D^\lambda H + \text{h.c.}\rb)$	\\
& XV-3	&	$ \bar{\chi}\gamma^\mu \chi i \lb(W^a_{\lambda\mu}  H^\dagger t^a D^\lambda H - \text{h.c.}\rb)$	\\
& XV-4	&	$ \bar{\chi}\gamma^\mu \chi i \lb(\widetilde{W}^a_{\lambda\mu} H^\dagger t^a D^\lambda H - \text{h.c.}\rb)$ \\[1ex]
\hline	 
8* & XVII-3	 & $\bar{\chi}\gamma^{\mu 5}\chi\  i \lb(W^a_{\lambda\mu} H^\dagger t^a D^\lambda H - \text{h.c.}\rb)$ &
\multirow{2}{\colwidth}{\justifying{Similar to the category just above, but $p$-wave components to the cross section cause suppression of the $s$-wave cross section and weaken line constraints above a few hundred GeV.}} \\[1ex]
& XVII-4	& 	$\bar{\chi} \gamma^{\mu 5} \chi\ i \lb(\widetilde{W}^a_{\lambda\mu} H^\dagger t^a D^\lambda H - \text{h.c.}\rb)$	\\
[1ex] \hline
9 & XIX-1 & $\bar{\chi}\gamma^{\mu\nu}\chi\ W^a_{\mu\nu} H^\dagger t^a H $  & 
Annihilation to continuum modes is dominant, and inclusive limits are
constraining up to a hundred GeV. \\ [1ex] \hline
10 & XIX-2	 & $\bar{\chi}\gamma^{\mu\nu}\chi\ \widetilde{W}^a_{\mu\nu} H^\dagger t^a H $ &
Annihilation to continuum modes dominates, but are $p$-wave at low DM masses. \\ [1ex]  \hline
11 & IX-2	& $\bar{\chi}i\gamma^5\chi\ H^\dagger H$	&
\multirow{2}{\colwidth}{\justifying{Only continuum annihilation modes exist: one of either the $f\bar{f}$ or $W^+W^-$ modes dominate at all masses.}}	\\
& XX-1   & $\bar{\chi}\gamma^{\mu\nu}\chi\ B_{\mu\nu}$  \\
[1ex] \hline
12 & X-1 & $\lb( \phi^\dagger \partial^\mu \phi + \text{h.c.} \rb) \lb(B_{\lambda\mu} Y_H H^\dagger D^\lambda H + \text{h.c.}\rb)$	&
Only $Zh$ $s$-wave modes are present. \\
& X-2   & $\lb( \phi^\dagger \partial^\mu \phi + \text{h.c.} \rb) \lb(W^a_{\lambda\mu} H^\dagger t^a D^\lambda H + \text{h.c.}\rb)$	\\
[1ex] \hline	
13 & XVI-1	&  $\bar{\chi}\gamma^{\mu 5} \chi\  \lb(B_{\lambda\mu} Y_H H^\dagger D^\lambda H + \text{h.c.}\rb)$	&
\multirow{2}{\colwidth}{\justifying{Only $Zh$ $s$-wave modes are present but
    there are also significant $p$-wave contributions.}} \\
& XVI-3   & $\bar{\chi}\gamma^{\mu 5} \chi\  \lb(W^a_{\lambda\mu} H^\dagger t^a D^\lambda H + \text{h.c.}\rb)$	\\[1ex]
\end{tabular}
\end{ruledtabular}
\end{table}

\renewcommand{\arraystretch}{1.0}

Broadly speaking, we find that the 34 operators with $s$-wave annihilation channels can be classified into thirteen qualitatively distinct classes on the basis of the limits which apply to each operator; we summarize these in \tabref{classification}. 

\begin{itemize}
\item {\bf Operators coupling to hypercharge gauge bosons; photon lines present (categories 1-3)}

For operators in the first 3 categories in \tabref{classification}, annihilation to final states giving lines has a large branching fraction as soon as the dark matter mass is larger than lowest kinematical threshold for such channels. This is a consequence of the fact that the $B$ field strength tensor has a dominant photon component so that couplings to the $Z$ are parametrically suppressed with respect to couplings to the $\gamma$ by a factor of $\tan^2 \theta_W \sim 0.3$ (this suppression may be partially invalidated by other large factors arising in the evaluation of the matrix element, but nevertheless serves as a useful rule-of-thumb). As a result, the operators in categories 1 and 2 are excluded at 95\% confidence by the Fermi-LAT \cite{FERMI1305} and H.E.S.S. \cite{HESS1301} line limits for most masses from the lowest kinematic threshold for a line final state up to a few TeV, except a) if the very conservative isothermal profile choice is made for the H.E.S.S. limits and b) the dark-matter mass is in the region 300 GeV and 500 GeV, which is not addressed by any of the line limits.  Depending on the profile choice, the required thermal cross section for these operators may also be in tension with the experimental upper limits from a few to 20 TeV. 

For operators in category 3, the total $s$-wave cross section drops away from the canonical thermal value of approximately $3\times 10^{-26}$cm$^3$s$^{-1}$ as $M^{-2}$ for $M$ above a few hundred GeV.  This phenomenon occurs generically for operators with fermion DM coupling via an axial-vector $J_{DM}$ current, and is due to the presence of $p$-wave components to the annihilation cross section which are enhanced over the $s$-wave by a factor of $s/m_V^2 \sim M^2$. This enhancement can be traced to a coupling in the $p$-wave (but not the $s$-wave) to the longitudinal component $\epsilon^\mu_L \sim k^\mu/ m_Z$ of the massive vector. Since both $s$- and $p$-wave components of $\sigma v$ can be important at freeze-out [\textit{cf.} Eq. \eqref{eq:relic_density}], having $a/b \sim M^{-2}$ naturally suppresses the present-day annihilation cross section for large $M$ by driving up the value of $\Lambda$ required to saturate the present-day average DM density. As a result, a significantly smaller range of $M$ is definitively excluded by, or put in tension with, the experimental limits compared to categories 1 and 2, especially for TeV-range DM. The constraints from the inclusive limits from Ref.~\cite{Hooper1209} are less stringent: in most cases, the limits are in tension with the thermal relic cross section for $M$ below one or two hundred GeV only for some choices of halo profile, except for operators with $\gamma\gamma$ modes where, below \textit{ca.}\ 100 GeV, the inclusive limits exclude such operators even for conservative profile choices.

\item {\bf Operators coupling to $SU(2)_L$ gauge bosons and Higgs; photon lines present (categories 6-8)}

For operators coupling to the $SU(2)_L$ gauge bosons and Higgs (entries 6--8 in \tabref{classification}), the $W^+W^-$ final state is always dominant for $M \gtrsim m_W$ since in this case couplings to the $Z$ and $\gamma$ are parametrically suppressed by $\cos^2\theta_W$ and $\sin^2\theta_W$, respectively. Consequently, for operators in categories 6 and 7, the line limits are only severely constraining between the lowest kinematic threshold for any line mode and $M \simeq m_W$; for $M$ greater than about $100$ GeV and less than a few TeV the experimental limits are typically either only in tension with the required thermal cross section for more aggressive choices of halo profile, or are completely unconstraining. Operators in category 8 again contain fermion DM with coupling via an axial-vector current and thus suffer the same falling $\sigma v$ at large $M$ as the operators in category 3, and as a result are even less constrained at large $M$ than the operators in categories 6 and 7. The inclusive limits from Ref.~\cite{Hooper1209} are somewhat stronger for all these operators than for the operators in categories 1--3, and may be in tension with the required thermal cross section up to a few hundred GeV depending on the profile choice. Nevertheless, taken together with the line limits, these operators are only weakly constrained for heavy DM. 

In addition, for most operators in categories 6--8, the shape-based ratio limits \cite{Wacker1207} are either very constraining (operators XI-3 and XVII-3), or are marginal in the sense that the current limit is very close to the predicted value from the operators.  We however caution the reader that the limits in Ref.\ \cite{Wacker1207} were obtained by assuming the line signal would account for the Fermi signal for some choice of DM mass, and allowing the relative strengths of the lines to float to the optimal value. The first assumption may or may not be applicable to our analysis depending on the operator considered; but the second is never applicable to our analysis since the relative strengths of all annihilation channels are fixed for any given operator. Therefore, marginal cases may or may not actually be constrained by these limits and a more careful analysis is needed here.

\item {\bf Operators with annihilation modes to fermions; photon lines present (categories 4,5,9,10)}

The analysis in Ref.~\cite{Hooper1209} as applied to our operators indicates that if the only available $s$-wave annihilation mode for $M<10$ GeV dark matter is to Standard Model fermions, there is a strong constraint which seems to rule out the annihilation cross section necessary for obtaining the thermal relic abundance for $M<10$ GeV.

The operators listed in entries 4, 5, 9 and 10 of \tabref{classification} all have annihilation modes to both fermions and line final states (amongst others). Of these, the operators coupling to the field-strength tensor (rather than its dual) have $s$-wave annihilation to $f\bar{f}$ final states and so can be excluded for $M$ below 10 GeV; they may also be in tension with the inclusive limits, depending on the profile choice, from 10 GeV to a few hundred GeV. 
Also, for these operators, the branching ratio to $f\bar{f}$ annihilation modes falls once the on-shell diboson modes become kinematically allowed. This is because the annihilation to fermions goes through $s$-channel gauge boson exchange and so requires an additional Higgs vev insertion and coupling of fermions to $\gamma/Z$ compared to the $Zh$ and $\gamma h$ modes (see the next paragraph for a comment on the $W^+W^-$ mode). On dimensional grounds this results in a suppression of the $f\bar{f}$ modes relative to the $Zh$ and $\gamma h$ modes by a factor $\sim m_{Z,W}^2/s$ (where $s \sim 4M^2$). For the operators coupling instead to the dual field strength tensor, the fermion cross section is $p$-wave so the inclusive limits are less constraining: they may be in tension with the required cross section from the $\gamma h$ threshold to a few hundred GeV, but only for fairly aggressive profile choices.

For the operators in categories 4 and 5 with couplings to hypercharge and Higgs, the photon line becomes strong above the kinematic threshold for the $\gamma h$ annihilation mode leading to strongly excluding line constraints, whereas for the operators in categories 8 and 9 with couplings to the $SU(2)_L$ gauge bosons and Higgs, the diboson annihilation modes are dominated by $W^+W^-$, which leads to line limits for these operators in tension with the required thermal cross section only if an aggressive profile choice is made. We note that this occurs notwithstanding the fact that for the $W^+W^-$ mode (as for the $f\bar{f}$ mode), both Higgs fields must be replaced by their vevs; the reason that this does not suppress this final state is that there is a cancellation of the suppression factor $\sim m_W^2/ s$ against an enhancement $\sim s/m_W^2$ from the longitudinal mode of the $W^\pm$ (the by-now-familiar suppression, relative to $W^\pm$, of couplings to $Z$ and $\gamma$ by weak-mixing-angle factors accounts for the relative strength of the $W^+W^-$ and $Zh$ or $\gamma h$ modes).


\item {\bf Operators without photon lines (categories 11-13)}

Finally, entries 11--13 in \tabref{classification} are operators which couple only to SM final states without photon lines.\footnote{Annihilation to monochromatic photons is also possible at subleading order, for example through loops or see Ref.\ \cite{Weiner:2012cb}; we do not include these subdominant modes.} The two operators in entry 11 have strong annihilation to $f\bar{f}$ either for all $M$ (operator XX-1)\footnote{Here we include annihilation modes purely through $s$-channel gauge boson exchange which explains why the suppression of the $f\bar{f}$ modes does not occur for large $M$ as it does for operators XVIII-1, XIX-1 and IX-2.} or at low $M$ (operator IX-2),\footnote{The explanation here for the suppression of the $f\bar{f}$ branching ratio for $M\in[m_W, m_t]$ or for $M \gg m_t$  is that this annihilation mode is via $s$-channel Higgs exchange which leads to a suppression from the $hf\bar{f}$ coupling $\sim m_f^2/s$.} but in either case this means that they are strongly constrained for $M \lesssim 10$ GeV even for conservative profile choices. Owing to the very flat (with changing $M$) nature of the inclusive upper limits on $\sigma v$, the astrophysical uncertainty gives a rather broad range of $M$ for which the operators may be in tension with the experimental limits depending on the profile choice: 10 GeV to a few hundred GeV. The operators in entries 12 and 13 have only $Zh$ final states: the former may be in tension with the astrophysical limits for aggressive profile choices for $M$ between 100 and 400 GeV, but they are unconstrained for more modest profile choices; the latter are unconstrained regardless of the profile chosen due to $s$-wave cross section suppression by the same enhanced--$p$-wave effect which was discussed at length above.
\end{itemize}

In addition to these indirect limits, we emphasize the point made in CKW that the `tensor' fermion DM currents $\chi \gamma^{\mu\nu} \chi$ (\textit{e.g.}, operators XVIII-n, XIX-n and XX-n in categories 4-5 and 9-11) give rise to magnetic or electric dipole couplings (if necessary, by replacing both Higgs fields with their vevs) which causes them to be very strongly constrained by direct detection experiments \cite{Banks:2010uq,Fortin:2011fk}. Given our assumptions about the DM being a cold thermal relic WIMP saturating the measured average density, the operators giving electric-dipole couplings (coupling to the dual field tensor) are excluded at at least 90\% confidence by both CDMS and XENON across the entire mass range from $M=10$ GeV to at least 20 TeV (probably somewhat larger), while the operators giving magnetic-dipole couplings (coupling to the field tensor, not its dual) are excluded at at least 90\% confidence from $M=10$ GeV to $M>20$ TeV, $M\simeq1$ TeV, or $M\simeq 400$ GeV for operators XX-1, XVIII-1, or XIX-1, respectively (with the exception of a very narrow mass window around $m_Z/2$ for the latter two of these).

We have presented all of our results under the assumption that fermionic DM is Dirac, however the compensating factors of 2 in Eqs.\ \eqref{eq:relic_density} and \eqref{eq:flux} imply that there is a uniform shifting of all $\sigma v$ values down by a factor of 2 if Majorana fermion DM was assumed instead; therefore, the same thermal cross section \emph{relative to all of our 95\% CL UL limits} would obtain, at least for all the operators which do not vanish identically for Majorana fermion DM.

\begin{center}
  {\bf $\gamma$-ray Line Signals}
\end{center}

As discussed in \sectref{Wenigerlimits} and \sectref{Cohen_line},
there is some evidence, first reported by \citet{Weniger1204}, for a
photon line in the Fermi-LAT data near 130 GeV (updated to
approximately 133 GeV in Ref.\ \cite{FERMI1305}) with a flux matching
DM annihilation into monochromatic photons with cross section $\sigma
v \approx 10^{-27}$ cm$^3$/s. Although the most recent official
Fermi-LAT collaboration analysis \cite{FERMI1305} finds less
significant evidence for this putative line-like feature, it cannot
fully explain it as a systematic effect, and cautions that more work
is needed to understand it.  If we however take this signal seriously,
we find that there are a number of operators (indicated in
\tabref{classification} by a superscript star) which could plausibly
explain it which are not yet excluded by either the upper limits on
the continuum emission or by the latest official Fermi-LAT limits on
lines.

The signal strength for the photon line is a factor of a few (up to an
order of magnitude, depending on the halo profile) below the thermal
relic cross section. Operators with annihilation into $SU(2)_L$ gauge
bosons (categories 6--8) naturally give a suppression of this size
simply from the mixing angle $\sin^2 \theta_W$. Numerical factors are
also important, and those operators with smaller branching ratios into
lines do not produce enough monochromatic photons if the cross section
is fixed by the matching onto the correct relic density. To put it
another way, these operators may be in some tension with continuum
limits or the constraint on the continuum-to-line ratio since the BR to 
such continuum modes is then larger.

Operators XV-3 and XV-4 are most promising as explanation of the line, but for
completeness we also note that almost any of the other operators in
categories 6--8 in \tabref{classification} could also work, including
VI-3, VI-4, VIII-3, VIII-4, XI-4, XIV-3, XIV-4, and XVII-4. The
operators XI-3 and XVII-3 have smaller branching ratios to lines, less
than 10$\%$, and so do not fit the line as well for
the reasons discussed above. However, given the systematic
uncertainties, it is not yet conclusive that these operators \textit{cannot}
explain the Fermi gamma-ray line.  (Finally, although the operators
XVIII-1 and XIX-2 also have lines of roughly the correct strength,
they are ruled out by direct detection experiments, as discussed
above.)

\begin{center}
  {\bf Other Indirect Constraints and Outlook}
\end{center}

Photon fluxes do not of course supply the only prospects for indirect detection of dark matter. At present, the IceCube experiment \cite{Aartsen:2013vn,Abbasi:2012ws} reports limits on $\sigma v$ from neutrino fluxes which are at best on the order of $10^{-22}-10^{-23}$cm$^3$s$^{-1}$, which are thus much too weak to be constraining. The absence of sharp spectral features in the AMS-02 positron fraction data provides a constraint  \cite{Bergstrom:2013ys} on annihilation modes giving a sharp edge in the positron fraction: we expect that these limits could be stronger than the GC inclusive limits for $M$ less than a few tens of GeV for operators with sizeable BR for annihilation to $e^+e^-$, and to a lesser extent $\mu^+\mu^-$, final states (operators XVIII-1, XIX-1, IX-2 and XX-1; the latter may be more constrained since $f\bar{f}$ dominates for all masses). A slightly different analysis \cite{Kopp:2013zr} constraining only the total number of positrons from various DM annihilation modes using a combination of Fermi-LAT and AMS-02 data confirms the approximate equivalence of the limits from positron measurements and the GC inclusive photon flux limits for the $\mu^+\mu^-$ annihilation mode, but indicates that constraints arising from positron measurements for other channels ($b\bar{b},\ W^+W^-,\ ZZ$) would be between one and two orders of magnitude weaker than the GC inclusive photon flux limits. Additionally, constraints from antiproton fluxes may be useful to consider. Ref.~\cite{Cotta:2012nj} found that the PAMELA antiproton limits are typically weaker than the inclusive GC photon flux limits for $W^+W^-$ or $b\bar{b}$ annihilation modes and do not typically have sufficient reach to exclude the thermal WIMP scenario for at least some of the operators we have considered; however, Ref.~\cite{Cirelli:2013ly} finds that antiproton bounds are roughly comparable with the GC inclusive photon limits from Ref.\ \cite{Hooper1209} for pure $b\bar{b}/t\bar{t}$ or $W^+W^-/ZZ$ final states for $M \gtrsim 100$ GeV (the antiproton limits are much weaker for smaller $M$ or for leptonic annihilation modes), but which have fairly large uncertainties arising from cosmic ray propagation models. One would have to fully reconstruct the analysis of Ref.\ \cite{Cirelli:2013ly} to take advantage of these comparable limits though, since independently comparing the individual mode cross sections to their respective limits would again be subject to the cost of a factor of BR$_f$, which is usually fairly small for $b\bar{b}$ or $t\bar{t}$ modes for our operators.

Looking forward, Ref.\ \cite{Bergstrom:2012kx} indicates that it would be appropriate to consider a factor of 3-5 improvement in the present best limits from the next generation of indirect detection experiments looking at photon fluxes (GAMMA-400, CTA, H.E.S.S.-II). This magnitude of improvement would be able to constrain a fair amount more of the parameter space for the mass of the thermal relic WIMP scenario even factoring in the astrophysical uncertainties, but would still not be able to completely rule out all DM masses for every EFT operator with $s$-wave annihilation channels which we have considered.
\\

\clearpage

\clearpage

\clearpage

\clearpage

\clearpage

\clearpage

\clearpage

\clearpage

\clearpage

\clearpage

\clearpage

\clearpage

\clearpage

\begin{figure}
\begin{center}
\includegraphics[width=0.95\textwidth]{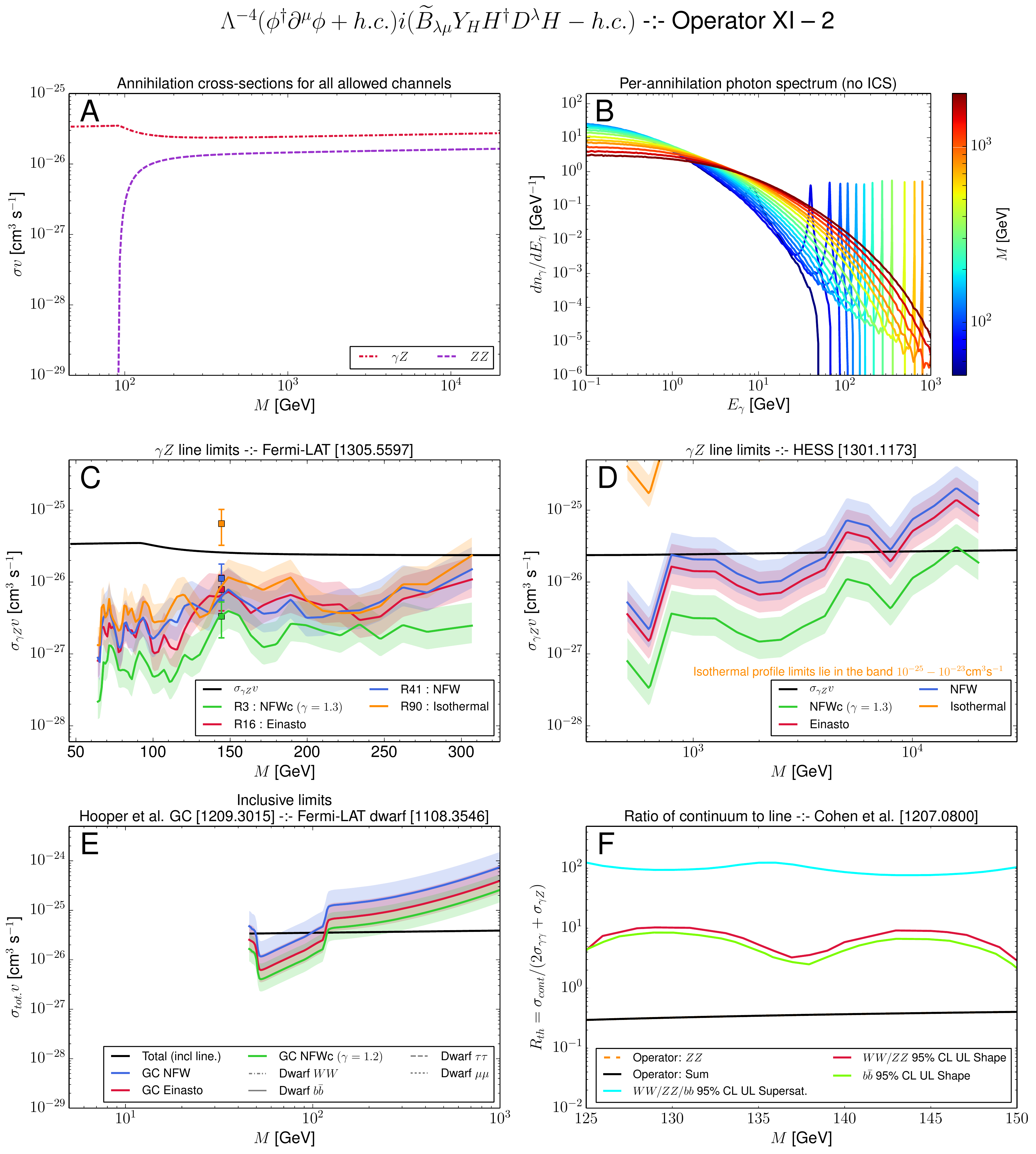}
\caption{\label{fig:TabXI_OP2} Figure captions are provided in \sectref{panela} through \sectref{panelf}.}
\end{center}
\end{figure} 
\clearpage

\begin{figure}
\begin{center}
\includegraphics[width=0.95\textwidth]{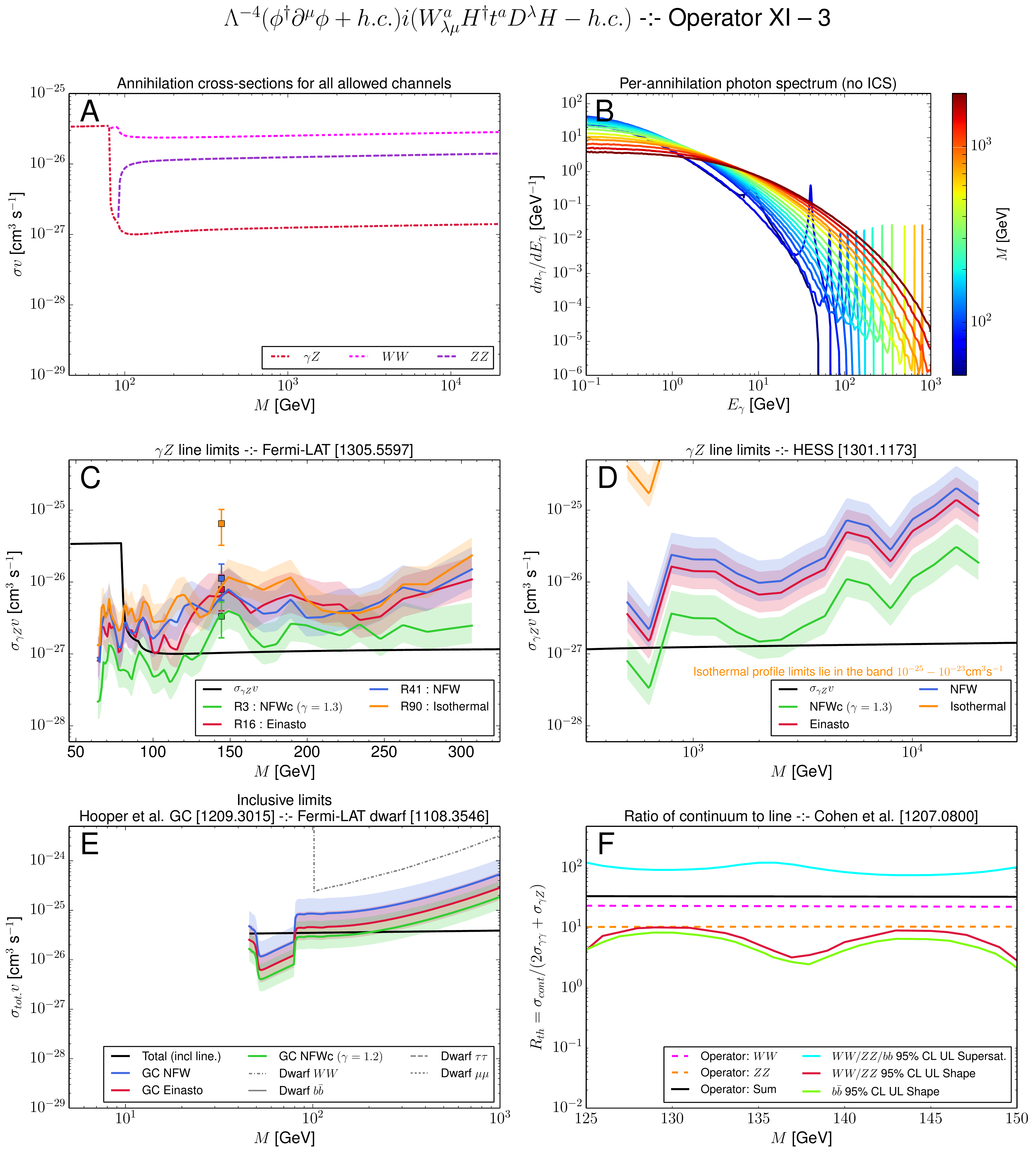}
\caption{\label{fig:TabXI_OP3} Figure captions are provided in \sectref{panela} through \sectref{panelf}. This operator may be compatible with present experimental limits and capable of accounting for the 130 GeV photon line for some profile choices, although the line BR is somewhat small. }
\end{center}
\end{figure} 
\clearpage

\begin{figure}
\begin{center}
\includegraphics[width=0.95\textwidth]{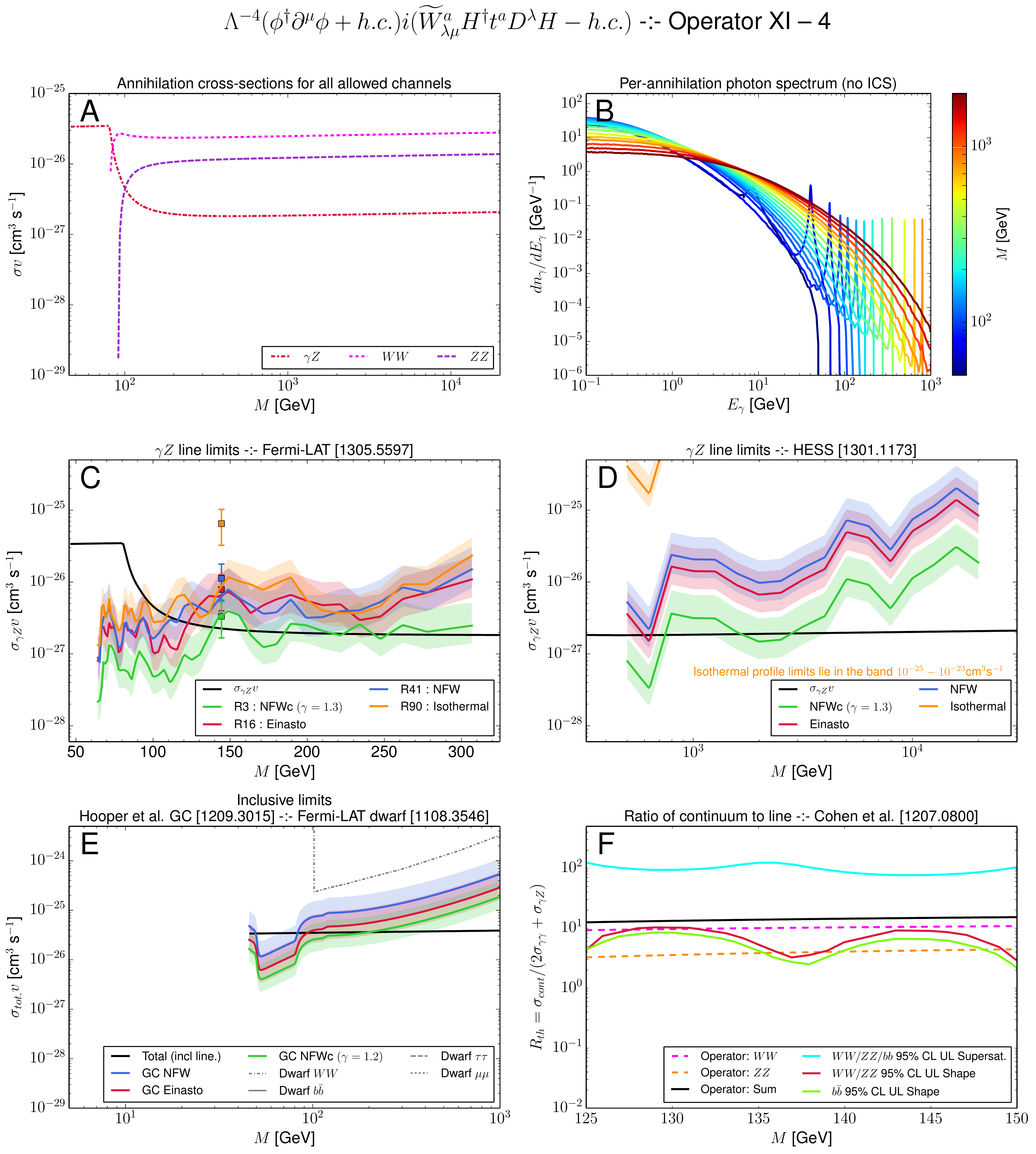}
\caption{\label{fig:TabXI_OP4} Figure captions are provided in \sectref{panela} through \sectref{panelf}. This operator may be compatible with present experimental limits and capable of accounting for the 130 GeV photon line for some profile choices.}
\end{center}
\end{figure} 
\clearpage

\begin{figure}
\begin{center}
\includegraphics[width=0.95\textwidth]{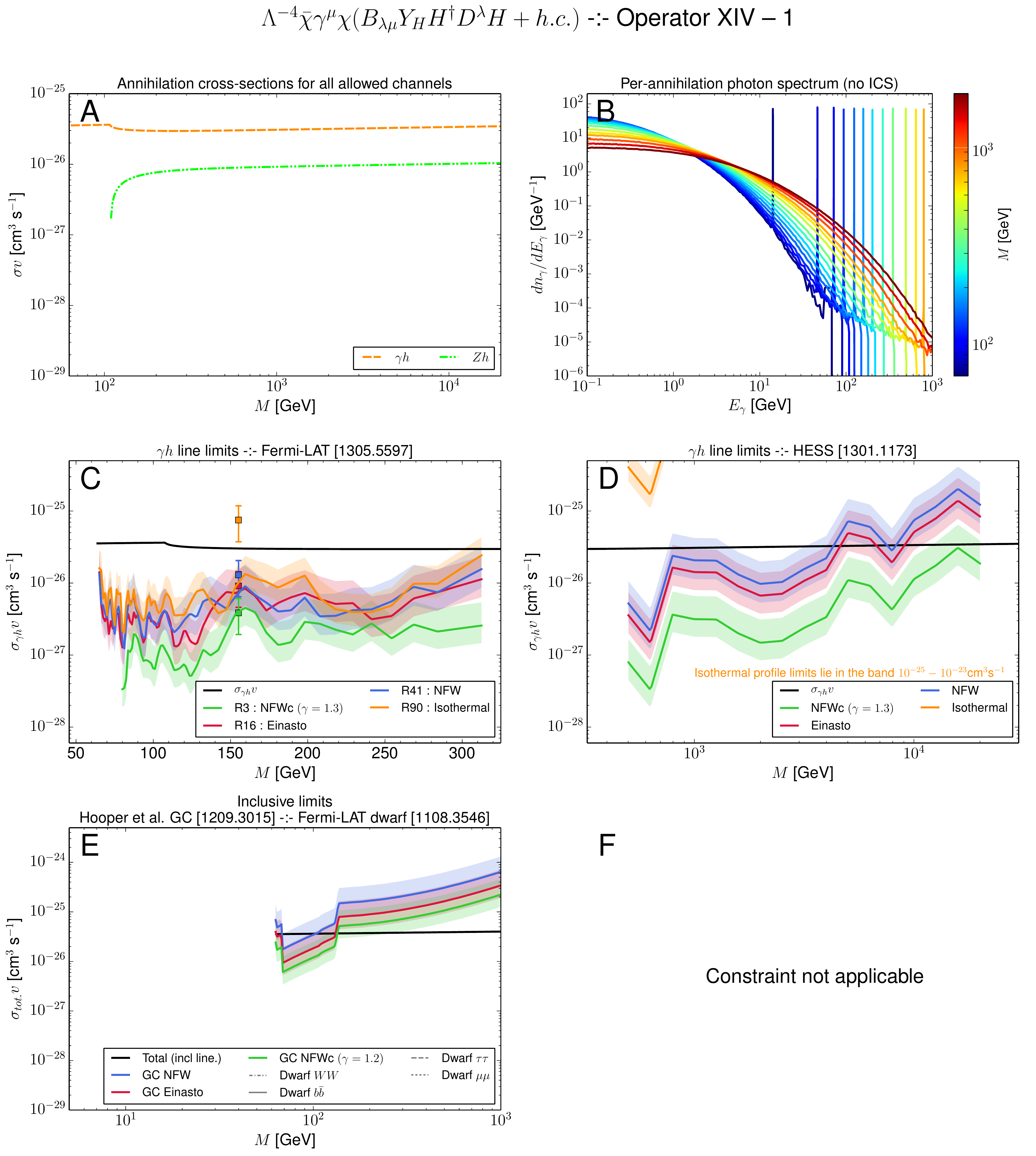}
\caption{\label{fig:TabXIV_OP1} Figure captions are provided in \sectref{panela} through \sectref{panelf}.}
\end{center}
\end{figure} 
\clearpage

\begin{figure}
\begin{center}
\includegraphics[width=0.95\textwidth]{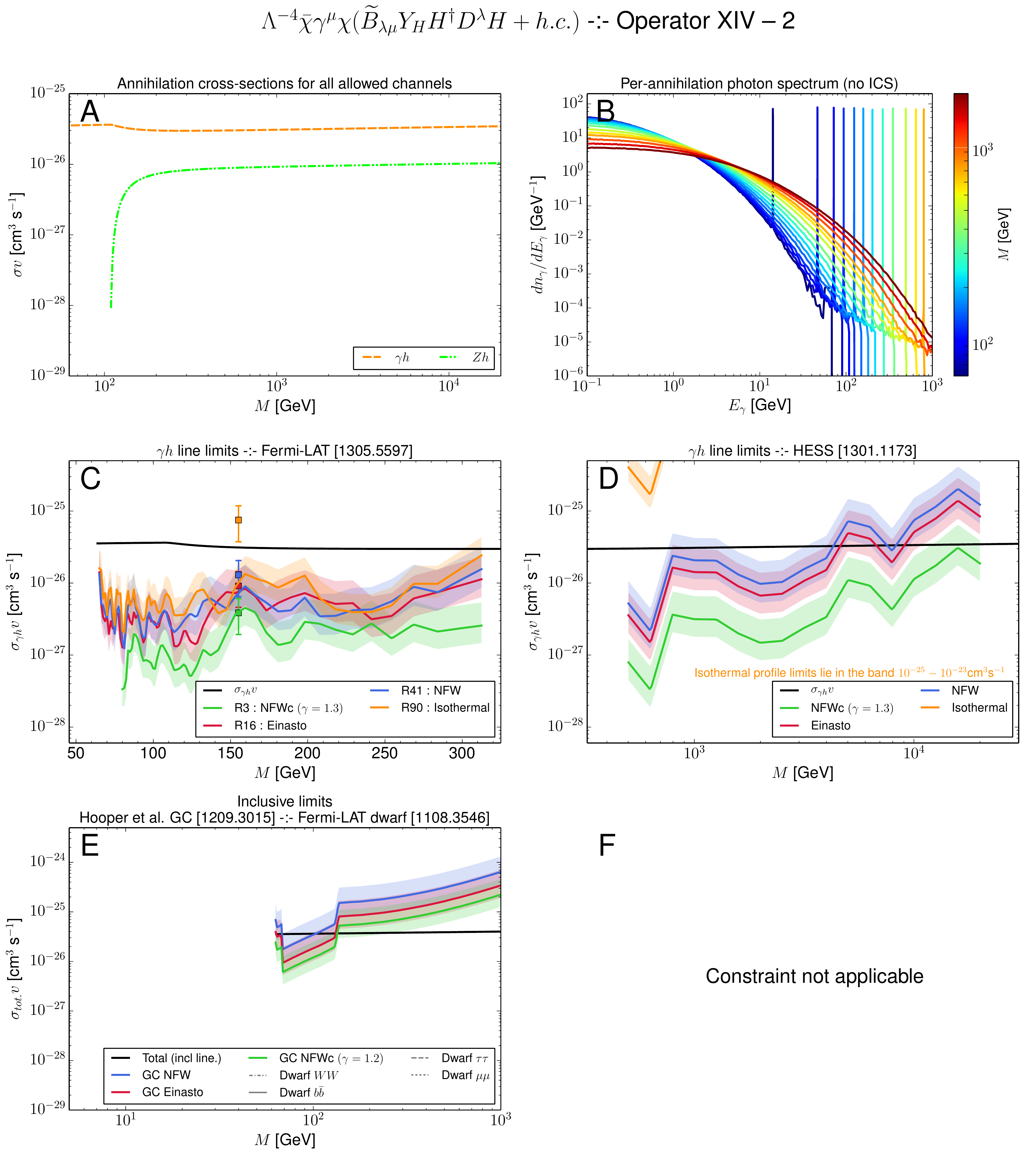}
\caption{\label{fig:TabXIV_OP2} Figure captions are provided in \sectref{panela} through \sectref{panelf}.}
\end{center}
\end{figure} 
\clearpage

\begin{figure}
\begin{center}
\includegraphics[width=0.95\textwidth]{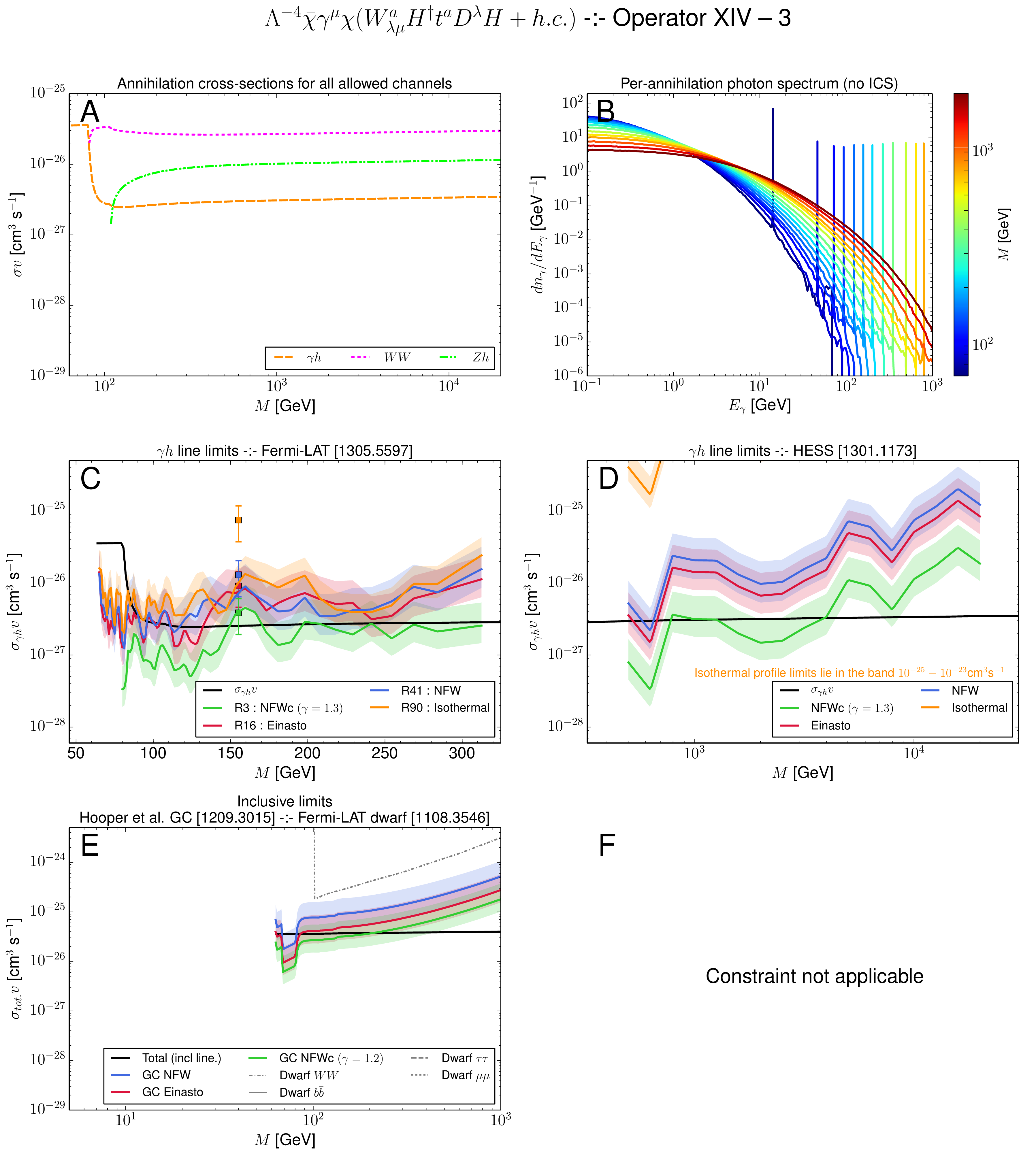}
\caption{\label{fig:TabXIV_OP3} Figure captions are provided in \sectref{panela} through \sectref{panelf}. This operator may be compatible with present experimental limits and capable of accounting for the 130 GeV photon line for some profile choices.}
\end{center}
\end{figure} 
\clearpage

\begin{figure}
\begin{center}
\includegraphics[width=0.95\textwidth]{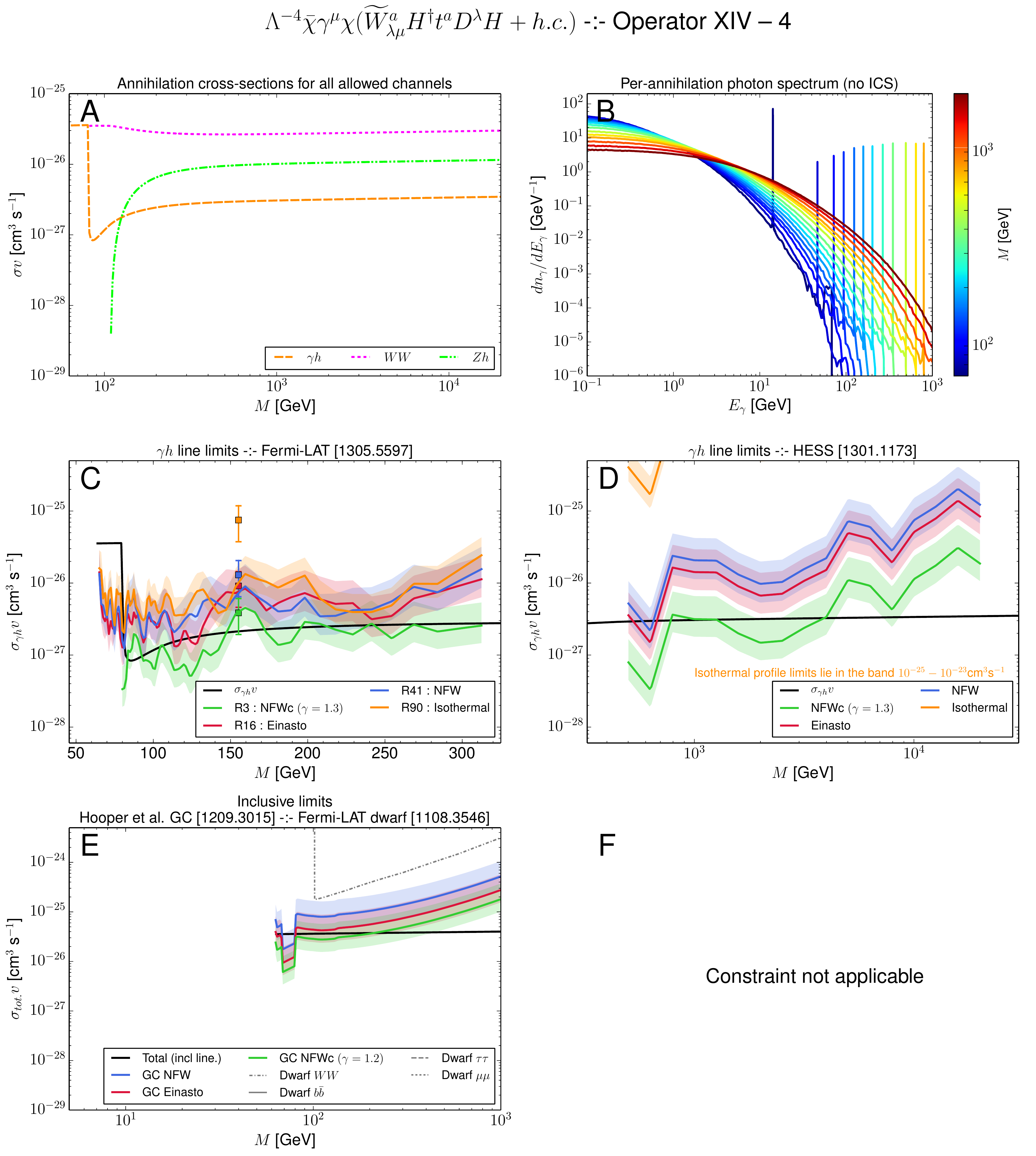}
\caption{\label{fig:TabXIV_OP4} Figure captions are provided in \sectref{panela} through \sectref{panelf}. This operator may be compatible with present experimental limits and capable of accounting for the 130 GeV photon line for some profile choices.}
\end{center}
\end{figure} 
\clearpage

\begin{figure}
\begin{center}
\includegraphics[width=0.95\textwidth]{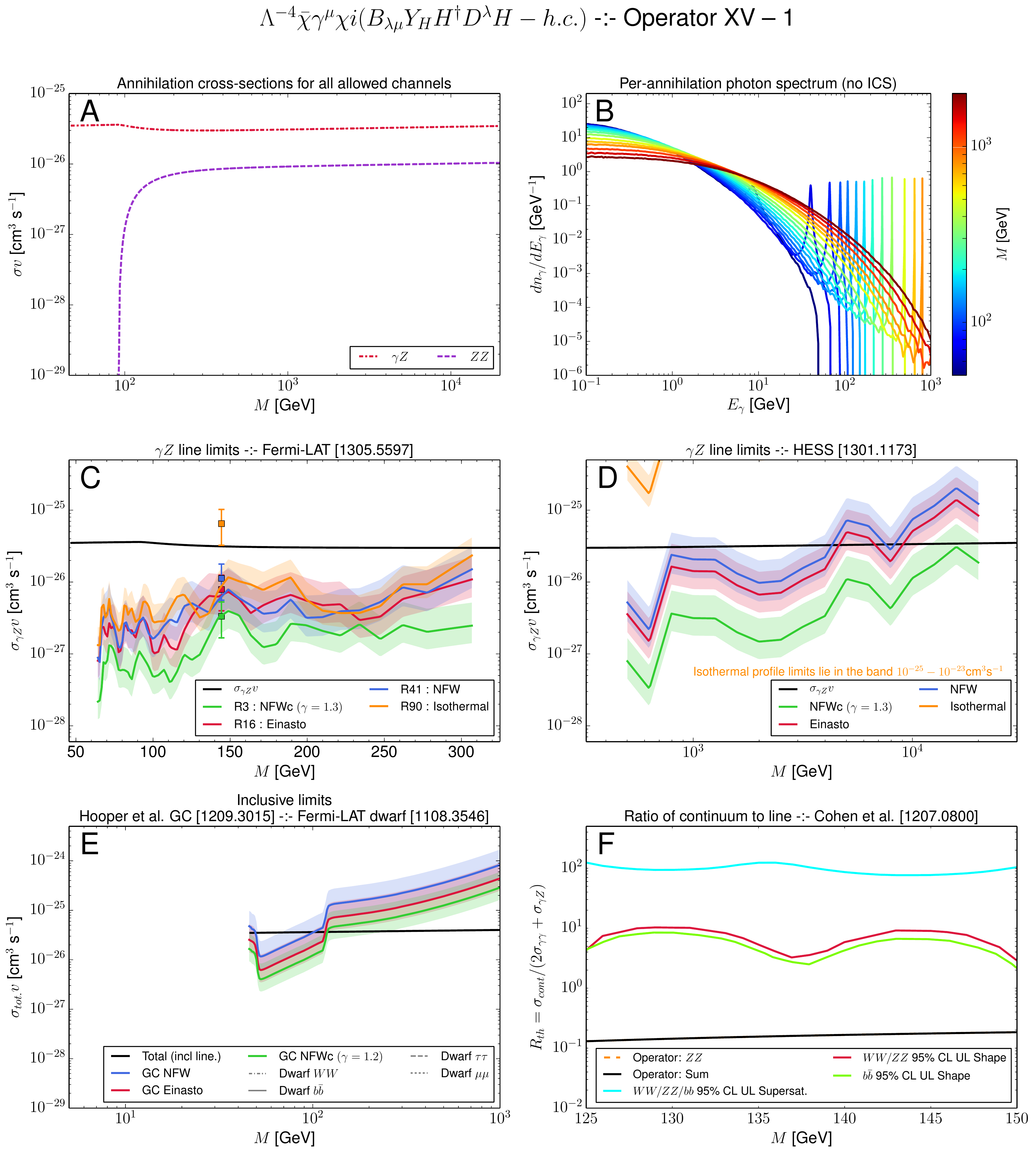}
\caption{\label{fig:TabXV_OP1} Figure captions are provided in \sectref{panela} through \sectref{panelf}.}
\end{center}
\end{figure} 
\clearpage

\begin{figure}
\begin{center}
\includegraphics[width=0.95\textwidth]{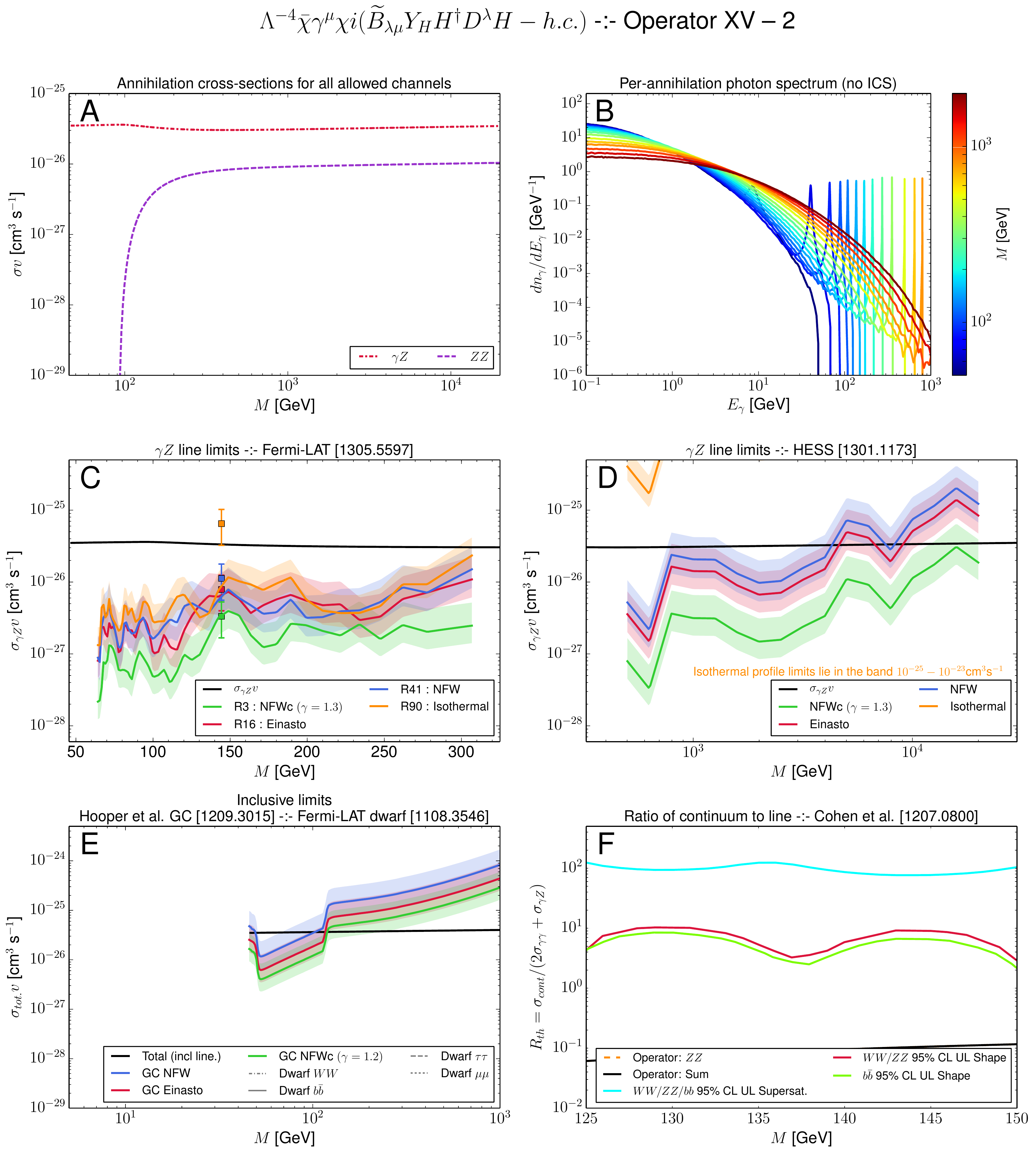}
\caption{\label{fig:TabXV_OP2} Figure captions are provided in \sectref{panela} through \sectref{panelf}.}
\end{center}
\end{figure} 
\clearpage

\begin{figure}
\begin{center}
\includegraphics[width=0.95\textwidth]{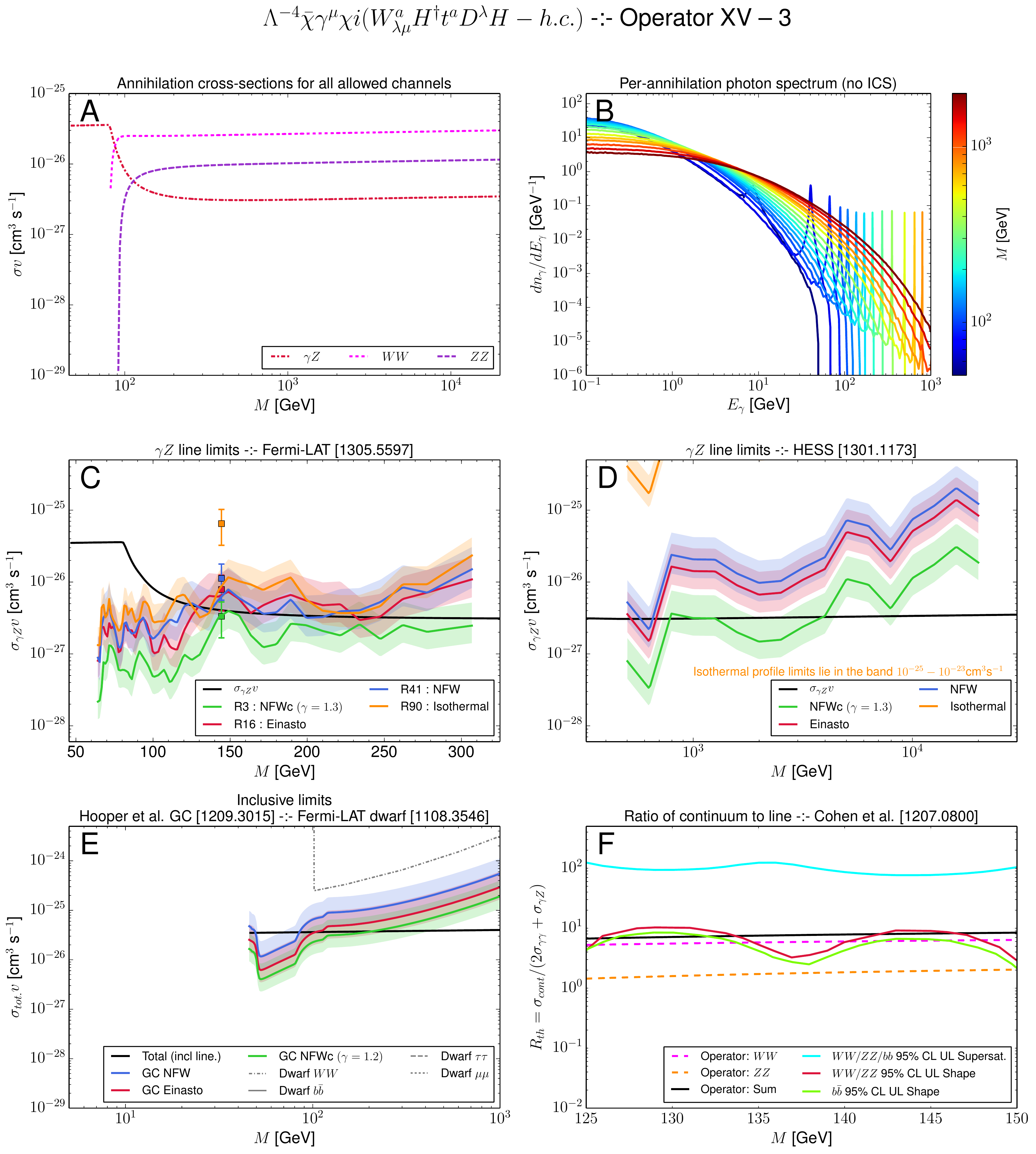}
\caption{\label{fig:TabXV_OP3} Figure captions are provided in \sectref{panela} through \sectref{panelf}. This operator may be compatible with present experimental limits and account well for the 130 GeV photon line for some profile choices.}
\end{center}
\end{figure} 
\clearpage

\begin{figure}
\begin{center}
\includegraphics[width=0.95\textwidth]{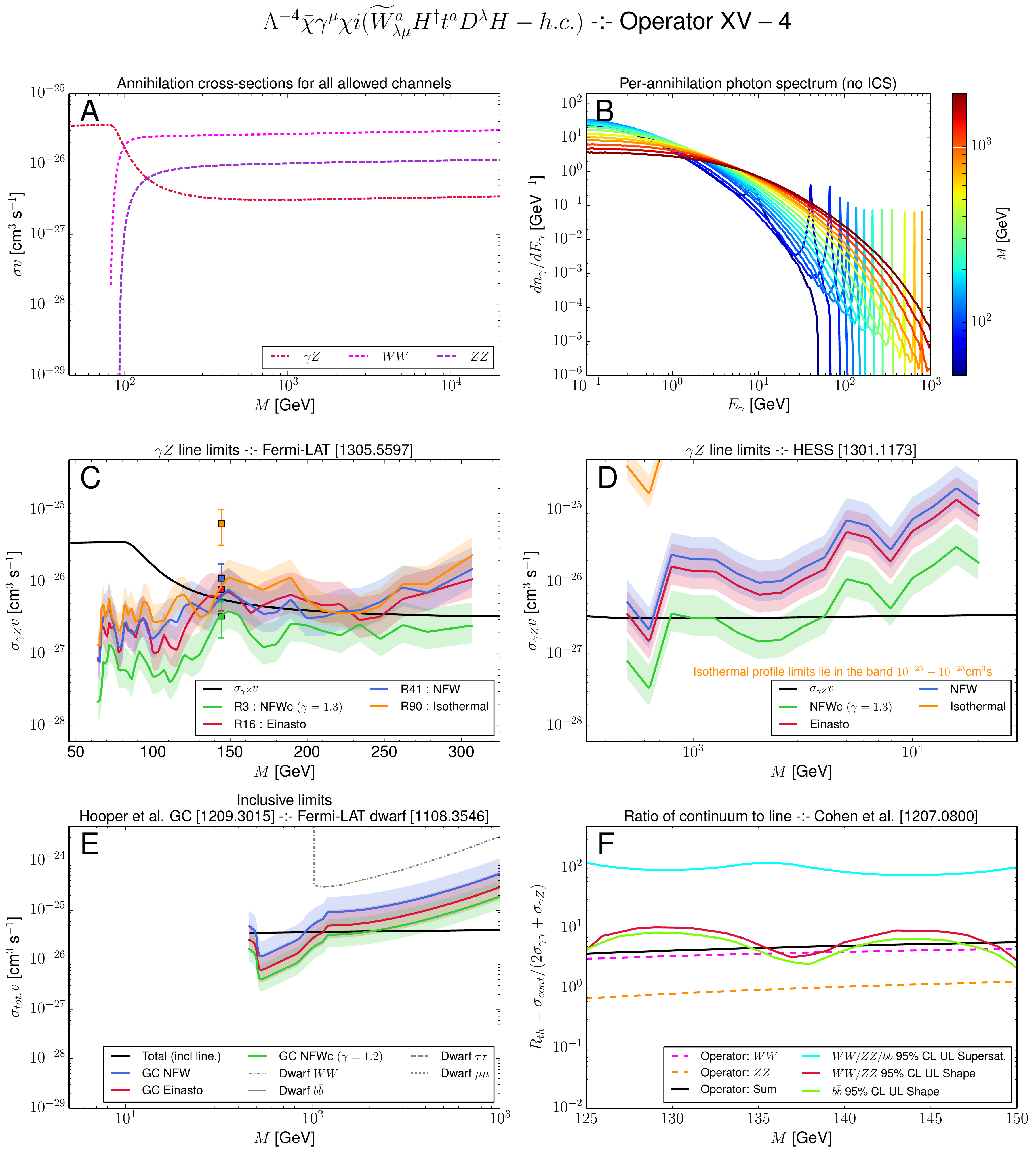}
\caption{\label{fig:TabXV_OP4} Figure captions are provided in \sectref{panela} through \sectref{panelf}. This operator may be compatible with present experimental limits and account well for the 130 GeV photon line for some profile choices.}
\end{center}
\end{figure} 
\clearpage

\begin{figure}
\begin{center}
\includegraphics[width=0.95\textwidth]{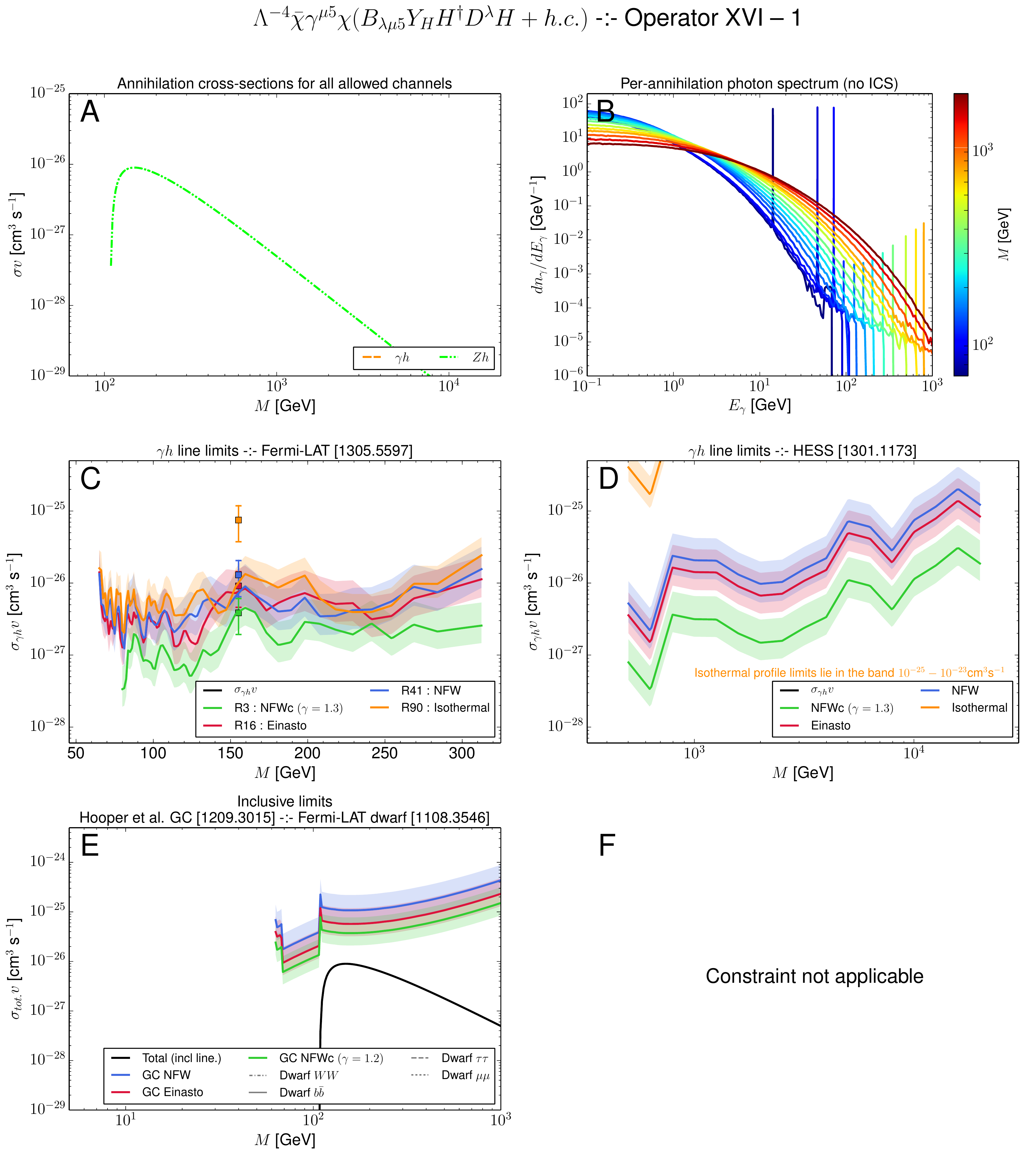}
\caption{\label{fig:TabXVI_OP1} Figure captions are provided in \sectref{panela} through \sectref{panelf}. The $\gamma h$ mode is $p$-wave suppressed and so $\sigma_{\gamma h}v$ is orders of magnitude below the range of cross sections shown in panels C and D.}
\end{center}
\end{figure} 
\clearpage

\begin{figure}
\begin{center}
\includegraphics[width=0.95\textwidth]{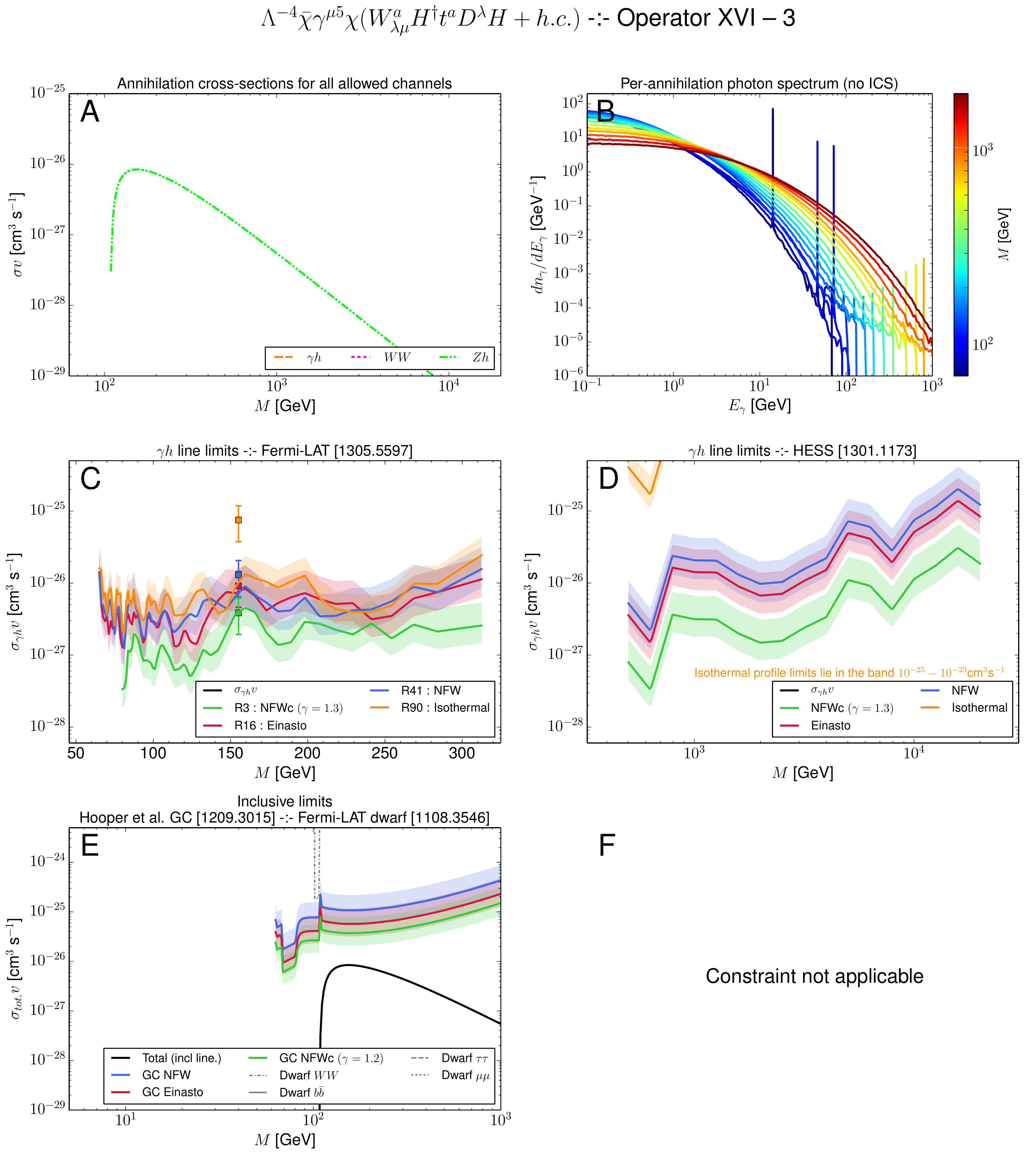}
\caption{\label{fig:TabXVI_OP3} Figure captions are provided in \sectref{panela} through \sectref{panelf}. The $\gamma h$ mode is $p$-wave suppressed and so $\sigma_{\gamma h}v$ is orders of magnitude below the range of cross sections shown in panels C and D.}
\end{center}
\end{figure} 
\clearpage

\begin{figure}
\begin{center}
\includegraphics[width=0.95\textwidth]{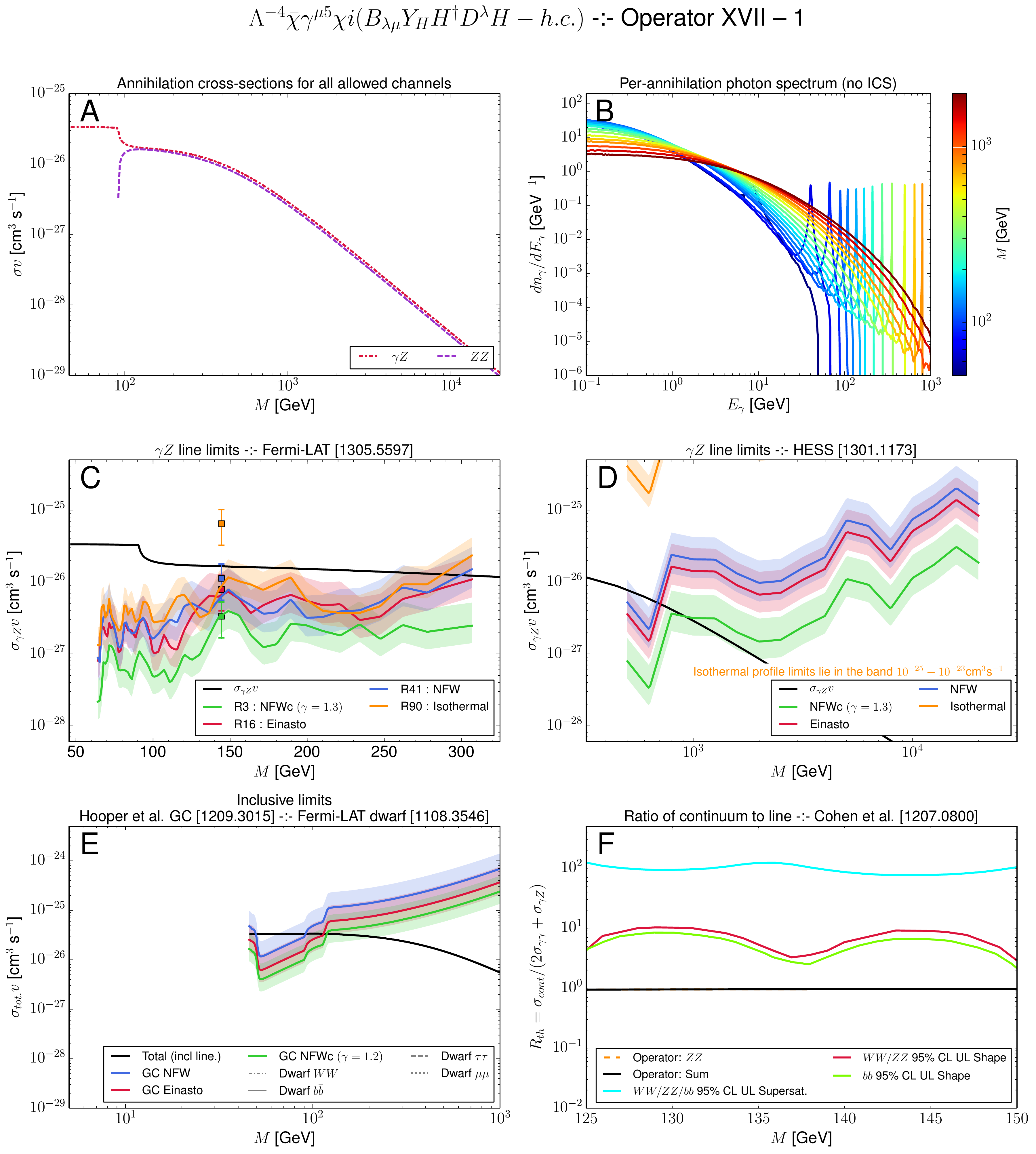}
\caption{\label{fig:TabXVII_OP1} Figure captions are provided in \sectref{panela} through \sectref{panelf}.}
\end{center}
\end{figure} 
\clearpage

\begin{figure}
\begin{center}
\includegraphics[width=0.95\textwidth]{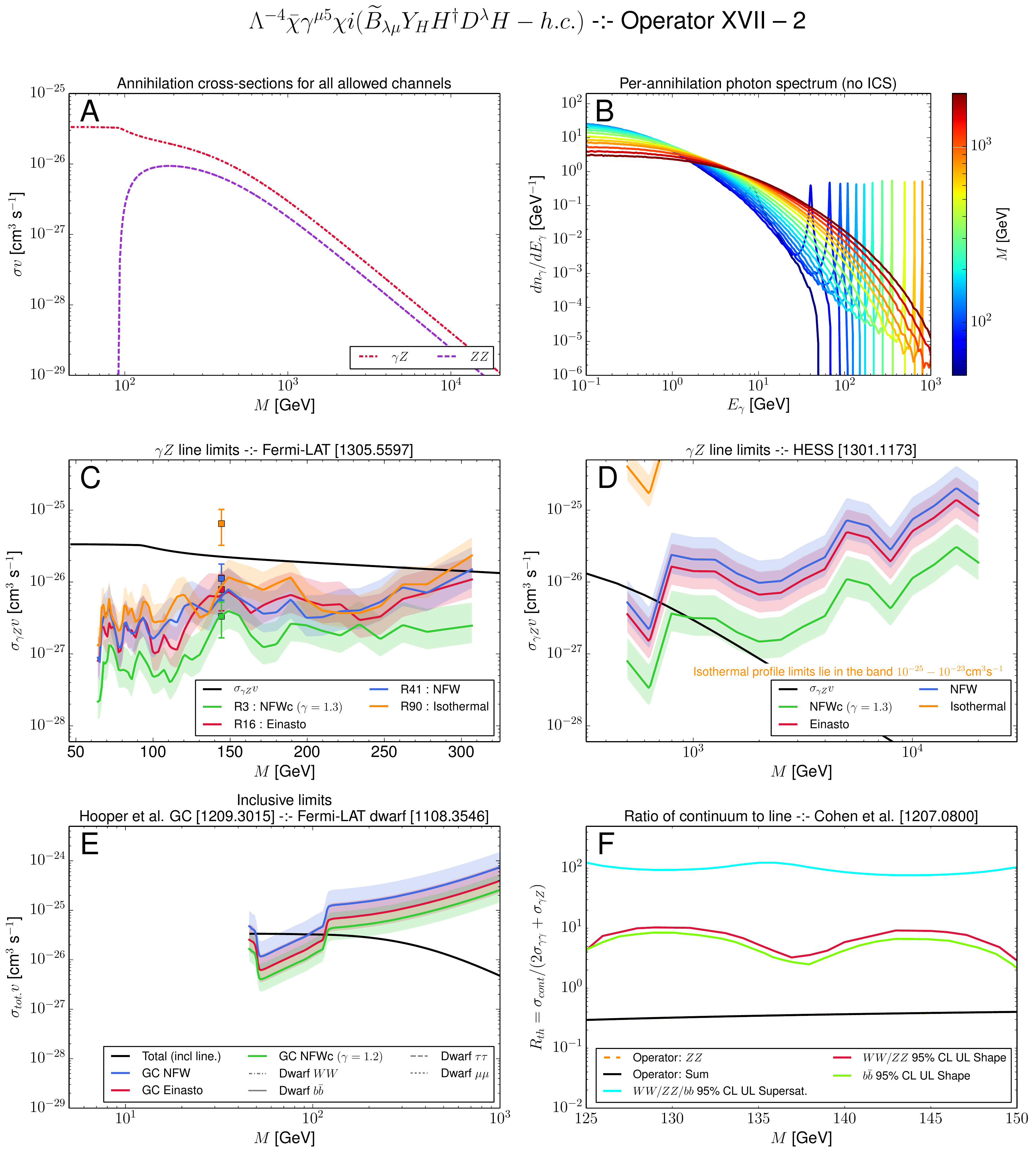}
\caption{\label{fig:TabXVII_OP2} Figure captions are provided in \sectref{panela} through \sectref{panelf}.}
\end{center}
\end{figure} 
\clearpage

\begin{figure}
\begin{center}
\includegraphics[width=0.95\textwidth]{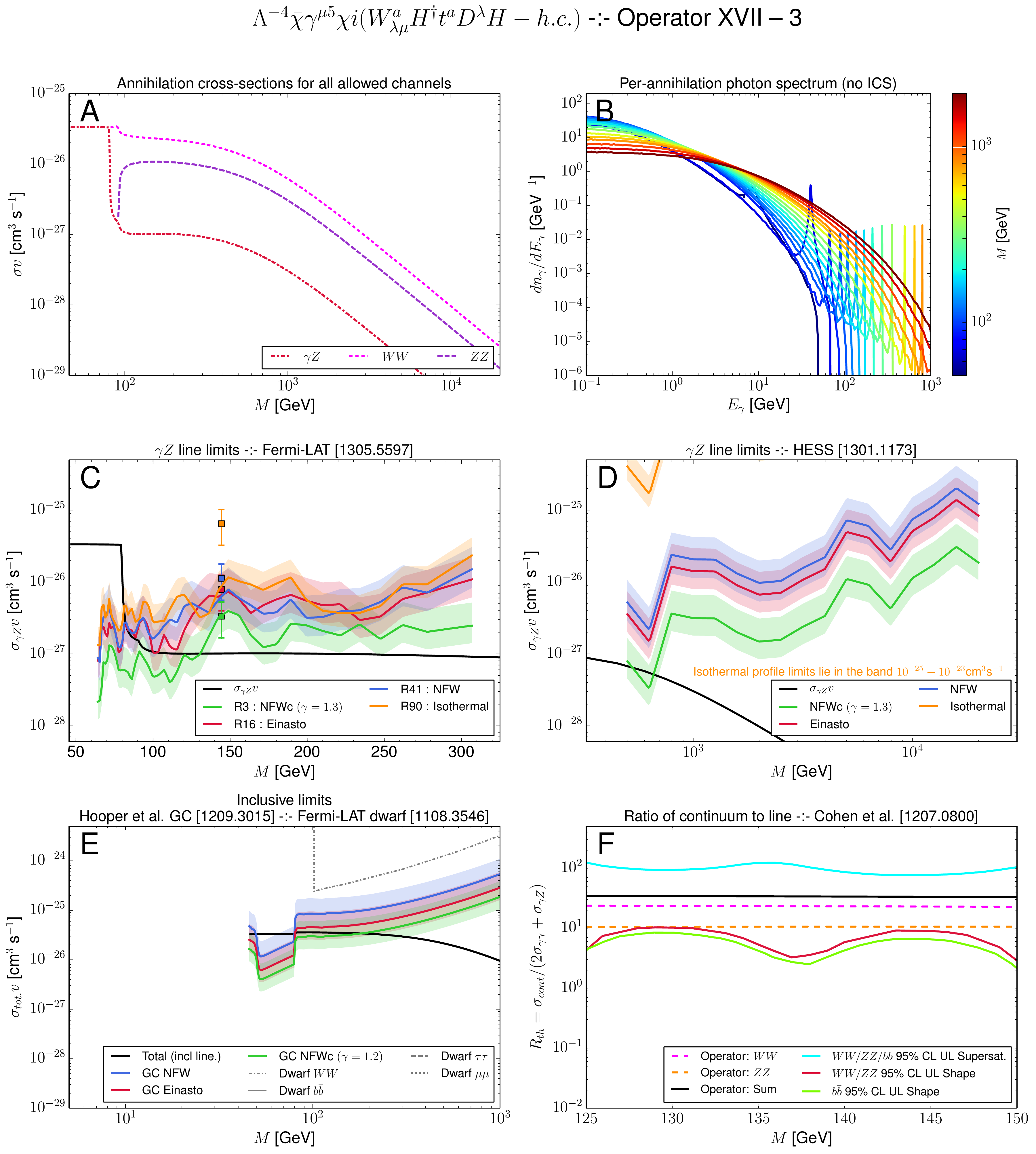}
\caption{\label{fig:TabXVII_OP3} Figure captions are provided in \sectref{panela} through \sectref{panelf}. This operator may be compatible with present experimental limits and capable of accounting for the 130 GeV photon line for some profile choices, although the line BR is somewhat small.}
\end{center}
\end{figure} 
\clearpage

\begin{figure}
\begin{center}
\includegraphics[width=0.95\textwidth]{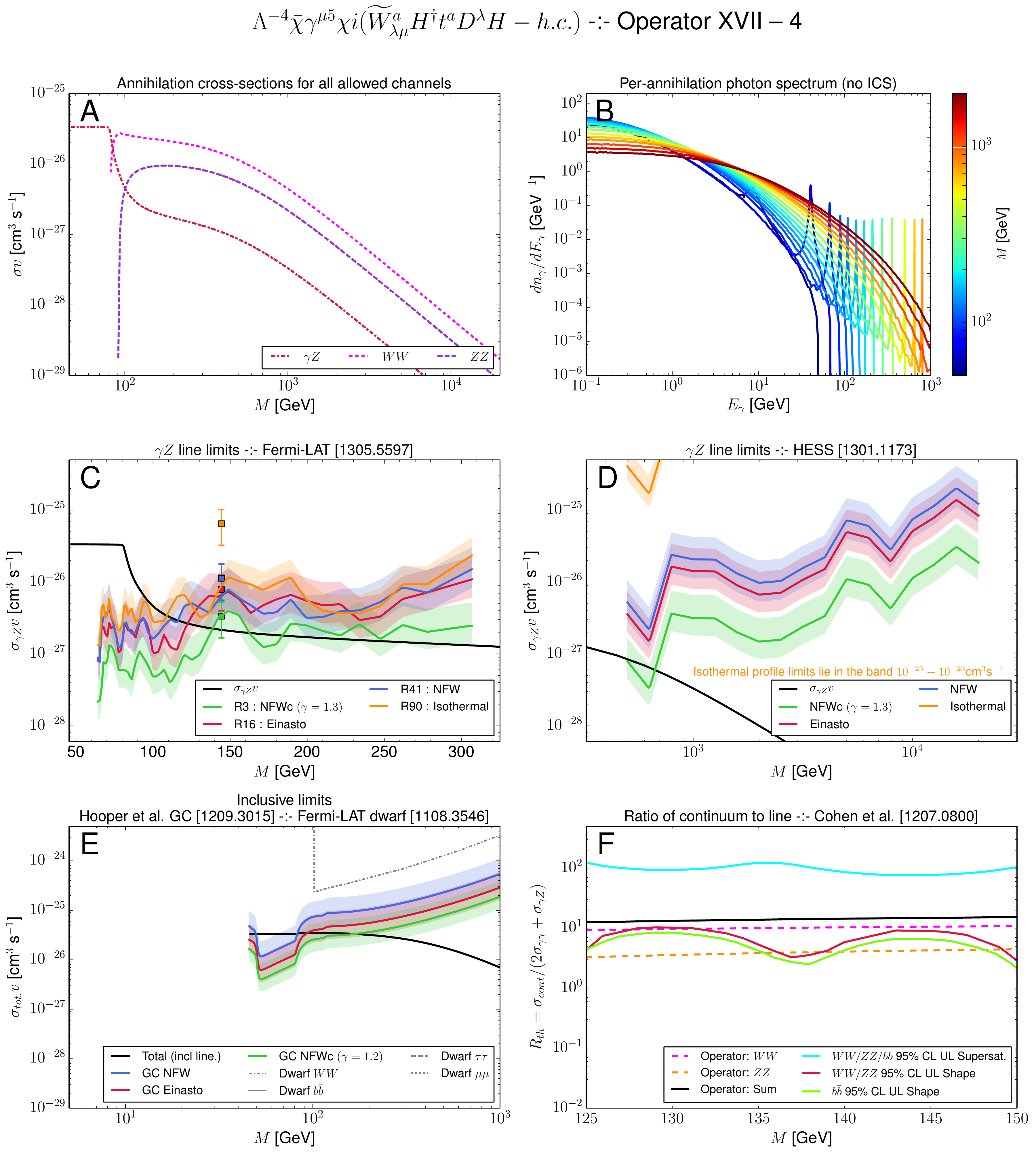}
\caption{\label{fig:TabXVII_OP4} Figure captions are provided in \sectref{panela} through \sectref{panelf}. This operator may be compatible with present experimental limits and capable of accounting for the 130 GeV photon line for some profile choices.}
\end{center}
\end{figure} 
\clearpage

\begin{figure}
\begin{center}
\includegraphics[width=0.95\textwidth]{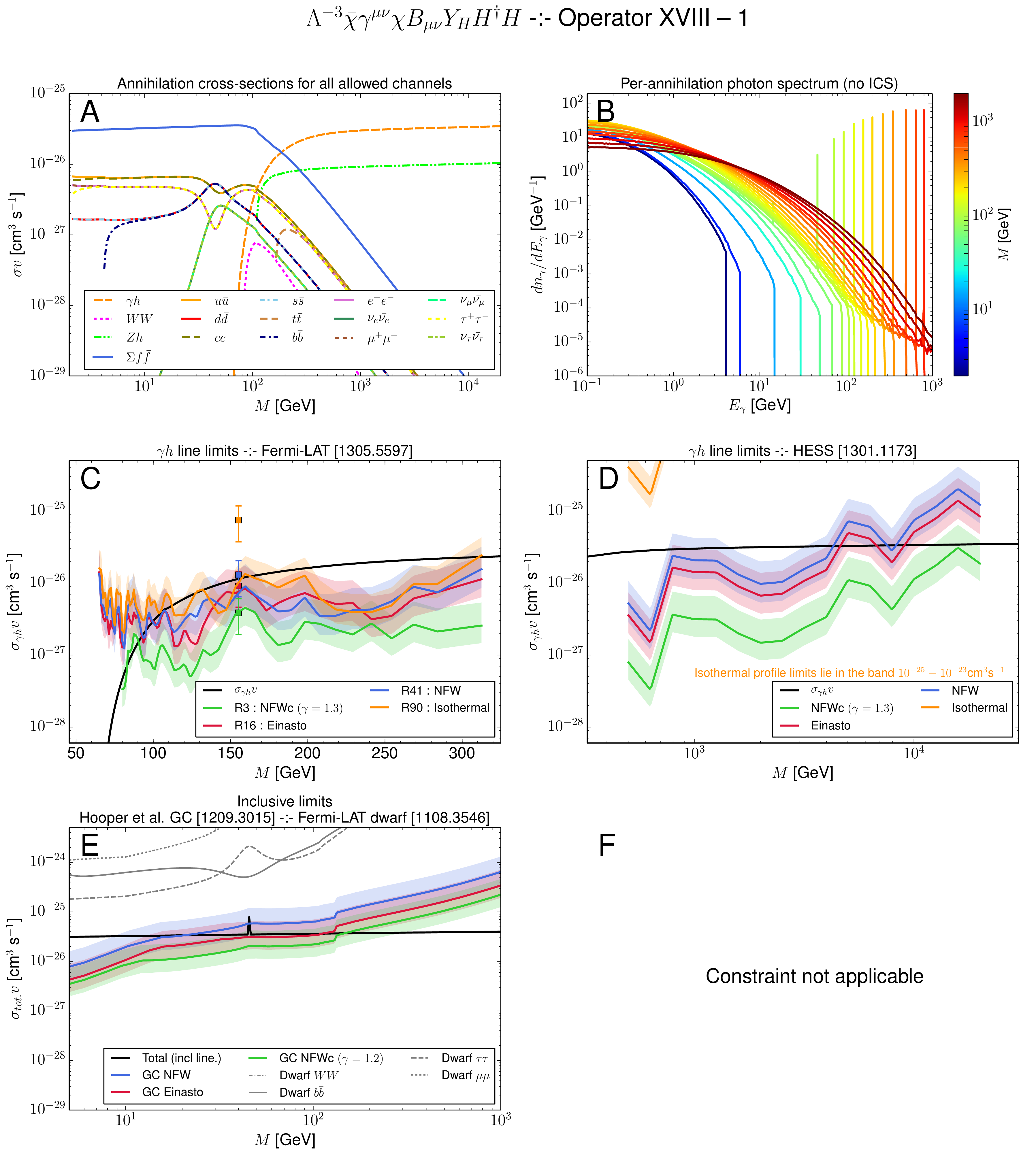}
\caption{\label{fig:TabXVIII_OP1} Figure captions are provided in \sectref{panela} through \sectref{panelf}. This operator may be compatible with present experimental limits and account well for the 130 GeV photon line for some profile choices; however, this operator gives rise to a magnetic dipole coupling of the DM and as such is excluded at 90\% confidence by the CDMS and XENON direct detection experiments \cite{Banks:2010uq,Fortin:2011fk} for $M$ from 10 GeV to \textit{ca.} 1 TeV.}
\end{center}
\end{figure} 
\clearpage

\begin{figure}
\begin{center}
\includegraphics[width=0.95\textwidth]{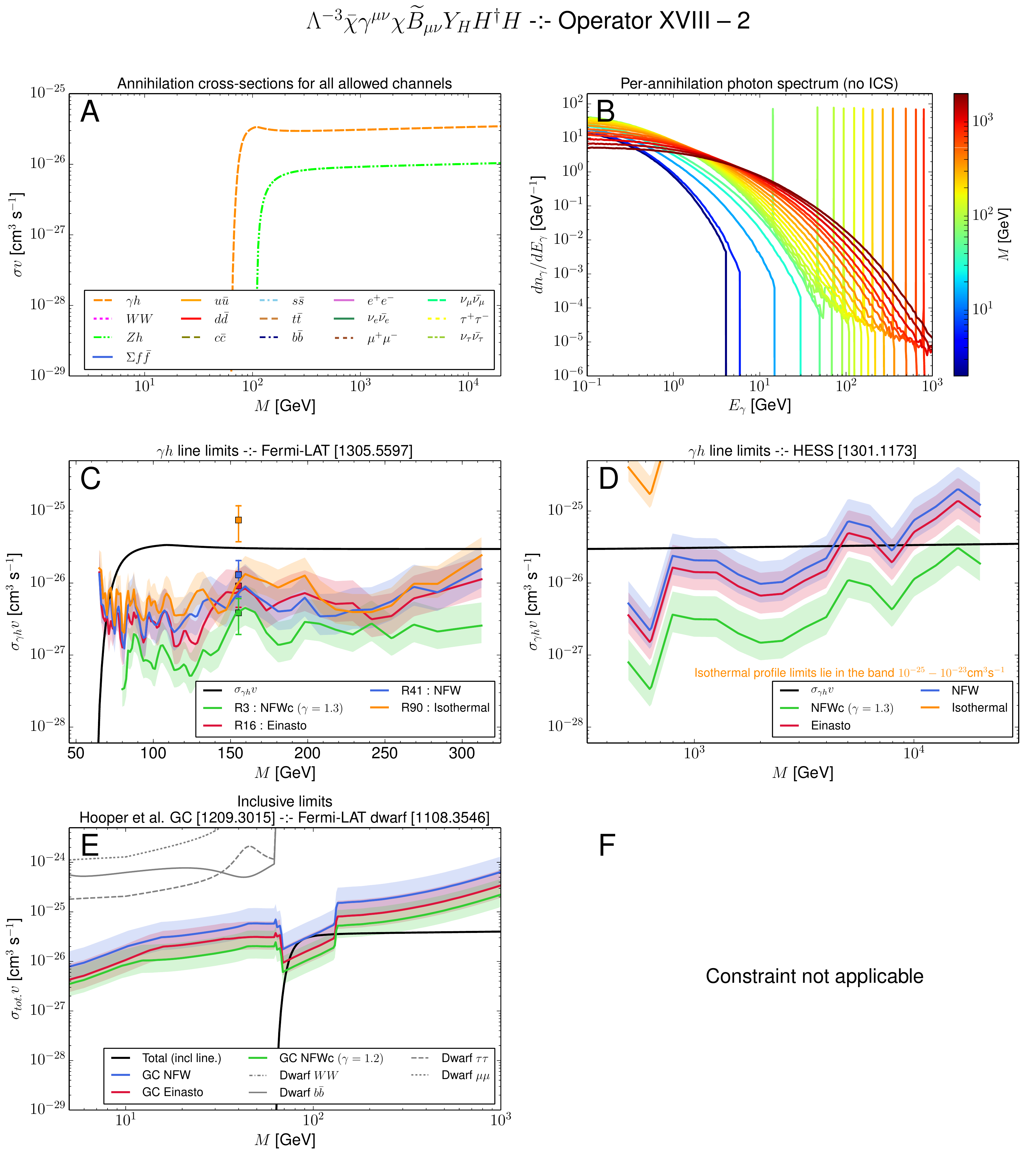}
\caption{\label{fig:TabXVIII_OP2} Figure captions are provided in \sectref{panela} through \sectref{panelf}. This operator gives rise to an electric dipole coupling of the DM and as such is excluded at 90\% confidence by the CDMS and XENON direct detection experiments \cite{Banks:2010uq,Fortin:2011fk} for the entire range of $M$ shown here.}
\end{center}
\end{figure} 
\clearpage

\begin{figure}
\begin{center}
\includegraphics[width=0.95\textwidth]{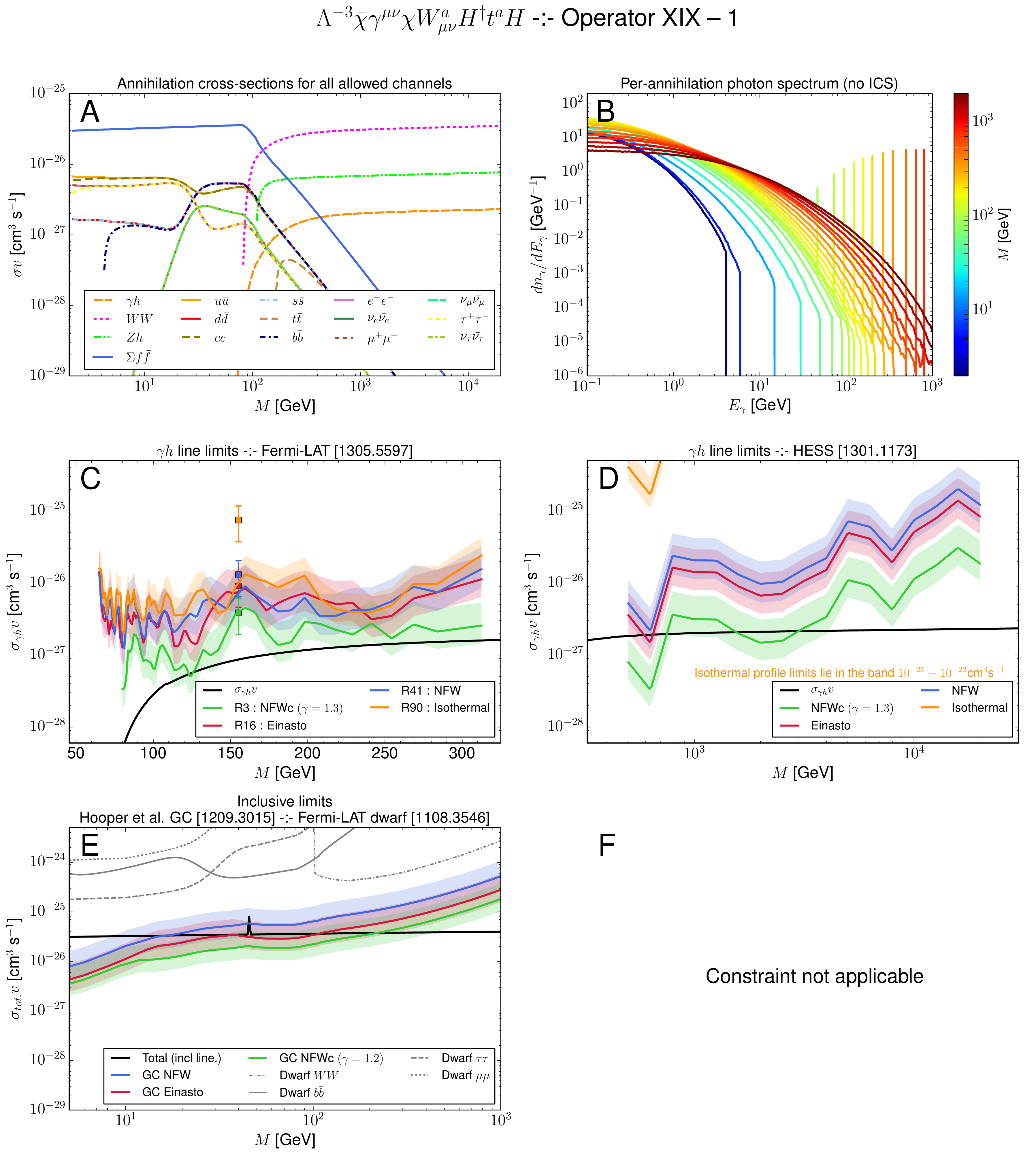}
\caption{\label{fig:TabXIX_OP1} Figure captions are provided in \sectref{panela} through \sectref{panelf}. This operator gives rise to a magnetic dipole coupling of the DM and as such is excluded at 90\% confidence by the CDMS and XENON direct detection experiments \cite{Banks:2010uq,Fortin:2011fk} for $M$ from 10GeV to \textit{ca.}\ 400GeV.}
\end{center}
\end{figure} 
\clearpage

\begin{figure}
\begin{center}
\includegraphics[width=0.95\textwidth]{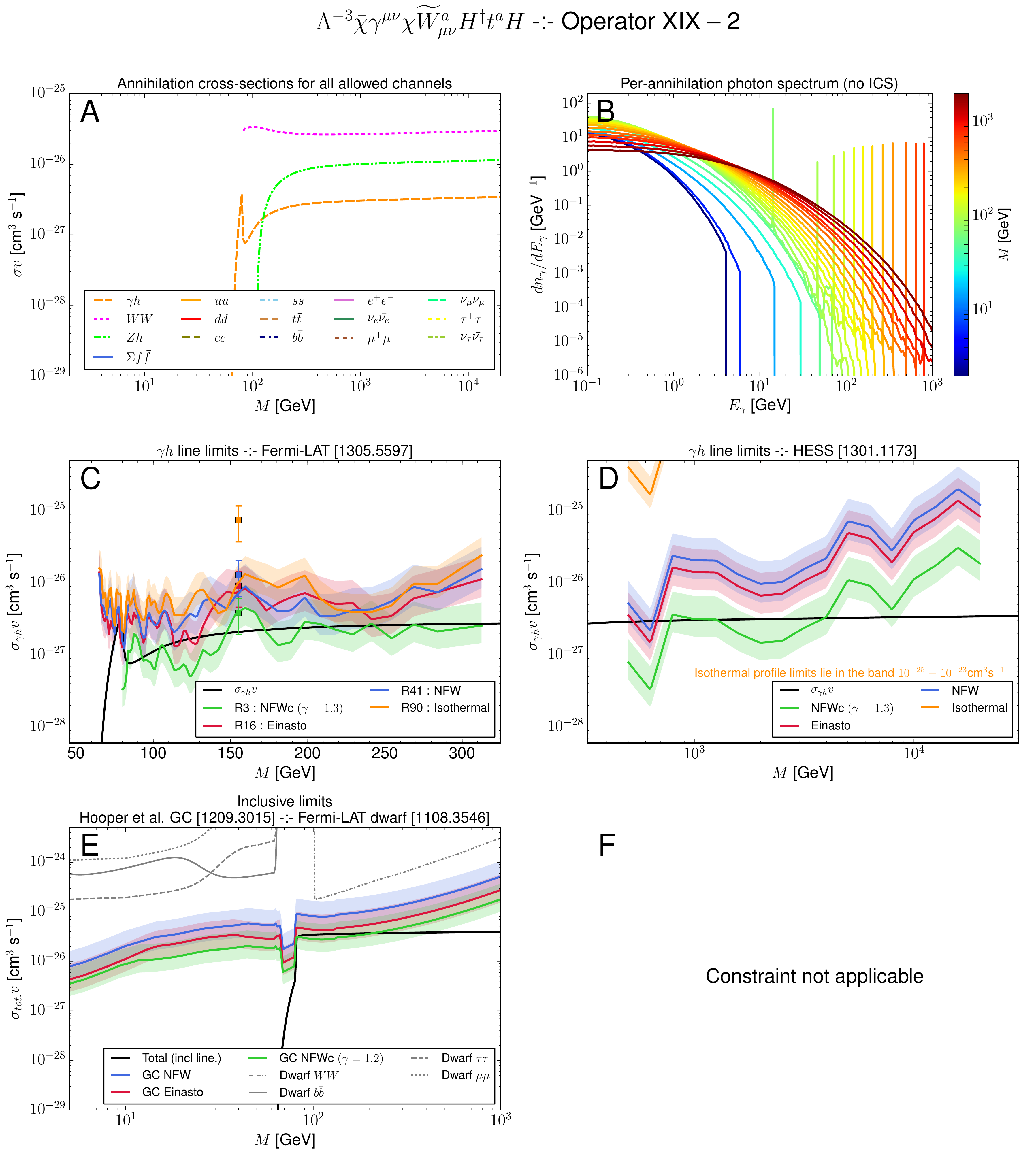}
\caption{\label{fig:TabXIX_OP2} Figure captions are provided in \sectref{panela} through \sectref{panelf}. This operator may be compatible with present experimental limits and capable of accounting for the 130 GeV photon line for some profile choices; however, this operator gives rise to an electric dipole coupling of the DM and as such is excluded at 90\% confidence by the CDMS and XENON direct detection experiments \cite{Banks:2010uq,Fortin:2011fk} for the entire range of $M$ shown here.}
\end{center}
\end{figure} 
\clearpage

\begin{figure}
\begin{center}
\includegraphics[width=0.95\textwidth]{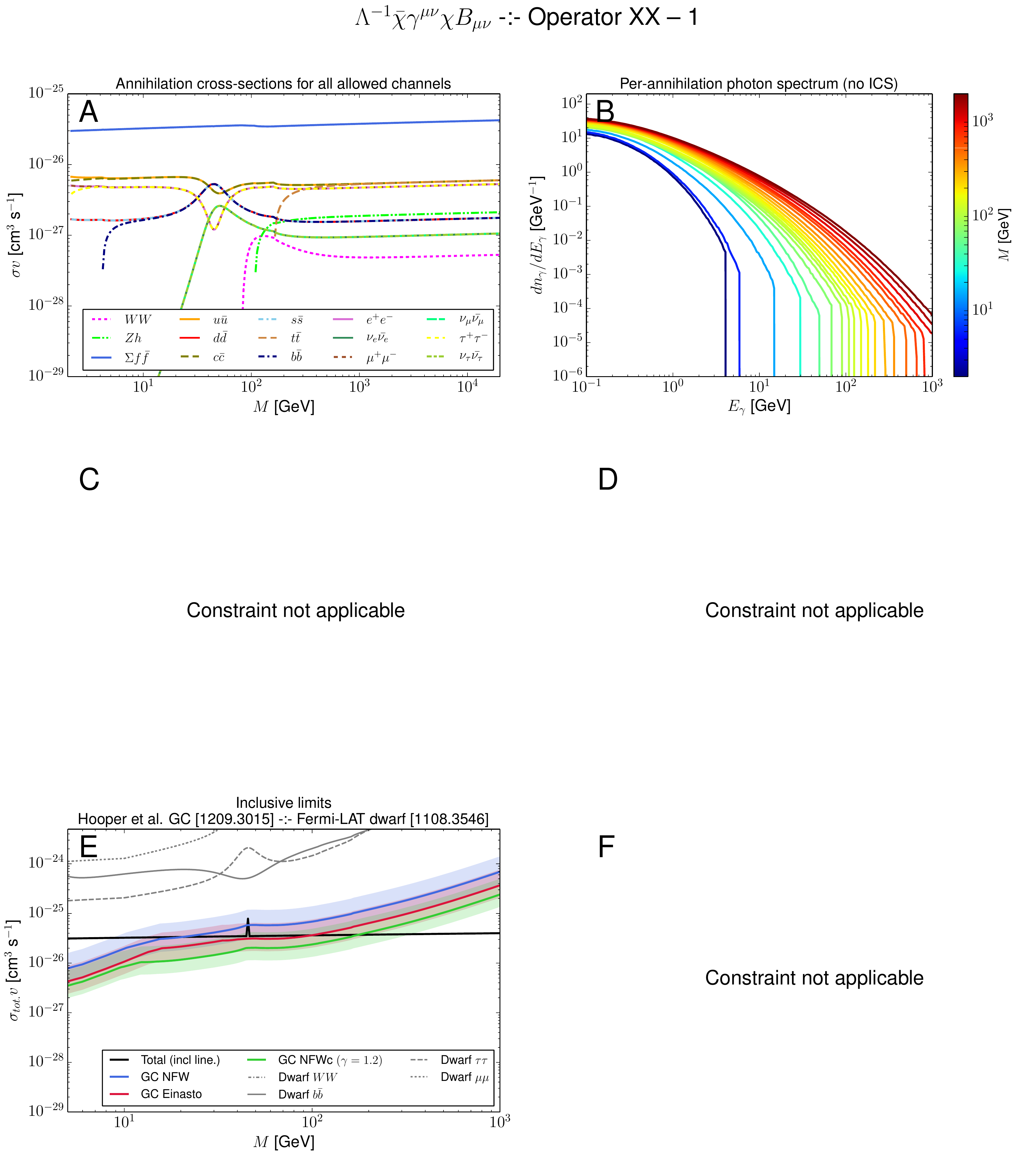}
\caption{\label{fig:TabXX_OP1} Figure captions are provided in \sectref{panela} through \sectref{panelf}.  This operator gives rise to a magnetic dipole coupling of the DM and as such is excluded at 90\% confidence by the CDMS and XENON direct detection experiments \cite{Banks:2010uq,Fortin:2011fk} for the entire range of $M$ shown here.}
\end{center}
\end{figure} 
\clearpage

\acknowledgments
This work was supported in part by the Kavli Institute for Cosmological Physics at the University of Chicago through grant NSF PHY-1125897 and an endowment from the Kavli Foundation and its founder Fred Kavli.  L.T.W.\ is supported by the DOE Early Career Award under grant de-sc0003930.  The work of E.W.K.\ is also supported by the Department of Energy.

\appendix

\section{Inverse Compton Scattering \label{app:ICS}}

In addition to prompt photons, there may be gamma rays resulting from
Inverse Compton Scattering (ICS) of any electrons/positrons injected
into the GC radiation field. ICS is most relevant for annihilation
modes which inject energy primarily in the form of $e^\pm$, in
particular the $e^+ e^-$ or $\mu^+ \mu^-$ modes.\footnote{For $\tau^+
\tau^-$, ICS is not as important since over $60\%$ of $\tau$ decays
go to hadronic final states.} For the heavy quark or diboson
primary final states relevant for most of the operators, the effect of ICS at
high energies is subdominant compared to the prompt gamma rays produced by
pion decay. Although the ICS contribution can dominate at lower photon
energies, the most constraining limits for continuum emission
typically come from the 10-100 GeV energy range, so we do not expect
significant changes to results for these operators.

There are some operators which have direct annihilation to charged
fermions.  However, even for these operators the charged lepton
annihilation modes are relevant only for masses below the thresholds
for diboson final states, and here the inclusive limits are already
constraining. For example, in the case of operator IX-2 
\figref{TabXI_OP2}, the fermion contributions are proportional to
$m_f^2$, so only the $\tau^+\tau^-$ annihilation mode is relevant, and
only for $M \lesssim 100$ GeV. For all other operators where fermion
final states are present the charged leptons typically comprise no
more than 40\% of the total charged fermion branching fraction, and
only are relevant below a few hundred GeV (with the exception of
Operator XX-1, \figref{TabXX_OP1}).  We thus do not expect a very
large correction to the limits we have set when the ICS photons are
included.

Here we demonstrate that including ICS will not change limits
significantly by giving a quantitative estimate for operators with
charged fermion final states. We will make use of the results of
\citet{PPPC4DMID}, which give pre-computed Green functions
$I_{\text{IC}}(E_s,E_\gamma,l,b)$ that allow one to convert from an
electron injection spectrum to an ICS photon spectrum, taking into
account the various $e^\pm$ energy-loss mechanisms in the GC.

It is important to note that the morphologies of the prompt and ICS
photon signals will not in general be exactly the same, since the ICS
signal depends on the galactic radiation fields as well as the energy
loss and diffusion of the injected electrons.  As detailed in
\sectref{Hooperlimits}, our primary source of inclusive spectrum
limits is \citet{Hooper1209} and the limits in that reference are
extracted for $E_\gamma \in [0.1,100]$ GeV from the upper limit of the
normalization of a prompt-photon signal template $\propto
\int_{\text{LOS}} \rho^2(r[s,l,b])ds$ smoothed over $0.5^\circ$
regions.  The limits set in that analysis are
therefore not applicable for photon signal components which do not
follow approximately the same morphology as prompt photons.  That
said, we find morphological differences give rise to only a factor of
a few variation in the ratio of the ICS to prompt signal when
considered over $\mathcal{O}(10^\circ)$ regions. We therefore take the
approximation of the limits in Ref.\ \cite{Hooper1209} being applicable to
the combined ICS and prompt photon signals.

\begin{figure}[b]
\includegraphics[width=0.47\textwidth]{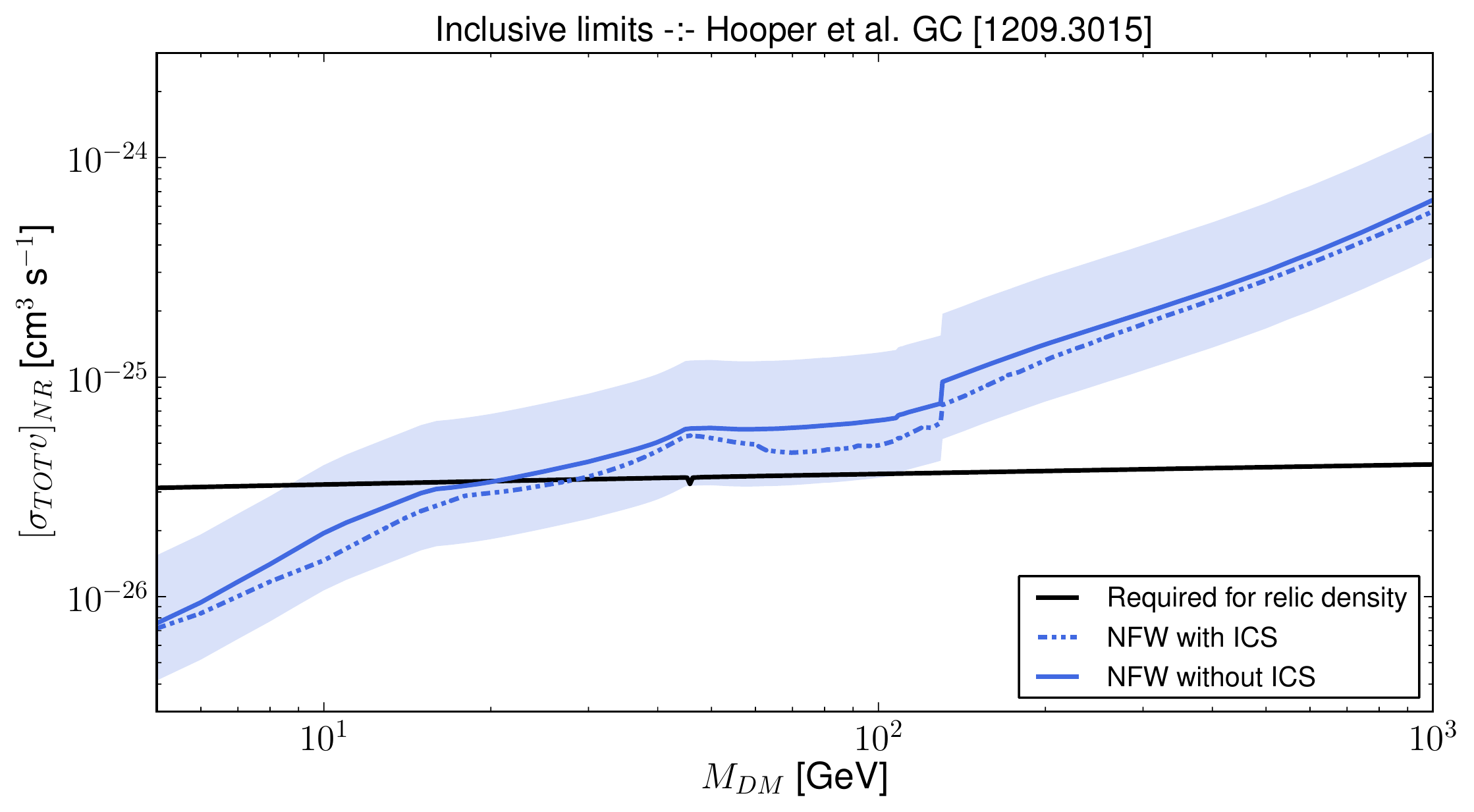}
\includegraphics[width=0.47\textwidth]{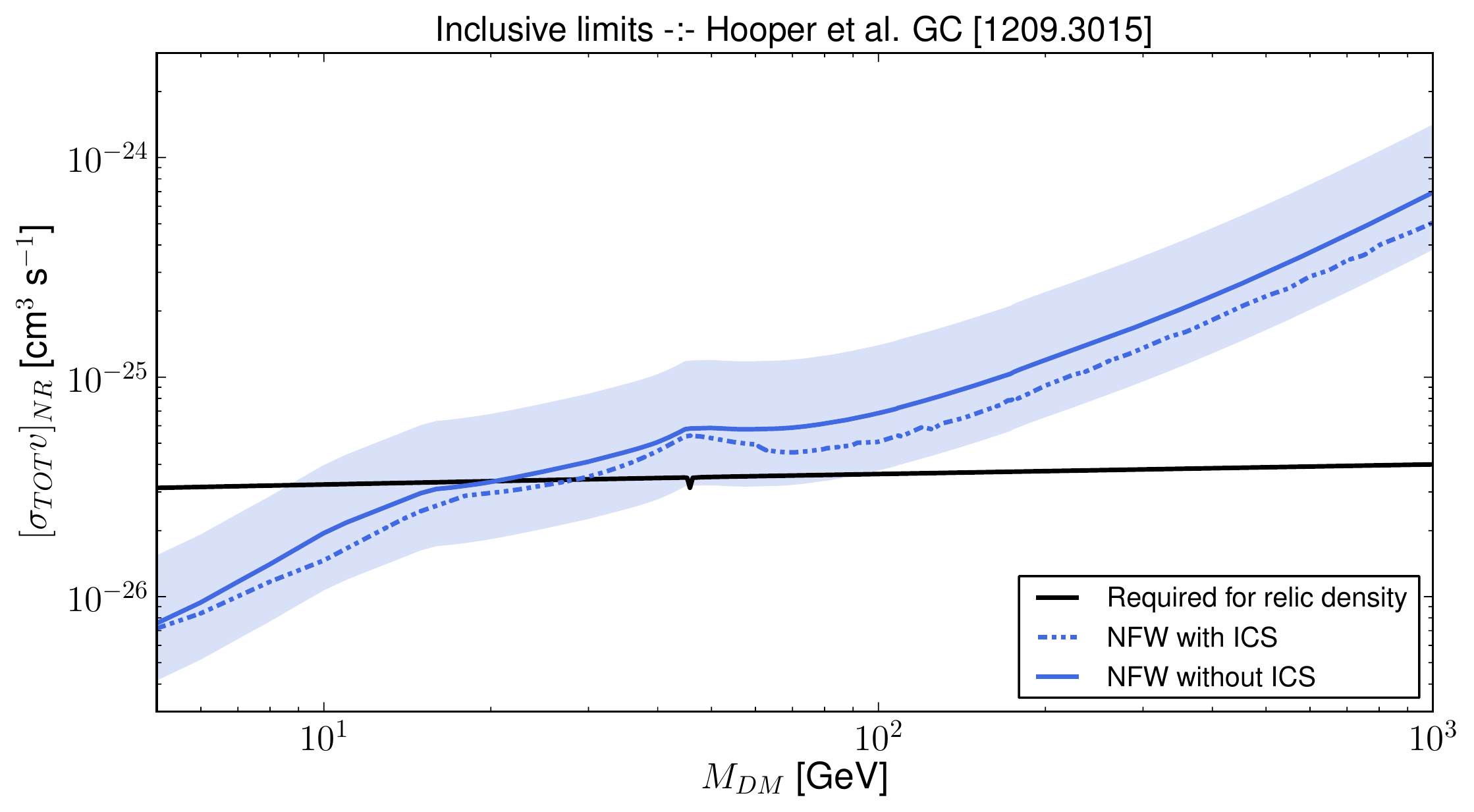}
\caption{\label{fig:effect_on_limits_TabXVIII_OP1} An example of the
  impact on the inclusive limits when ICS is included (subject to the
  approximations discussed in the text). These results are for operators XVIII-1 (left panel) and XX-1 (right panel) and are representative of the most dramatic
  effect seen for any operator: even so, the effect on the limits is
  well within the astrophysical uncertainties.} 
\end{figure}

Since we are not however reproducing the template subtraction
analysis, we must make a choice for the ROI over which to extract the
effective ICS spectrum in order to make a direct comparison with the
cross section limits in Ref.\ \cite{Hooper1209}. Motivated by the cuspy
nature of the expected signal morphologies, and the fact that the
photon signal used in setting the limits in that reference is
mostly contained within a region of about $3^\circ$ around the GC, we
chose to utilize a $5^\circ$ radius circular region centered
on the GC in extracting the effective ICS spectra. Furthermore, at
very high injection energies, the high-energy ICS signal is more constrained to
the galactic center, since electrons must be at high energy to create
high energy photons via single Klein-Nishina-regime Compton scatters,
and this only happens near the point of production as the electrons
lose energy as they propagate out of the GC region.

We integrate the ICS signal over this ROI and normalize by the line of
sight integral for that region:
\begin{align}
\lb.\frac{dN^{\text{ICS}}_\gamma}{dE_\gamma} \rb|_{\text{ROI}} & =  \frac{1}{E_\gamma^2} \times \lb[ \int_{m_e}^{M} \!\!\!\!\! dE_s \ \lb(\sum_f \text{Br}_f \frac{dN^{f\text{, inj. }}_{e^\pm}}{dE_s}\rb) \ \int_{\text{ROI}} d\Omega\  I_{\text{IC}}(E_s,E_\gamma,l,b)  \rb] \times \lb[ \int_\text{LOS-ROI}  d\Omega\frac{ds}{r_\odot} \lb(\frac{\rho\lb(r\rb)}{\rho_\odot}\rb)^2 \rb]^{-1}.
\label{eq:dNdEICS}
\end{align}
This gives an effective ROI-averaged per-annihilation spectrum of
up-scattered ICS photons, in terms of the constituent electron
injection spectra $dN^{f \text{, inj.}}_{e^\pm} / dE_s$, for final
state $f$, and where $E_s\in [m_e,M]$ is the energy of the injected
electron, $E_\gamma$ is the energy on the up-scattered ICS photon and
$l,b,s,r$ are coordinates and distances as defined in the main text
A fairly conservative estimate for the systematic uncertainty here is
$\pm50$\% due to the choice of ROI. Note finally that the diffusion setups MIN, MED, MAX (discussed in \citet{PPPC4DMID}) give largely similar results for an ROI of this size near the GC.

Explicitly checking the shift in the limits set using the results of
Hooper \textit{et al.}, we find that the limits are strengthened by no more
than a factor of 1.4 at any $M$ for any operator, which is well within
the astrophysical uncertainty due to the normalization of the halo
profile. The resulting effect on the limit is shown for a
representative case in Fig.\ \ref{fig:effect_on_limits_TabXVIII_OP1}.

\section{Estimate of systematic uncertainties \label{app:systematics}}

It is useful to have an approximate knowledge of the size of various systematic uncertainties which impact our results and which we have not explicitly included in the main body of the paper. We summarize the systematics in \tabref{systematics}. It is relatively clear that the astrophysical (halo profile normalization) uncertainties dominate the systematics, but depending on the result and mass range in question, it may be prudent to take the total systematic uncertainty as something up to twice as large as the astrophysical uncertainty.

\renewcommand{\arraystretch}{1.5}
\begin{table}
\caption{ \label{tab:systematics} Estimated maximum systematic uncertainties from various effects, quoted as the percentage variation in the $\sigma v$ limit:  $f \equiv 100 \times  \lb(\lb[ \sigma v \rb]^{\text{95\% CL UL}}_{\text{effect accounted for}} - \lb[ \sigma v \rb]^{\text{95\% CL UL}}_{\text{limit as presently set}} \rb)/ \lb[ \sigma v \rb]^{\text{95\% CL UL}}$ (so $f<0$ means a more stringent limit with the effect properly accounted for, while $f>0$ means the limit as currently set is too stringent). Values estimated by us are marked with a $^\star$; values taken from the relevant literature source are not marked. For comparison, the typical variation between the most conservative normalization for the most conservative profile choice and the most aggressive normalization for the cuspiest profile is about an order of magnitude, so relative to the mid-point the astrophysical uncertainties can cause the limit to vary by about $+100\%$ / $-75\%$.}
\begin{ruledtabular}
\begin{tabular}{p{3cm}p{10cm}p{3cm}}
Result(s) affected & \multicolumn{1}{c}{Description of effect} & Variation in result $f$ \\ \hline
All & Halo rescaling or $J$ factor computation & $\pm 5\% ^\star$ \\ \hline
\citet{FERMI1305} line limits & Variations in the fitted signal strength due to the finite spacing between tested values of the scanned line peak energy and estimate of the energy resolution & $+12\%$ / $-7\%$	\\
\citet{FERMI1305} line limits & Variations in the effective exposure used in converting the fitted number of 
photons in a line to a flux & $\pm 16\%$	\\
\citet{FERMI1305} line limits	& Fake or masked signals: for low WIMP masses (a few tens of GeV), up to roughly the size of the statistical uncertainties, which are about a factor of 2. & $+100\%$ / $-50\%$ for low masses, dropping rapidly as mass increases		\\
\citet{FERMI1305} line limits	& Correctly accounting for the $\gamma Z$ photon linewidth & $+7\%$	\\
\citet{FERMI1305} line limits	&
Mis-modelling of the photon line shape for the cases with both $\gamma\gamma$ and $\gamma Z$ peaks in the vicinity of $M\sim 80$GeV and $M\sim160$ GeV.	&
$+10\% ^\star$	\\ \hline \citet{HESS1301} line limits	& Overall systematics estimate	& 
$+50\%$ / $-50\%$	\\ \hline
\citet{Hooper1209} inclusive limits & 
Subtraction of additional background associated with Galactic Ridge or central gamma-ray point source &
$-50\%$ \\ \citet{Hooper1209} inclusive limits	&
Uncertainty in the exact value of the ``fudge-factor'' of 2.8 necessary to resolve the discrepancy noted in the main body	& $\pm10\% ^\star$ \\
\citet{Hooper1209} inclusive limits &
Including ICS photons in setting the limit (estimated using the $5^\circ$ ROI normalization;
the systematic uncertainty on the normalization of ICS spectrum itself is about $\pm 50\%$ owing to our approximate treatment) & $-40\% ^\star$	\\ \hline
\citet{FERMI1108} continuum limits &
Excluding from the limit computation the two most uncertain spherical dwarf measurements &$+50\%$ \\ \hline
\citet{Weniger1204} line fit & Overall systematics estimate & $+15\%$ / $-20\%$
\end{tabular}
\end{ruledtabular}
\end{table}

\begin{table}
\caption{\label{tab:Wacker_Jfactors} The $J$ factors for the
  $1^\circ-3^\circ$ annular ROI defined in \citet{Wacker1207}. Unless
  otherwise indicated, these results assume the central normalization
  and parameter values from \tabref{norms}.}
\begin{ruledtabular}
\begin{tabular}{cd}  
Halo Profile & \multicolumn{1}{c}{$J$ factor [10$^{21}$ GeV$^{2}$ cm$^{-5}$]} \\ \hline
NFWc ($\gamma = 1.3$)	&	32.50		\\
Einasto				&	15.05		\\
NFW					&	9.02			\\
Isothermal				&	0.361		\\ \hline
Einasto with $R_\odot = 8.5$ kpc and $\rho_\odot = 0.4$ GeV/cm$^3$	&	21.41
\end{tabular}
\end{ruledtabular}
\end{table}

\renewcommand{\arraystretch}{1.0}

\section{Exposure for line fit in annular region}
\label{app:exposure}

In \sectref{Cohen_line} we describe the application of the line-fit
results of \citet{Wacker1207} to operators with both $\gamma \gamma$
and $\gamma Z$ lines. However, Cohen, \textit{et al.}\ supplied only fitted
values of $N_{\gamma\gamma} + N_{\gamma Z}$. In order to apply their
results, a necessary ingredient is the exposure: the mission-time
integrated effective observing area multiplied by the observing time,
averaged over the ROI. In this Appendix we summarize our calculation
of the exposure for their reported data set.

We have reproduced their data-extraction using the Fermi ScienceTools
\texttt{v9r31p1} software\footnote{Available online at
  \url{http://fermi.gsfc.nasa.gov/ssc/data/analysis/software/}.} with
the publicly available Fermi-LAT weekly data files.\footnote{Available
  online at \url{http://fermi.gsfc.nasa.gov/ssc/data/access/lat/}.} To
be explicit, we have considered data on the time interval
239557447---356400002 (in Mission Elasped Time), with photon energy
range 5-200 GeV in an annular region centered on the GC with inner and
outer radii of $1^\circ$ and $3^\circ$ respectively. We use the
\texttt{P7\_V6\_ULTRACLEAN} data set and filter the data using option
2 as recommended for a diffuse analysis.\footnote{Described at
  \url{http://fermi.gsfc.nasa.gov/ssc/data/analysis/documentation/Cicerone/Cicerone_Likelihood/Exposure.html}}
Histogramming the photon counts in the annular ROI we reproduce almost
exactly\footnote{The only discrepancies seem to be associated with the
  exact boundaries of the energy bins or to a slightly different
  choice of time interval.} the counts presented in Tables II and III
of Ref.\ \cite{Wacker1207}.

We compute the full-sky exposure map as a function of photon energies
near $125-150$ GeV, and subselect the annular ROI from the full-sky
exposure map (exposure values in angular bins). The ROI-averaged exposure is given by
\begin{align}
\mathcal{E}_{\text{ROI}} = \frac{1}{\Delta \Omega} \int_{\Delta \Omega} \mathcal{E}(\Omega) d\Omega \approx \frac{1}{\Delta \Omega} \sum_i \mathcal{E}_i \cos b_i \Delta l \Delta b,
\label{eq:E_ROI}
\end{align}
where $\Delta \Omega = 7.65\eten{-3}$sr, $\mathcal{E}_i$ is the
exposure value in angular bin $i$, and the sum runs over all angular
bins (which have angular size $\Delta l \times \Delta b = 0.1^\circ
\times0.1^\circ$). Strictly speaking, one should weight the exposure
by the signal morphology before performing the integration in Eq.\
\eqref{eq:E_ROI}; however, the very small size of the ROI together
with the almost uniform exposure over the ROI make neglecting this
subtlety a good approximation. We find that the exposure does not vary
significantly over the energy range $125-150$ GeV, and takes a value of
$\mathcal{E}_{\text{ROI}} = 1.05\eten{11}$cm$^2$s.

The photon flux corresponding to the line is given by
$({N_{\gamma\gamma} + N_{\gamma Z}})/{ \mathcal{E}_{\text{ROI}}}$. We
find the total normalization $N_{\gamma\gamma} + N_{\gamma Z}$ for
each operator using a 2D polynomial interpolation of the contour map
given in Fig.\ 2 of Ref.\ \cite{Wacker1207}, extracting the value of
$N_{\gamma\gamma} + N_{\gamma Z}$ on the operator-specific curves
indicated in \figref{Wacker_line}. Finally, we compute the $J$ factors
necessary to extract cross sections from these fluxes given the
central normalization values in \tabref{norms}; the results are in
\tabref{Wacker_Jfactors}.
Applying Eq.\ \eqref{eq:cohenlinelimit}, we thus interpret the fit
results in terms of the annihilation cross section to lines,
$(2\sigma_{\gamma\gamma} + \sigma_{\gamma Z}) v$; we plot this
quantity in our result plots on the regions of $M$ parameter space
where the delta-log-likelihood for the double line fit is $\leq
3\sigma$, $\leq 2\sigma$ and $\leq 1\sigma$ (see
\figref{Wacker_line}).

To check our work, we compare with the analysis of a single line
presented in Ref.\ \cite{Weniger1204}. We compute the
value\footnote{The coefficient of 8 rather than 16 is correct here; we
  are comparing to a limit set under the assumption of self-conjugate
  DM.} of $\sigma_{\gamma\gamma}v = (8\pi M^2/2J)   N_{\gamma\gamma}/ \mathcal{E}_{\text{ROI}} $ assuming
$\theta_{\gamma z/\gamma\gamma} = 0$, $N_{\gamma\gamma} = 33$ and $M =
130$ GeV, along with the Einasto profile $J$ factor at the bottom of
\tabref{Wacker_Jfactors} which has been computed using the indicated
parameter values taken from Ref.\ \cite{Weniger1204}. We obtain
$\sigma_{\gamma\gamma}v = 3.2\eten{-27}$cm$^3$s$^{-1}$, which is a
factor of 3 larger than the central value ($\sigma_{\gamma\gamma} v =
1.1\times 10^{-27}$cm$^3$s$^{-1}$) reported in
Ref.\ \cite{Weniger1204}. There is thus some tension between these two
analyses.

\bibliography{bib_fklw}

\end{document}